\documentclass[journal=jacsat,manuscript=article]{achemso}
\setkeys{acs}{doi = true}

\usepackage{mathtools}

\usepackage[version=3]{mhchem} 

\usepackage{aligned-overset}
\usepackage{xcolor}
\usepackage[hidelinks]{hyperref}
\usepackage{xr-hyper}
\newcommand{\unquart}{\frac{1}{4}}

\newcommand{\ts}[2]{t^{#1}_{#2}}
\newcommand{\td}[4]{t^{#1#2}_{#3#4}}

\newcommand{\tXs}[2]{\overset{\textsc{x}}{t}{}^{#1}_{#2}}

\newcommand{\tYs}[2]{\overset{\textsc{y}}{t}{}^{#1}_{#2}}

\newcommand{\tZs}[2]{\overset{\textsc{z}}{t}{}^{#1}_{#2}}

\newcommand{\tbXs}[2]{{\overset{{\textsc{x}}}{\bar{t}^{#1}_{#2}}}}

\newcommand{\tbYs}[2]{{\overset{{\textsc{y}}}{\bar{t}^{#1}_{#2}}}}

\newcommand{\tbZs}[2]{{\overset{{\textsc{z}}}{\bar{t}^{#1}_{#2}}}}

\newcommand{\bra}[1]{\langle #1  | }
\newcommand{\ket}[1]{  |  \text{#1} \rangle}

\author{Xiang Yuan}
\affiliation[PhLAM]{Univ. Lille, CNRS, UMR 8523 - PhLAM - Physique des Lasers Atomes et Molécules, F-59000 Lille, France}
\alsoaffiliation[Vrije Universiteit Amsterdam]
{Department of Chemistry and Pharmaceutical Sciences, Faculty of Science, Vrije Universiteit Amsterdam, 1081 HV Amsterdam, The Netherlands}

\author{Loïc Halbert}
\affiliation[PhLAM]{Univ. Lille, CNRS, UMR 8523 - PhLAM - Physique des Lasers Atomes et Molécules, F-59000 Lille, France}

\author{Lucas Visscher}
\affiliation[Vrije Universiteit Amsterdam]
{Department of Chemistry and Pharmaceutical Sciences, Faculty of Science, Vrije Universiteit Amsterdam, 1081 HV Amsterdam, The Netherlands}

\author{André Severo Pereira Gomes}
\email{andre.gomes@univ-lille.fr}
\affiliation[PhLAM]{Univ. Lille, CNRS, UMR 8523 - PhLAM - Physique des Lasers Atomes et Molécules, F-59000 Lille, France}

\title[An \textsf{achemso} demo]
  {A Comparison of Relativistic Coupled Cluster and Equation of Motion Coupled Cluster Quadratic Response Theory}

\abbreviations{IR,NMR,UV}
\keywords{American Chemical Society, \LaTeX}

\begin{document}

\begin{abstract}
We present the implementation of relativistic coupled cluster quadratic response theory (QR-CC), following our development of relativistic equation of motion coupled cluster quadratic response theory (QR-EOMCC) [X.\ Yuan \textit{et al.}, \textit{J.\ Chem.\ Theory Comput.} \textbf{2023}, \textit{19}, 9248--9259]. These codes, which can be used in combination with relativistic (2- and 4-component based) as well as non-relativistic Hamiltonians, are capable of treating both static and dynamic perturbations for electric and magnetic operators. We have employed this new implementation to revisit the calculation of static and frequency-dependent first hyperpolarizabilities of hydrogen halides (HX, X=F-Ts) and the Verdet constant of heavy noble gas atoms (Xe, Rn, Og) {and of selected hydrogen halides (HF to HI)}, in order to investigate the differences and similarities of QR-CC and the more approximate QR-EOMCC. Furthermore, we have determined the relative importance of scalar relativistic effects and spin-orbit coupling to these properties, through a comparison of different Hamiltonians, and extended our calculations to superheavy element species (HTs for hyperpolarizabilities, Og for the Verdet constant). Our results show that as one moves towards the bottom of the periodic table, QR-EOMCC can yield rather different results (hyperpolarizabilities) or perform rather similarly (Verdet constant) to QR-CC. These results underscore the importance of further characterizing the performance of QR-EOMCC for heavy element systems.
\end{abstract}

\section{Introduction}
\label{intro}

External perturbations such as those induced by electric or magnetic fields allow us to probe the electronic structure of atoms, molecules and materials, with different perturbations providing us with complementary information about the system~\cite{norman_principles_2018,Jaszuski2015}. The first-order (linear) responses to such a perturbation are particularly sought-after quantities, since from them one can obtain information about electronically or vibrationally~\cite{bishop_molecular_1990} excited states though processes involving a single photon, or magnetic properties such as NMR shieldings or indirect spin-spin coupling constants~\cite{helgaker1999ab,vaara2007theory,helgaker2008quantum} to name but a few. 

Properties arising from the interaction with two or more photons~\cite{papadopoulos_non-linear_2006,he2008multiphoton} at a time are also of great interest be it for technological applications in non-linear optics~\cite{Norman2006,boyd2020nonlinear} or for speciation through two-photon absorption~\cite{cronstrand_multi-photon_2005}, as the differences in selection rules compared to one-photon processes allow for probing different states.

Theoretical models are indispensable to interpret experimental results, and the last fifty years have seen the development~\cite{helgaker_recent_2012,helgaker_molecular_2014} of approaches based on \textit{ab initio} molecular electronic structure theory gain popularity in applications, thanks in no small part to the development of response theory~\cite{norman_perspective_2011,Pedersen2015,Jaszuski2015} and its application to different models such as multi-configurational self-consistent field theory~\cite{dalgaard1980time,olsen_linear_1985,Hettema1992}, density functional theory~\cite{Casida1996}, and coupled cluster theory~\cite{Monkhorst2009, dalgaard_aspects_1983,Koch1990a,Koch1990b,Koch1994,kobayashi1994calculation,kondo_orthogonally_1995,piecuch_property_1995}. The use of the time-averaged quasienergy formalism~\cite{langhoff1972aspects,rice1991calculation,sasagane1993higher,Koch1994,Christiansen1995a,Christiansen1995b,Christiansen1996,Hattig1997,coriani_coupled_1997,Christiansen1998,Gauss1998,kauczor2013communication,khani_uv_2019,hattig_cc2_2000} has provided not only a unified framework, with which to derive response for both variational and non-variational approaches, but also a way to enable the development of embedding approaches in which variational and non-variational electronic structure methods are combined~\cite{Hofener2012,Hofener2012b,Hofener2013,Hofener2016,Niemeyer2025,Myhre2014,Myhre2016,Goletto2021} to calculate molecular properties in large systems within a fully quantum mechanical description in a cost-effective manner.

In the case of coupled cluster theory, one can identify two main approaches, one which employs an exponential parametrization for the time evolution of the cluster operators (CC response)\cite{Monkhorst2009, dalgaard_aspects_1983,Koch1990a,Koch1990b,Koch1994,kobayashi1994calculation,Christiansen1995a,Christiansen1995b,Christiansen1996,Hattig1997,coriani_coupled_1997,Christiansen1998,kondo_orthogonally_1995,piecuch_property_1995,spirko_molecular_1996,piecuch_molecular_1996,piecuch_convergence_1997}, and another based on the equation of motion formalism (EOMCC response)\cite{Stanton1993,Stanton1993b,Sekino1995,gauss_coupledcluster_1995,Rozyczko1997,Sekino1999,Crawford2009,Nanda2015,nanda_static_2016,nanda_communication_2018} for which a linear parametrization is employed to obtain molecular properties. 

From the outset, key differences between the two approaches have been identified with respect to separability~\cite{Koch1994,Stanton1994,caricato2009difference}, with as consequence that CC transition moments are size-intensive whereas EOMCC ones are not. More recently the interconnections between these two approaches were further investigated from a formal point of view~\cite{pawlowski_molecular_2015, coriani_molecular_2016}. 

In spite of such differences, the calculation of response properties within the EOMCC formalism remains appealing from both the implementation and the application points of view~\cite{faber2018resonant,andersen2022cherry,andersen2022probing}, as simplifications in the equations also translate into a somewhat more modest computational cost. This calls for an understanding of how the two approaches differ in practical calculations. In the literature comparisons 
between linear response CC (LR-CC) and EOMCC (LR-EOMCC) properties~\cite{Sekino1994,Sekino1999,Crawford2009,coriani_molecular_2016} so far suggest that, in spite of the lack of size-extensivity of EOM-CC transition moments, in practice the two approaches yield quite similar results. 

Similar conclusions have been reached when comparing quadratic response CC (QR-CC) and EOMCC (QR-EOMCC) results for 
dipole hyperpolarizabilties~\cite{Rozyczko1997,Sekino1995,coriani_molecular_2016}, though comparisons were initially limited to a few systems. More recently, an extensive benchmark study on excited state transition moments~\cite{irek2025}, which also probe quadratic response functions, seem to confirm the tendencies observed previously. These previous evaluations were carried out for valence properties of molecules containing light elements, that is, occupying the first two rows of the periodic table. For properly representing the molecular properties of heavier systems it is necessary to account for relativistic effects\cite{Pyykko1988a,saue_relativistic_2011,Pyykk2012,Autschbach2012a,liu2017handbook,Schwerdtfeger2020,Smits2023} and to the best of our knowledge there is currently no data in the literature assessing the differences between the two approaches beyond the second row of the periodic table.

Response theory can be used in combination with relativistic Hamiltonians\cite{Saue2002_PostDirac}, and implementations for mean-field approaches (Hartree-Fock and DFT) are for example available for the calculation of frequency-dependent linear and non-linear response properties for 4-component-based Hamiltonians\cite{visscher_4-component_1997,Saue:Jensen:JCP2003,norman_quadratic_2004,salek2005linear,Henriksson:2005,Henriksson_JCP2006,Tellgren_JCP2007,bast:JCP2009,Bast_ijqc2009,Bast2009b,Villaume_JCP2010}. To go beyond mean-field, relativistic coupled cluster theory~\cite{Liu2021,visscher1996formulation,pototschnig_implementation_2021,Yuwono2024,Chamoli2025,Brandejs2025} offers a flexible framework for obtaining accurate energies and properties which treats electron correlation and relativistic effects on the same footing. It can thereby be used to investigate systems of interest across the periodic table, and benchmark more approximate methods. 

In recent years there have been extensive efforts by different groups to address states beyond closed-shell ground states through EOMCC theory for excitation and ionization energies, and electron affinities\cite{Pathak:2016ck,Pathak:2015he,Pathak:2014gv,Pathak:2014dm,Pathak:2016dk,Klein:2008hx,Yang:2012gd,Wang:2015jw,Epifanovsky:2015hsa,Cao:2016fx,Cao:2017id,Zhang:2017jj,Akinaga:2017hv,Wang:2016hx,shee2018equation,Koulias2019,Halbert2021,Halbert2023,Zhang2024,Koulias2019,Zhang2024b,Uhlov2024,Yuwono2025,Li2025,Banerjee2025,Mukhopadhyay2025}. While some of these implementations allow for the calculation of transition moments or static properties (through the application of finite fields or via approximate expectation values), full-fledged implementations of CC or EOMCC response theory are much more scarce. \citet{Yuan_Xiang_LR} have reported an implementation capable of carrying out frequency-dependent LR-CC and LR-EOMCC calculations for different (from 1- to 4-component based) Hamiltonians for electric, magnetic and mixed electric-magnetic properties. A comparison of CC and EOMCC showed that for dipole polarizabilities the two approaches do yield similar results as relativistic effects become more pronounced, but for indirect spin-spin coupling constants there is growing difference between the two. ~\citet{Chakraborty2025} reported an efficient implementation of LR-CC for the spin-free eXact 2-Component Hamiltonian (X2C) and applied it to investigate the static and dynamic dipole polarizabilities of an extensive set of molecular systems.

Beyond linear response, the analytical treatment of quadratic response coupled cluster theory would be particularly useful in applications in which finite field calculations are not straightforward, easy to converge or inapplicable, such for frequency-dependent calculations of non-linear optical properties or in the study of birefringences. 
However, general implementations beyond linear response are scarcer still, with our report of relativistic QR-EOMCC~\cite{Yuan_Xiang_QR} being, to the best of our knowledge, the first to appear in the literature. This implementation allowed us to gauge the performance of more approximate approaches such as TDDFT and TDHF in the calculation of hyperpolarizabilies, two-photon absorption cross-sections and Verdet constants. However, as discussed above QR-CC and QR-EOMCC are not equivalent, and it is desirable to further characterize the behavior of the latter with respect to the former across the periodic table, before carrying out more extensive applications with the more approximate QR-EOMCC scheme.

The aim of the current work is therefore to report our implementation of quadratic response coupled cluster for relativistic Hamiltonians, capable of handling for both static and frequency-dependent perturbations involving electric or magnetic operators. We showcase it by providing a first comparison of how QR-CC and QR-EOMCC behave as one goes towards the bottom of the periodic table. We thereby chose to re-investigate the first dipole hyperpolarizabilities of hydrogen halides, and the Verdet constants of heavy noble gas atoms as they show significant relativistic effects~\cite{Yuan_Xiang_QR} {and include an exploration of the Verdet constants of selected hydrogen halides}. Going further than this previous work, we carried out a more extensive exploration on the effect of the Hamiltonian on these properties, and investigated the properties of systems made up of superheavy elements, such as the Tennessine hydride molecule and the Oganesson atom. 

Tennessine and Oganesson, whose existence has been recognized by IUPAC and IUPAP in 2015, are the two heaviest elements known. They have been synthesized at the Joint Institute for Nuclear Research in Dubna by nuclear fusion upon bombardment of Berkelium and Californium targets by between atoms of Calcium-48 ions, respectively\cite{Karol2016a,Karol2016b}. 
While they are expected to be members of groups 17 and 18, the difficulties in their production--only a few atoms are generated at a time, and their lifetimes are in the order of milliseconds--present formidable challenges to the experimental characterization of their physico-chemical properties~\cite{Smits2023b}, making theoretical work an indispensable tool to learn about their electronic structure~\cite{Smits2023}. In view of their large polarizabilties and more pronounced spin-orbit coupling than that in 6p, 5d and 5f block elements, they are also of potential interest as probes for any subtle difference between CC and EOMCC in the calculation of response properties.

\section{Theory and Implementation}
\label{Theory}

\subsection{Quadratic Response Theory}

As was the case in our previous work on coupled cluster linear response\cite{Yuan_Xiang_LR} (LR-CC) and EOM-CC quadratic response~\cite{Yuan_Xiang_QR} (QR-EOMCC), our work is carried out within the framework of the quasienergy formulation of response theory~\cite{Christiansen1998}. For the sake of brevity, we only review the formalism for CC quadratic response (QR-CC), and refer to the SI for the working equations of QR-EOMCC as presented in~\cite{Yuan_Xiang_QR}. Throughout the text, the CC acronym will imply the use of coupled cluster singles and doubles (CCSD) wavefunctions.

We shall remain within the no-pair approximation~\cite{visscher1996formulation,liu_going_2013,Almoukhalalati2016}, with the electronic part of the unperturbed molecular Hamiltonian given by
\begin{equation}
    \hat{H}_0 = 
    \sum_{pq} h^p_q \left\lbrace \hat{a}^{\dagger}_{p} \hat{a}_{q} \right\rbrace +
    \frac{1}{4} \sum_{pqrs} V^{pq}_{rs} \left\lbrace \hat{a}^{\dagger}_{p}  \hat{a}^{\dagger}_{q}\hat{a}_{s}\hat{a}_{r} \right\rbrace
    \label{eq:unperturbed_hamiltonian}
 \end{equation}
where and $p, q, r, s$ denote general (occupied or positive energy virtual) indexes, $h_{pq} = \bra{p} \hat{h} \ket{q}$ denotes matrix elements of the one-body operator $\hat{h}$, and $ V^{pq}_{rs} = \bra{pq} \hat{g} \ket{rs} - \bra{pq} \hat{g} \ket{sr}$ the anti-symmetrized matrix elements of the two-body operator $\hat{g}$. In the calculations reported in this work, $\hat{g}$ corresponds to Coulomb operator. In our current implementation, Gaunt or Breit interactions are assumed to be treated in an effective manner\cite{Schimmelpfennig.Gropen.1998f8,sikkema2009molecular,iliavs2007infinite,saue2002four}. External time-(in)dependent perturbations will be represented by one-body operators 
\begin{equation}
    \hat{Y} = 
    \sum_{pq} Y^p_q \left\lbrace \hat{a}^{\dagger}_{p} \hat{a}_{q} \right\rbrace
    \label{eq:pertubing-operators}
 \end{equation}
where $Y^p_q = \bra{p} \hat{Y} \ket{q}$.

Due to the fact that in CC response theory the time-dependent quasienergy Lagrangian~\cite{Christiansen1998}
is made stationary with respect to the wavefunction parametrization (CC amplitudes and multipliers), we can use the ($2n+1$) and ($2n+2$) rules in perturbation theory\cite{Christiansen1998,helgaker_molecular_2014}. The calculation of a third-order property such as quadratic response, thus only requires the determination of the unperturbed and first-order perturbed CC amplitudes and Lagrange multipliers. 

The unperturbed CC amplitudes $T$ and Lagrange multipliers $\bar{T}$ are defined as
 \begin{align}
 T = T_1 + T_2 &= \sum_{ai} \ts {a}{i}  \left\lbrace \hat{a}^{\dagger}_{a} \hat{a}^{}_{i} \right\rbrace  + \unquart \sum_{abij} \td abij \left\lbrace \hat{a}^{\dagger}_{a} \hat{a}^{\dagger}_{b} \hat{a}^{}_{j} \hat{a}^{}_{i} \right\rbrace = \sum_{ai} \ts {a}{i} \hat{\tau}_1+ \unquart \sum_{abij} \td abij \hat{\tau}_2   \label{eq:t_amplitudes} \\
 \bar{T} = \bar{T}_1 + \bar{T}_2 &=  \sum_{ia} \bar{t}^{i}_{a} \left\lbrace \hat{a}^{\dagger}_{i} \hat{a}^{}_{a} \right\rbrace + \unquart \sum_{ijab} \bar{t}^{ij}_{ab} \left\lbrace \hat{a}^{\dagger}_{i} \hat{a}^{\dagger}_{j} \hat{a}^{}_{b} \hat{a}^{}_{a} \right\rbrace = \sum_{ia} \bar{t}^{i}_{a} \hat{\tau}^\dagger_1 + \unquart \sum_{ijab} \bar{t}^{ij}_{ab} \hat{\tau}^\dagger_2 
    \label{eq:bar_t_amplitudes}
 \end{align}
with $a,b$ for particle lines and $i,j$ for hole lines. As a shorthand notation, in the following we shall use greek letters $\left\lbrace \mu, \nu, \sigma \right\rbrace$ to represent the strings of second quantization operators, with subscripts to represent single or double excitation; for instance, $\hat{\tau}_{\mu_1}\equiv \hat{\tau}_1=\hat{a}^{\dagger}_{a} \hat{a}^{}_{i}$ whereas $\hat{\tau}_{\mu}$ will represent a generic (either single or double) excitation operator. Conversely, $\hat{\tau}_{\mu}^{\dagger}$ will denote a de-excitation operator, which is biorthogonal to operator $\hat{\tau}_{\mu}$, that is, $\bra{R}\hat{\tau}_{\mu}^{\dagger}\hat{\tau}_{\nu}\ket{R} = \delta_{\mu\nu}$, with $\ket{R}$ representing the reference state, here a single Slater determinant made up of molecular orbitals--or molecular spinors whenever the Hamiltonian accounts for spin-orbit coupling. Also, $|\mu\rangle$ will denote an excited determinant ($|\mu\rangle = \hat{\tau}_{\mu} \ket{R}$) contained within the singles and doubles manifold. 

From the exponential ansatz for the CC wavefunction~\cite{shavitt2009many,bartlett2007coupled}
\begin{align}
\ket{CC} = \exp({T}) \ket{R},
\end{align}
the time-independent Dirac (Schr\"odinger) wave equation
\begin{equation}
\hat{H}_0 \ket{CC} = E\ket{CC} \label{cc-wave-eq-gs}
\end{equation}
and the $\Lambda$ operator 
\begin{equation}
\bra{\Lambda} = \bra{\mathrm{R}} + \sum_\mu \bar{t}_{\mu} \bra{\mu}\exp(-T),
\end{equation}
we have the unperturbed amplitudes and multipliers obtained from the equations
\begin{align}
\left\langle \mu \right| \bar{\mathbf{H}} \left| \mathrm{R} \right\rangle &= 0 \\
\bar{t}\;\bar{\mathbf{H}} + \eta &= 0
\end{align}
where $\mathbf{\bar{H}}$ %
is matrix representation of the normal-ordered similarity transformed Hamiltonian. The definitions of $\mathbf{\mathbf{\bar{H}}}$ and $\eta$ are given in~\autoref{tab:ListOfTensor}. Further details on the evaluation of these equations in our implementation can be found in~\cite{pototschnig_implementation_2021}.

After solving for the unperturbed wave function parameters, the first-order parameters are next obtained by solving the response equations~\cite{Christiansen1998,Yuan_Xiang_LR,Yuan_Xiang_QR}
\begin{subequations}
\label{eq:PerturbatedLinearEquations}
\begin{align}
\left( \mathbf{\bar{H}} - \omega_{Y}\mathbf{I} \right) \tYs{}{}(\omega_Y) + \xi^Y= & 0 \label{eq:tPerturbated}\\
	\tbYs {}{} \left(\omega_Y\right)\left(\mathbf{\bar{H}}+\omega_Y \mathbf{I}\right)+\eta^Y+\mathbf{F} \tYs {}{} \left(\omega_Y\right)=&0,
 \label{eq:tBarPerturbated}
\end{align}
\end{subequations}
where $\mathbf{I}$ is the identity matrix. We employ the shorthand notation $\tYs{}{}, \tbYs{}{}$ to describe the first-order perturbed amplitudes and multipliers, obtained for perturbation $\hat{Y}$ at frequency $\omega_Y$, respectively. We note that, in contrast to the first-order response equations in QR-EOMCC, the equations for the first-order perturbed multipliers depend on the first-order perturbed amplitudes. The definitions of terms $\mathbf{F}$, $\xi^Y, \eta^Y$ are also given in~\autoref{tab:ListOfTensor}. 

The frequency-(in)dependent quadratic response function is given by the expression
\begin{align}
	\langle\langle Z; X, Y\rangle\rangle & {}_{\omega_X, \omega_Y} 
	= \frac{1}{2} C^{\pm \omega} P\left( X\left(\omega_{X}\right), Y\left(\omega_{Y}\right), Z\left(\omega_{Z}\right)\right) \nonumber \\
	& \times\left\{ \left[\frac{1}{2} \mathbf{F}^{X}+\frac{1}{6} \mathbf{G} \tXs {}{} \left(\omega_{X}\right)\right] \tYs{}{} \left(\omega_{Y}\right) +\tbXs{}{}\left(\omega_X\right)\left[\mathbf{A}^{Y}+\frac{1}{2} \mathbf{B} \tYs{}{}\left(\omega_{Y}\right)\right]\right\} \tZs{}{} \left(\omega_{Z}\right)\ ,
    \label{eq:QuadraticResponse}
\end{align}
where the perturbing frequencies satisfy the relationship $\omega_Z = -(\omega_X + \omega_Y)$. The operator $P\left( X\left(\omega_{X}\right), Y\left(\omega_{Y}\right), Z\left(\omega_{Z}\right)\right)$  takes care of the permutation of the perturbing operators, and 
\begin{equation}
C^{\pm\omega}f^{XYZ}(\omega_X,\omega_Y,\omega_Z) = f^{XYZ}(\omega_X,\omega_Y,\omega_Z) + [f^{XYZ}(-\omega_X,-\omega_Y,-\omega_Z)]^*
\label{eq:symmetrization}
\end{equation}
is used for the symmetrization of the response functions~\cite{Christiansen1998}, which requires the solution of response $f^{XYZ}(\omega_X,\omega_Y,\omega_Z)$ also for the negative of the frequencies (inverting the frequencies followed by complex conjugation). \autoref{eq:QuadraticResponse} can also be rewritten as 
\begin{align}
	   \langle\langle X, Y, Z\rangle\rangle_{\omega_Y, \omega_Z} 
	   &= \frac{1}{2} C^{\pm \omega} P\left( X\left(\omega_{X}\right), Y\left(\omega_{Y}\right), Z\left(\omega_{Z}\right)\right) \times\left\{ 
        \left[O^{X,Y}\right] +\left[O^{\bar{X},Y}\right]\right\} \tZs{}{} \left(\omega_{Z}\right),
        \label{eq:PropertyWith_O_Expression}
\end{align}
where  $O^{X,Y}$ and $O^{\bar{X},Y}$ group the terms involving contractions only with perturbed amplitudes or with both perturbed amplitudes and multipliers, respectively:
\begin{subequations}
    \label{eq:ExpressionOfSubEquationWithO}
    \begin{align}
        O^{X;Y} = & \left[ \frac{1}{2} \mathbf{F}^{X}+\frac{1}{6} \mathbf{G} \tXs{}{} \left(\omega_{X}\right) \right] \tYs{}{} \left(\omega_{Y}\right)  = \frac{1}{2} \mathbf{F}^{X} \tYs{}{} \left(\omega_{Y}\right) +\frac{1}{6} \mathbf{G} \tXs {}{} \left(\omega_{X}\right) \tYs {}{} \left(\omega_{Y}\right) \label{eq:Ott}   \\
        O{}^{\bar{X};Y} =&\ \tbXs {}{} \left(\omega_X\right)\left[ \mathbf{A}^{Y}+\frac{1}{2} \mathbf{B} \tYs {}{} \left(\omega_{Y}\right) \right]  = \tbXs{}{} \left(\omega_X\right) \mathbf{A}^{Y}+ \frac{1}{2} \tbXs{}{} \left(\omega_X\right)  \mathbf{B} \tYs {}{} \left(\omega_{Y}\right) \label{eq:Otbart}
    \end{align} 
\end{subequations}
and that for all combinations allowed by the $C^{\pm \omega} P  $ product.

\begin{table}
\centering
\begin{tabular}{l l}
\hline
    $\mathbf{\eta}_{\nu}$ & $\bra{{\mathrm{R}}}[\hat{H}_0,\hat{\tau}_{\nu}]|\mathrm{CC}\rangle$\\
    $\mathbf{\bar{H}}_{\mu \nu}$ & $\langle\bar{\mu}| [\hat{H}_0,\hat{\tau}_{\nu}]|\mathrm{CC}\rangle$ \\
    $\mathbf{F}_{\mu \nu}$  & $     \bra{\Lambda} [[ \hat{H}_0,\hat{\tau}_{\mu} ],\hat{\tau}_{\nu} ] \ket{CC} $ \\
    $\boldsymbol{\eta}^{Y}_{\mu}$ & $ \bra{\Lambda}[\hat{Y},\hat{\tau}_{\mu}]\ket{CC}$ \\
    $\boldsymbol{\xi}^{Y}_{\mu} $ &$  \bra{\bar{\mu}}\hat{Y}\ket{CC}$ \\
    $  \mathbf{F}^X_{\mu \nu}$ & $\langle\Lambda  |  [[\hat{X},\tau_{\mu}],\tau_{\nu}]  |  \mathrm{CC}\rangle $ \\
     $\mathbf{G}_{\mu \nu \sigma}$ & $\langle\Lambda  |  [[[\hat{H}_0,\tau_{\mu}],\tau_{\nu}],\tau_{\sigma}]  | \mathrm{CC}\rangle$  \\
    $\mathbf{A}^Y_{\mu \nu}$  & $ \langle\bar{\mu}  | [\hat{Y},\tau_{\nu}]  |  \mathrm{CC}\rangle$  \\
	$\mathbf{B}_{\mu \nu \sigma}$ & $\langle\bar{\mu}  |  [[\hat{H}_0,\tau_{\nu}],\tau_{\sigma}]  | \mathrm{CC}\rangle $ \\

    \hline
\end{tabular}
\caption{Expressions for the tensors appearing in the coupled cluster quadratic response functions as well as in the linear response equations determining perturbed amplitudes and Lagrange multipliers. Here $\ket{R}$ denotes the reference state in the definition of the coupled cluster wave function $  
\ket{CC} = \exp(T) \ket{R}$ and for the $\bra{\Lambda} = \bra{\mathrm{R}} + \sum_{\mu}\bar{t}_{\mu}\bra{\bar{\mu}}$ operator, with $\bra{\bar{\mu}}=\bra{\mathrm{R}}\hat{\tau}_{\mu}^{\dagger}\exp(-T)$. Finally, $\hat{X}$ and $\hat{Y}$ denote one-body perturbation operators. We note that $\mathbf{\bar{H}}$ is strictly equivalent to the $\mathbf{A}$, the CC Jacobian defined by~\citet{Christiansen1998}.}
\label{tab:ListOfTensor}
\end{table}

As is the case for $\mathbf{\bar{H}}$ and $\mathbf{F}$~\cite{Yuan_Xiang_LR,Yuan_Xiang_QR}, the $\mathbf{F^X}$, $\mathbf{A^Y}$, $\mathbf{G}$ and $\mathbf{B}$ tensors, given by the expression in ~\autoref{tab:ListOfTensor}, are never constructed explicitly but rather as contractions with the perturbed amplitudes and/or multipliers.  The programmable expressions required to assemble the $\mathbf{G} \tXs{}{} \tYs{}{} \tZs{}{}$ term are presented in the SI. The programmable expressions for $\mathbf{F^X}$, $\mathbf{A^Y}$ and $\mathbf{B}$ can be obtained from minor modifications of those already presented in our previous work~\cite{Yuan_Xiang_LR,Yuan_Xiang_QR}.

\subsection{Implementation details}\label{implementation}

Our QR-CC implementation has been carried out in the ExaCorr module~\cite{pototschnig_implementation_2021} of the relativistic quantum chemistry program DIRAC\cite{saue2020dirac}. Below we provide details on the implementation of the QR-CC response function calculation, and refer the reader to prior works concerning (a) the solution of ground-state amplitudes and multipliers~\cite{pototschnig_implementation_2021}; and (b) the solution of first-order response equations~\cite{Yuan_Xiang_LR}, which leveraged our prior implementation of a generalized Davidson eigenvalue solver\cite{hirao1982generalization,shee2018equation} to implement the solution of systems of linear equations~\cite{olsen1988solution}. We recall that through the use of the TAL-SH tensor library~\cite{githubRepo2023}, a modular component of the ExaTENSOR library~\cite{lyakh_domainspecific_2019}, all tensor operations in our code are parallelized via OpenMP and can be offloaded to NVIDIA or AMD GPUs whenever these are available.

In our implementation, we divide the calculation into different levels of subroutines: (a) a top-level driver to organize calculations for $\omega$ and $-\omega$ and to apply the symmetrization operator as in~\autoref{eq:symmetrization}; (b) a driver that calculates $\xi^Y, \eta^Y$ for all operators under consideration, solves all corresponding linear response equations for either QR-CC and QR-EOMCC, and calls the driver to calculate the appropriate response function. 

In the case of QR-CC, the response function driver carries out the six permutations between the triplet of $X, Y, Z$ operators; it calls a subroutine that calculates the terms within the curly braces in~\autoref{eq:PropertyWith_O_Expression} for a given pair or operators $X, Y$ and stores it on set of 2-/4-dimension tensors (singles and doubles manifolds) and performs the contraction with the $\tZs{}{}$ perturbed amplitudes. This allows us to, if desired, extract the contributions from each of the four terms (for each triplet of operators) and further break these down in terms of the contributions from the singles and doubles manifolds. 

{The correctness of our QR-CC implementation was validated through a comparison of non-relativistic results (for the HF, HCl and HBr systems) with (a) the analytic implementation of QR-CC in the Dalton~\cite{aidas2014d,Dalton2020p1} code (for $\alpha_{zz}, \beta_{zzz}, \beta_{zxx}$, see ~\autoref{SI-tab:verification-analytic-hf},~\autoref{SI-tab:verification-analytic-hcl} and~\autoref{SI-tab:verification-analytic-hbr}); and (b) finite-field calculations with the ExaCorr code, where the z-component of the electric dipole was employed as perturbing operator (with different perturbation strenghts) and included to the coupled cluster energy calculations. With that, we ensure that the numerical derivatives values (for $\alpha_{zz}, \beta_{zzz}$, see~\autoref{SI-tab:compare-nrdz-hf-hbr}) are consistent with the orbital-unrelaxed formulation of CC response. We also employed the analytic implementation of LR-CC in the CFOUR code~\cite{matthews2020coupled} to provide an additional check for $\alpha_{zz}$, and to investigate the relative importance of orbital relaxation for these systems (at the time of writing the QR-CC functionality in CFOUR is not part of the public release). In these calculations, carried out for HF, HCl and HBr, we employed contracted Dunning basis sets, a point-charge nucleus, and correlated all electrons and virtual orbitals. We note that we have at the same time validated the correctness of the code to calculate the left and right-hand side quadratic response equations, by calculating the components of the dipole polarizability with the asymmetric formulation of linear response theory (see e.g.~\cite{Yuan_Xiang_LR}) using the perturbed amplitudes and multipliers that go in the calculation of the QR-CC response functions. The results and computational procedures associated to these validations are found in the SI.} 

Our code currently lacks the functionality to obtain QR-EOMCC molecular properties from  finite-field calculations with the EOMCC model, and implementing such functionality is beyond the scope of this work. As explained in~\citet{Yuan_Xiang_QR}, we had verified our QR-EOMCC implementation against static $\beta_{zzz}$ values reported by~\citet{coriani_molecular_2016}, obtained with a QR-EOMCC implementation in the Dalton code (which, at the time of writing, is not publicly available). For convenience, a comparison between codes for the HF molecule is presented in the SI. 

\section{Computational details}\label{computational}

All QR-CC, QR-EOMCC 
calculations were carried out with a development version (revision number \texttt{8ef365b7}) of the DIRAC code\cite{saue2020dirac,DIRAC25}. All coupled cluster calculations have been carried out in GPU-accelerated nodes. 

As done in our prior EOM-QR calculations~\cite{Yuan_Xiang_QR}, we have used some of the functionality connected to the Cholesky-decomposition approach in ExaCorr\cite{pototschnig_implementation_2021} to avoid the explicit construction of the 2-electron repulsion integral tensor in AO basis in the integral transformation step and with that reduce the memory footprint of our calculations. We employed a threshold of 10$^{-5}$ for all systems except HAt and HTs, for which the threshold was set to 10$^{-4}$. The outcome of response calculations is not significantly affected by these approximations.

{In ref.~\citet{Yuan_Xiang_QR}, which describes our QR-EOMCC implementation, we investigated for the HF molecule the effect of basis set quality and virtual space truncation on the hyperpolarizabilities, and arrived at the conclusion that with a setup consisting of triply augmented triple-zeta basis sets (the Dunning basis set~\cite{kendall1992electron,woon1993gaussian,wilson1999gaussian} for H and the Dyall basis sets~\cite{dyall2006relativistic, basis-Dyall-TCA2012-131-3962} for all other atoms) was sufficiently converged for investigating the trends across the periodic table. In some cases and for elements up to bromide the t-aug-cc-pVTZ set was used to make a direct comparison to our prior EOM-QR calculations~\cite{Yuan_Xiang_QR} possible. The diffuse functions in question were automatically generated by DIRAC as an even-tempered sequence based on the most diffuse exponents for all angular momenta primitives in the original basis. In this manuscript we provide additional elements to show this choice of basis set was justified by carrying out a basis set study for the HF, HCl and HBr molecules.}

In addition to the non-relativistic {Levy-Leblond Hamiltonian (LL)}~\cite{levy1967nonrelativistic,visscher2000approximate}(as activated by the \texttt{.Levy-Leblond} keyword) and eXact 2-Component (X2C) relativistic\cite{iliavs2007infinite}, in the coupled cluster calculations we also investigated the molecular mean-field X2C method based on the Dirac-Coulomb ($^{2}$DC$^M$) and Dirac-Coulomb-Gaunt ($^{2}$DCG$^M$) Hamiltonians, as well as the spin-free (SF) variants of the X2C and $^{2}$DCG$^M$ Hamiltonians. In what follows, we shall use the term orbital as shorthand for both spinors and spin-orbitals, depending on the Hamiltonian used in the calculation. In X2C calculations including spin-orbit coupling, we make use of atomic-mean field SOC integrals, obtained from a calculation employing the Douglas-Kroll-Hess Hamiltonian.

For the correlation spaces, these included the valence electrons (for a total 8 electrons corresponding to those from the 1s shell of hydrogen, the ns$^2$np$^5$ shells for the halides and the ns$^2$np$^6$ shells for noble gases) and virtual orbitals with energies up to 5 a.u., {a value that was also established in ref.~\citet{Yuan_Xiang_QR} on the basis of calculations for HF. To illustrate the effect of this choice on the results, For HF, HCl and HBr we have also explored larger virtual cutoffs (10 and 100 a.u.) and no cutoffs (all virtuals correlated), and carried out calculations correlating all electrons and all virtual orbitals employing the $^{2}$DC$^M$ Hamiltonian. Additionally, in the case of the noble gases (Xe, Rn, Og) we also explored the correlation of the $(n-1)$d shell.}

The relativistic and non-relativistic calculations have been carried out with the Gaussian type\cite{visscher1997dirac} and point charge nucleus model, respectively. 

The molecular structures employed in our calculations have been taken from~\citet{huber1979constants} for HX (X=F, Cl, Br, I), from \citet{pereira_gomes_influence_2004} for HAt, and from~\citet{deFarias2017} for HTs, so respectively: H-F (0.91680 \AA), H-Cl (1.27455 \AA), H-Br (1.41443 \AA), H-I (1.60916 \AA), H–At (1.722 \AA) and H-Ts (1.9676  \AA). As the choice of orientation in these systems may affect the signs of the electric dipole and first hyperpolarizability (but not their absolute value), we note our results have been tabulated following the convention that the heavy atoms are placed at the origin of the coordinate system, with {the hydrogen atom placed on the positive} 
z-axis of the coordinate system.

EOM-CC excitation energy calculations exploiting point-group symmetry were carried out for the noble gas atoms using the RELCCSD module of DIRAC with the $^{2}$DC$^M$ and the same basis sets and correlating spaces as employed for the QR-CC calculations.

All inputs, outputs and structures used in this paper are available in the Zenodo repository as supplemental information.

\section{Results and discussion}
\label{results}

In the sections below we discuss our results for the first hyperpolarizability and Verdet constant for systems across the periodic table. We note that the choice of computational setup concerning the basis sets and correlation space, as well as a comparison to the B3LYP functional, have been discussed in detail in our work discussing the implementation of QR-EOMCC~\cite{Yuan_Xiang_QR}. Therefore, in the following we will mostly focus on a 
comparison between QR-CC and QR-EOMCC for different Hamiltonians.

\subsection{First hyperpolarizability of hydrogen halides (HX, X=F--Ts)}

Due to the symmetry of the systems under consideration, we restrict ourselves to the calculation of the parallel component of the static first hyperpolarizability ($\beta_{||}$)\cite{shelton_measurements_1994} 
\begin{equation}
    \beta_{||} = \frac{1}{5}\sum_{i=x,y,z}(\beta_{iiz}+\beta_{izi}+\beta_{zii})
\end{equation}
where the $\beta_{X;Y,Z}$ element is given by 
$\beta_{X;Y,Z}= - \left\langle \left\langle X;Y,Z \right\rangle \right\rangle_{\omega_{X};\ \omega_{Y},\ \omega_{Z}}$.   

\begin{figure}
    \centering
    \includegraphics[width=1\linewidth]{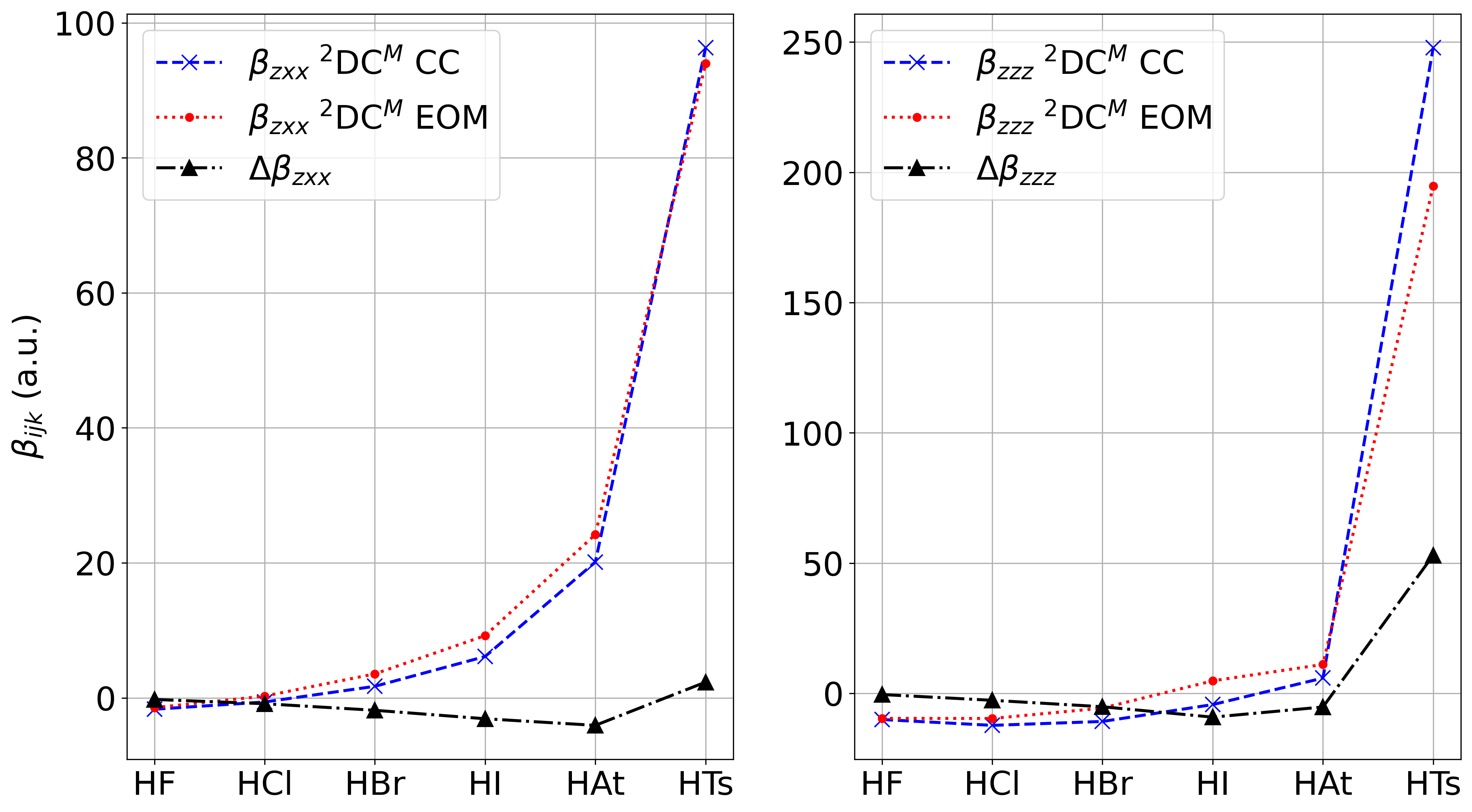}
    \caption{QR-CC and QR-EOMCC dipole hyperpolarizabilities components ($\beta_{zxx}$ and $\beta_{zzz}$, in a.u.) and the respective differences ($\Delta\beta_{zxx}$ and $\Delta\beta_{zzz}$, in a.u.) calculated with the $^{2}$DC$^M$ Hamiltonians for the HX series, upon correlating the $n$s$n$p shells (and with virtual threshold at 5 a.u.). [Adapted from~\citet{zenodo.17215085}, available under the Creative Commons Attribution 4.0 International license. Copyright X.\ Yuan, L.\  Halbert, L. Visscher, A.\ S.\ P.\ Gomes, 2025].}
    \label{fig:beta-components-static}
\end{figure}

In the following we only calculated two unique components, $\beta_{zzz}$ and $\beta_{zxx}$, so that $\beta_{||}$ and $\beta_\text{aniso}$ are calculated from
\begin{align}
    \beta_{||} =&\frac{3}{5}(\beta_{zzz}+2\beta_{zxx}) \\
    \beta_\text{aniso} =& \beta_{zzz} -  3\beta_{zxx}
\end{align}
respectively. Our results for the frequency-independent case are summarized in~\autoref{tab:hyperpolar-table3} for $\beta_{||}$ and in~\autoref{tab:hyperpolar-table4} for $\beta_\text{aniso}$. To illustrate the trends for the individual components, in~\autoref{fig:beta-components-static} we present results for the SF-$^2$DC$^M$ and $^2$DC$^M$ Hamiltonians, while in~\autoref{SI-tab:hyperpolar-table1} and~\autoref{SI-tab:hyperpolar-table2} we report the values for the $\beta_{zxx}$ and $\beta_{zzz}$ components respectively for all Hamiltonians under consideration. 

We recall that our use of a Gaussian nuclear model and Dyall bases in all calculations makes for a slightly different setup than that of~\citet{Yuan_Xiang_QR}. For ease of comparison, we also provide results obtained with the Dunning basis set on the halogens, and note small discrepancies for calculations with the Levy-Leblond Hamitonian arising from the difference in nuclear model.

Focusing first on the QR-EOMCC and QR-CC results for the different Hamiltonians for both $\beta_{||}$ and $\beta_\text{aniso}$, we observe that for HF and HCl scalar relativistic effects are dominant, with spin-orbit effects representing less than 1\% of the total response, something to be expected for these molecules which contain only light elements. For HBr we see that spin-orbit coupling is of some importance--in absolute terms, the hyperpolarizabilities increase with respect to spin-free calculations by about 5\% (20\%)  for $\beta_{||}$ QR-CC (QR-EOMCC) and reach about 8\% for HI and 10\% for HAt. For HTs, on the other hand, spin-orbit coupling is the dominant effect, with nearly a twofold increase in value compared to spin-free calculations.  For HF to HI the absolute values for $\beta_\text{aniso}$ are typically rather small, and become larger for HAt and HTs.

\begin{table}[H]
    \centering
    \setlength{\tabcolsep}{2.2mm}
{
    \begin{tabular}{clrrrrrr} 
    \hline
Method&	Hamiltonian	&	HF	&	HCl	&	HBr	&	HI	&	HAt	&	HTs	\\
    \hline
CC	&	LL	&	-8.0441	&	-8.2236	&	-5.5096	&	1.5590	&	9.9337	&	46.4544	\\
	&	LL$^\dagger$	&	-8.0370	&	-8.7902	&	-5.8960	&	nc	&	nc	&	nc	\\
	&	SF-X2C	&	-8.0483	&	-8.0824	&	-4.1250	&	6.3168	&	35.6445	&	173.6980	\\
	&	SF-$^{2}$DC$^M$	&	-8.0483	&	-8.0821	&	-4.1253	&	6.3164	&	35.6457	&	173.9446	\\
	&	X2C	&	-8.0537	&	-8.1031	&	-4.3473	&	5.7901	&	39.5212	&	476.3573	\\
	&	X2C$^\dagger$	&	-8.0407	&	-8.6627	&	-4.7106	&	nc	&	nc	&	nc	\\
	&	$^{2}$DC$^M$	&	-8.0498	&	-8.0942	&	-4.3343	&	5.7341	&	39.7956	&	495.1556	\\
&	${}^2$DC$^{M(1)}$	&	-7.9730	&	-8.1041	&	-4.3898	&	nc	&	nc	&	nc	\\
&	${}^2$DC$^{M(2)}$	&	-7.9327	&	-8.0703	&	-4.3439	&	nc	&	nc	&	nc	\\
&	${}^2$DC$^{M(3)}$	&	-7.9328	&	-8.0703	&	-4.3441	&	nc	&	nc	&	nc	\\
&	$^{2}$DC$^{M,*}$	&	-7.9142	&	-8.0537	&	-4.7243	&	nc	&	nc	&	nc	\\
	&	$^{2}$DCG$^M$	&	-8.0494	&	-8.0918	&	-4.3212	&	5.7574	&	39.8091	&	493.6418	\\
	&		&		&		&		&		&		&		\\
EOM	&	LL	&	-7.4972	&	-5.6034	&	-0.2535	&	11.1413	&	23.5288	&	66.9526	\\
	&	LL$^\dagger$	&	-7.5315	&	-6.2493	&	-0.6946	&	nc	&	nc	&	nc	\\
	&	SF-X2C	&	-7.4993	&	-5.4499	&	1.1741	&	15.6862	&	46.4367	&	164.0866	\\
	&	SF-$^{2}$DC$^M$	&	-7.4994	&	-5.4502	&	1.1731	&	15.6839	&	46.4288	&	164.1359	\\
	&	X2C	&	-7.5046	&	-5.4712	&	0.9285	&	14.8533	&	45.5480	&	405.8769	\\
	&	X2C$^\dagger$	&	-7.5333	&	-6.1101	&	0.5181	&	nc	&	nc	&	nc	\\
	&	$^{2}$DC$^M$	&	-7.5009	&	-5.4627	&	0.9445	&	14.8124	&	45.8277	&	423.0797	\\
&	${}^2$DC$^{M(1)}$	&	-7.6062	&	-5.5944	&	0.7430	&	nc	&	nc	&	nc	\\
&	${}^2$DC$^{M(2)}$	&	-7.6026	&	-5.5532	&	0.7973	&	nc	&	nc	&	nc	\\
&	${}^2$DC$^{M(3)}$	&	-7.6025	&	-5.5534	&	0.7966	&	nc	&	nc	&	nc	\\
&	$^{2}$DC$^{M,*}$	&	-7.6050	&	-5.6405	&	-0.4133	&	nc	&	nc	&	nc	\\
	&	$^{2}$DCG$^M$	&	-7.5016	&	-5.4658	&	0.9341	&	14.7813	&	45.8455	&	422.7453	\\
	&		&		&		&		&		&		&		\\
$\Delta \beta_{||}$	&	LL	&	-0.5469	&	-2.6203	&	-5.2561	&	-9.5822	&	-13.5951	&	-20.4982	\\
	&	LL$^\dagger$	&	-0.5055	&	-2.5409	&	-5.2014	&	nc	&	nc	&	nc	\\
	&	SF-X2C	&	-0.5490	&	-2.6324	&	-5.2991	&	-9.3694	&	-10.7921	&	9.6114	\\
	&	SF-$^{2}$DC$^M$	&	-0.5490	&	-2.6319	&	-5.2985	&	-9.3675	&	-10.7831	&	9.8088	\\
	&	X2C	&	-0.5491	&	-2.6319	&	-5.2758	&	-9.0631	&	-6.0268	&	70.4803	\\
	&	X2C$^\dagger$	&	-0.5074	&	-2.5526	&	-5.2287	&	nc	&	nc	&	nc	\\
	&	$^{2}$DC$^M$	&	-0.5490	&	-2.6316	&	-5.2787	&	-9.0782	&	-6.0321	&	72.0758	\\
&	${}^2$DC$^{M(1)}$	&	-0.3668	&	-2.5097	&	-5.1328	&	nc	&	nc	&	nc	\\
&	${}^2$DC$^{M(2)}$	&	-0.3301	&	-2.5171	&	-5.1412	&	nc	&	nc	&	nc	\\
&	${}^2$DC$^{M(3)}$	&	-0.3303	&	-2.5169	&	-5.1407	&	nc	&	nc	&	nc	\\
&	$^{2}$DC$^{M,*}$	&	-0.3092	&	-2.4132	&	-4.3109	&	nc	&	nc	&	nc	\\
	&	$^{2}$DCG$^M$	&	-0.5478	&	-2.6261	&	-5.2553	&	-9.0239	&	-6.0364	&	70.8965	\\
    \hline
\end{tabular}
}
    \caption{QR-CCSD (CC) and QR-EOMCCSD (EOM) values for the $\beta_{||}$ component (in a.u.) of the static first hyperpolarizability of hydrogen halides, as well as their difference ($\Delta\beta_{||}$). All calculations correlate eight (valence) electrons virtuals up to 5 a.u., except those marked by ``$^*$'', in which all electrons and virtuals are correlated; $^\dagger$calculated with Dunning basis sets for the halogens; nc: not calculated; ${}^2$DC$^{M(1)}$, ${}^2$DC$^{M(2)}$ and ${}^2$DC$^{M(3)}$ denote valence electrons correlated and virtual space truncations at 10 a.u., 100 a.u. and no truncation, respectively.}
    \label{tab:hyperpolar-table3}
\end{table}   

In all cases, there are no significant differences between the (SF-)X2C and (SF-)$^{2}$DC$^M$ Hamiltonians, in spite of the more approximate starting point of the X2C calculations, and the effect of the Gaunt interaction is only non-negligible for the HAt and HTs molecules (though it only contributes to about 1\%). 

We have noticed small differences between our results and those previously reported~\cite{Yuan_Xiang_QR}, the latter showing larger (in absolute value) hyperpolarizabilties. We have traced this difference to the use of Dyall basis sets for the current calculations, as opposed to the Dunning basis sets. This difference can be understood as a consequence of the bias towards somewhat less diffuse exponents as part of the process of obtaining  relativistic energy-optimized basis sets.

Focusing now first on QR-CC values for $\beta_{||}$, for HF to HBr we observe negative values (similar in magnitude for HF and HCl, but less negative for HBr), and positive for HI, HAt, and HTs, with rather large values for the latter molecule. These trends result from the fact that the clearly negative $\beta_{zzz}$ component for HF initially becomes less negative as the halogen becomes heavier, changes sign for HAt, and becomes strongly positive for HTs. The $\beta_{zxx}$ component shows a similar trend, with positive values from HBr onwards, but is smaller in magnitude than $\beta_{zzz}$ for most systems. These trends are qualitatively the same for QR-EOMCC across the series, with as important difference that the sign change in $\beta_{||}$ upon going down the series already occurs for HBr. 

Quantifying the differences between QR-CC and QR-EOMCC by computing ($\Delta\beta_{||}$), we see that differences are small for HF but become non-negligible already for HCl and remain sizable for the other species. In relative terms, the differences for $\beta_{||}$ are about 6\% for HF, but grow to about 30\% for HCl, 120 \% for HBr and 150\% for HI. For the very large positive hyperpolarizability of HTs, the difference is in absolute sense the largest, but amounts to only 15\% of the $^2$DCG$^M$ value. For the more accessible HAt molecule, the effect is similar but less pronounced. We note the situation is a bit different for $\beta_\text{aniso}$, where we observe that apart from HTs, the difference between is QR-CC and QR-EOMCC is fairly modest across the series.

{The results above have been obtained with a computational setup (triply augmented basis sets and virtual spaces truncated at 5 a.u.) established on the basis of a verification of basis set and correlation effects for the HF molecule in our paper describing the implementation of QR-EOMCC~\cite{Yuan_Xiang_QR}, and aimed at consistently addressing the whole hydrogen halide series. It is nevertheless important to have an idea of how close this setup is to the converged basis set and correlation treatment. To this end, we have investigated the effect of varying the number of diffuse functions and the cardinal number of the basis sets for HF, HCl and HBr, employing the $^2$DC$^M$ Hamiltonian.}

{From our results, found in~\autoref{SI-tab:InfluenceOfDiffuseSI}, we have that for triple-zeta basis sets, the triply and quadruply augmented basis sets typically show agreement better than 0.1 a.u.\ for all molecules, effectively validating our choice and in line with prior work on the light hydrogen halides~\cite{Fernandez1998}. We consider nevertheless that doubly augmented triple-zeta basis set can still be a good choice whenever it is not possible to employ larger basis sets for heavier elements. Furthermore, as expected singly augmented triple zeta bases are not sufficiently flexible, but are definitely much better suited than any double-zeta basis.}

{Another surprising but very interesting behavior revealed by this benchmark is that QR-CC and QR-EOMCC results are actually very close to each other for double-zeta basis sets without diffuse functions--including for HCl and HBr--and as soon as diffuse functions are added the differences between the two methods start to grow. To the best of our knowledge, this point has not been explored in the literature in detail, and possibly is the reason why our results seem at first at odds with the claims in the literature that CC and EOMCC yield results which are typically very close to each other.}

{With respect to the convergence of the correlation space, we observe from~\autoref{tab:hyperpolar-table3} and~\autoref{tab:hyperpolar-table4} that the 5 a.u.\ threshold is not a poor choice, but it is clear that with it the virtual space is not converged. Upon exploring other truncations while correlating only the valence electrons, we see that at 10 a.u.\ results are not quite converged either, whereas with a truncation at 100 a.u.\ results are essentially equivalent to those correlating all virtuals. An important finding is that, in qualitative terms, there is no difference in the trends across the halogen series for all virtual truncations employed. Finally, for HBr core correlation is not negiglible, accounting for an increase of 4\% and 18\% for $\beta_{||}$ calculated with QR-CC and QR-EOMCC. This change reduces the difference between the two approaches by nearly 1~a.u., but in any case the discrepancy remains significant (larger than 4~a.u.). It is important to keep in mind that such change is driven primarily by an increase in $\beta_{zzz}$.}

\begin{figure}
    \centering
    \includegraphics[width=1\linewidth]{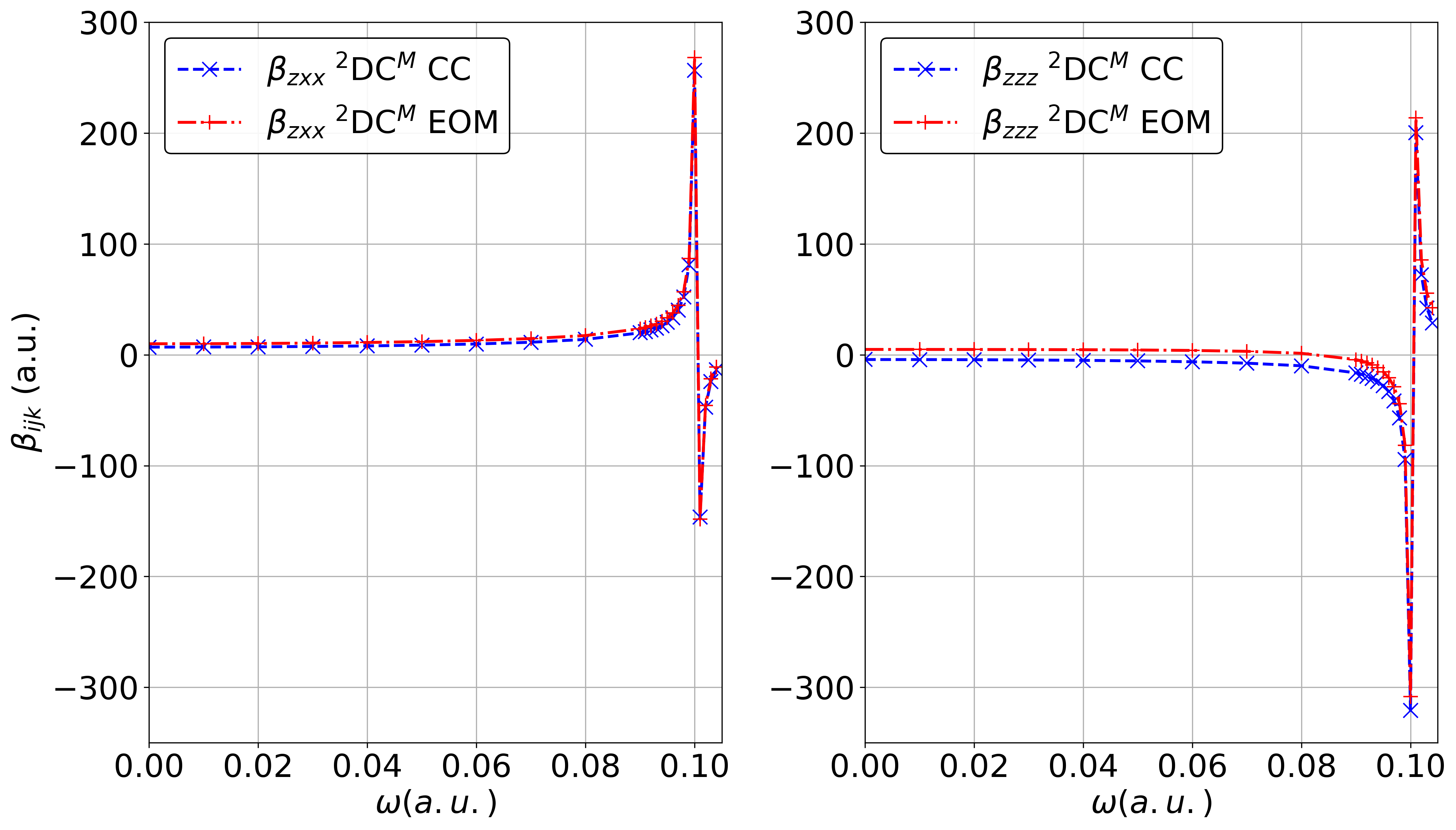}
    \includegraphics[width=1\linewidth]{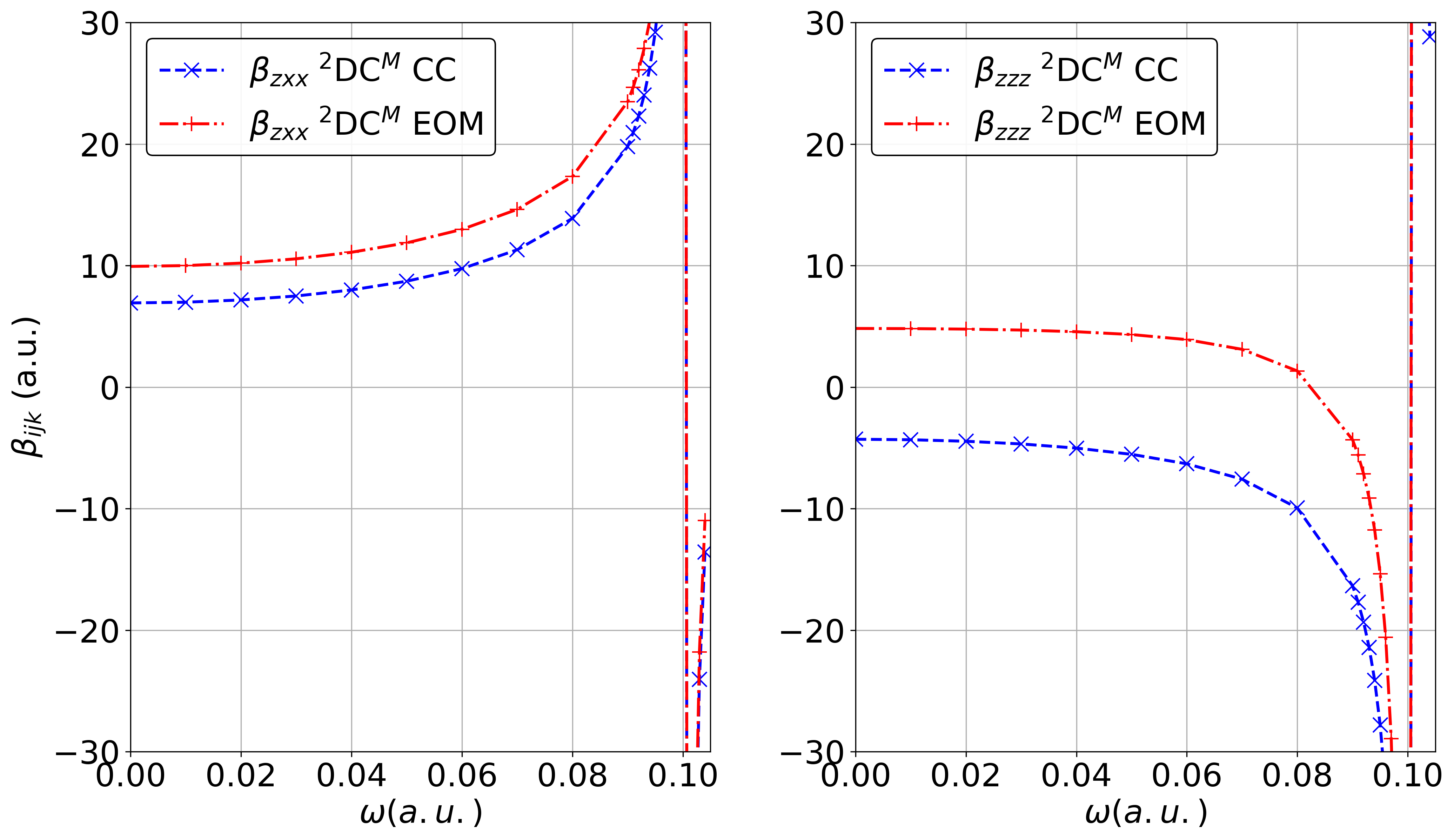}
    \caption{QR-CC and QR-EOMCC $^2$DC$^M$ dispersion curves for the $\beta_{zzz}, \beta_{zxx}$ components (top: general view; bottom: detailed view of the low-frequency region) of the hyperpolarizability of the HI molecule for second harmonic generation (SHG) calculations ($\omega = \omega_B = \omega_C$, in a.u.). We note the pole at around 0.10 a.u.\, corresponding to half of the excitation energy towards the $a^{3}\Pi_{0+}$ state~\cite{Yuan_Xiang_QR}. [Adapted from~\citet{zenodo.17215085}, available under the Creative Commons Attribution 4.0 International license. Copyright X.\ Yuan, L.\  Halbert, L. Visscher, A.\ S.\ P.\ Gomes, 2025].}
    \label{fig:HI-SHG}
\end{figure}

Going beyond the static case, in~\autoref{fig:HI-SHG} we present first the results for second harmonic generation (SHG) calculations on HI, which corresponds to the case in which $\omega_B = \omega_C = \omega$. As this property has been discussed in detail in our work presenting QR-EOMCC~\cite{Yuan_Xiang_QR}, we refer the reader to it for further details on the behavior of the dispersion curves. We nevertheless highlight the fact that that in the region around $\omega = 0.1$ a.u., corresponding to half the excitation energy of the spin--orbit allowed transition towards the $a^{3}\Pi_{0+}$ state,  we see the expected pole for the quadratic response functions. Beside that, we observe that across the range of frequencies sampled, the difference between QR-CC and QR-EOMCC results amounts to a nearly constant shift for low frequencies. For the region close to the resonance, on the other hand, we see a relatively small decrease in the difference between QR-CC and QR-EOMCC.

{In~\autoref{tab:pockels-table} we present a second dynamic property, the electro-optical Pockels effect~\cite{boyd2020nonlinear,norman_principles_2018}, which corresponds to the condition $\omega_B = \omega, \omega_C = 0$.  The Pockels effect can be understood as capturing the change in the dynamic polarizability due to an applied static electric field. A first consequence of the condition $\omega_C = 0$ is that the resonances now only occur near the first excited states (e.g.\ for HI, at about $\omega = 0.2$ a.u.) instead of at smaller frequencies such as in SHG. This means that {for} the systems under study we are very much off-resonance and consequently the $\beta_{||}(\omega,0)$ values remain very similar to those in the static case. Interestingly, we see a (very small) decrease in the difference between QR-CC and QR-EOMCC as the perturbation frequency increases, as found for the SHG case, but that in no way is capable of reversing the growing difference between QR-CC and QR-EOMCC across the halide series, and that these differences are driven by the $\beta_{zzz}$ component.}

\begin{table}[H]
    \centering
    \setlength{\tabcolsep}{2.2mm}
{
    \begin{tabular}{clrrrrrr}
    \hline
Method&	Hamiltonian	&	HF	&	HCl	&	HBr	&	HI	&	HAt	&	HTs	\\
    \hline
CC	&	LL	&	-4.9459	&	-10.3349	&	-14.5008	&	-16.3776	&	-18.5536	&	-50.1339	\\
	&	LL$^\dagger$	&	-4.6309	&	-10.0969	&	-14.3693	&	nc	&	nc	&	nc	\\
	&	SF-X2C	&	-4.9455	&	-10.4108	&	-15.4252	&	-20.1042	&	-41.2161	&	-191.5107	\\
	&	SF-$^{2}$DC$^M$	&	-4.9454	&	-10.4091	&	-15.4184	&	-20.0927	&	-41.1858	&	-191.6326	\\
	&	X2C	&	-4.9536	&	-10.4424	&	-16.0964	&	-24.8101	&	-83.2791	&	-599.5770	\\
	&	X2C$^\dagger$	&	-4.6278	&	-10.1948	&	-15.9223	&	nc	&	nc	&	nc	\\
	&	$^{2}$DC$^M$	&	-4.9467	&	-10.4348	&	-16.0811	&	-25.0659	&	-84.7735	&	-618.4277	\\
&	${}^2$DC$^{M(1)}$	&	-5.1487	&	-10.5333	&	-16.2143	&	nc	&	nc	&	nc	\\
&	${}^2$DC$^{M(2)}$	&	-5.1251	&	-10.5378	&	-16.2212	&	nc	&	nc	&	nc	\\
&	${}^2$DC$^{M(3)}$	&	-5.1252	&	-10.5381	&	-16.2215	&	nc	&	nc	&	nc	\\
&	$^{2}$DC$^{M,*}$	&	-5.1300	&	-10.6165	&	-16.5150	&	nc	&	nc	&	nc	\\
	&	$^{2}$DCG$^M$	&	-4.9478	&	-10.4450	&	-16.1219	&	-25.1406	&	-84.9810	&	-617.7646	\\
	&		&		&		&		&		&		&		\\
EOM	&	LL	&	-5.2227	&	-10.2724	&	-14.6625	&	-15.0853	&	-16.4412	&	-43.4775	\\
	&	LL$^\dagger$	&	-4.9146	&	-9.9864	&	-14.4596	&	nc	&	nc	&	nc	\\
	&	SF-X2C	&	-5.2236	&	-10.3699	&	-15.8135	&	-19.9561	&	-43.5784	&	-195.9931	\\
	&	SF-$^{2}$DC$^M$	&	-5.2234	&	-10.3687	&	-15.8100	&	-19.9571	&	-43.5978	&	-196.3284	\\
	&	X2C	&	-5.2315	&	-10.4017	&	-16.4911	&	-24.6949	&	-85.2353	&	-553.0503	\\
	&	X2C$^\dagger$	&	-4.9126	&	-10.1052	&	-16.2428	&	nc	&	nc	&	nc	\\
	&	$^{2}$DC$^M$	&	-5.2248	&	-10.3948	&	-16.4807	&	-24.9707	&	-86.8113	&	-570.8098	\\
&	${}^2$DC$^{M(1)}$	&	-5.3962	&	-10.4594	&	-16.5684	&	nc	&	nc	&	nc	\\
&	${}^2$DC$^{M(2)}$	&	-5.3543	&	-10.4722	&	-16.5866	&	nc	&	nc	&	nc	\\
&	${}^2$DC$^{M(3)}$	&	-5.3544	&	-10.4723	&	-16.5867	&	nc	&	nc	&	nc	\\
&	$^{2}$DC$^{M,*}$	&	-5.3591	&	-10.5605	&	-16.7843	&	nc	&	nc	&	nc	\\
	&	$^{2}$DCG$^M$	&	-5.2252	&	-10.4008	&	-16.5012	&	-24.9851	&	-86.8673	&	-570.0000	\\
	&		&		&		&		&		&		&		\\
$\Delta \beta_{\text{aniso}}$	&	LL	&	0.2768	&	-0.0624	&	0.1618	&	-1.2923	&	-2.1124	&	-6.6564	\\
	&	LL$^\dagger$	&	0.2838	&	-0.1105	&	0.0903	&	nc	&	nc	&	nc	\\
	&	SF-X2C	&	0.2780	&	-0.0409	&	0.3883	&	-0.1482	&	2.3622	&	4.4824	\\
	&	SF-$^{2}$DC$^M$	&	0.2781	&	-0.0404	&	0.3915	&	-0.1357	&	2.4120	&	4.6959	\\
	&	X2C	&	0.2780	&	-0.0407	&	0.3947	&	-0.1151	&	1.9562	&	-46.5267	\\
	&	X2C$^\dagger$	&	0.2848	&	-0.0896	&	0.3205	&	nc	&	nc	&	nc	\\
	&	$^{2}$DC$^M$	&	0.2781	&	-0.0400	&	0.3997	&	-0.0952	&	2.0379	&	-47.6178	\\
&	${}^2$DC$^{M(1)}$	&	0.2475	&	-0.0739	&	0.3541	&	nc	&	nc	&	nc	\\
&	${}^2$DC$^{M(2)}$	&	0.2292	&	-0.0657	&	0.3654	&	nc	&	nc	&	nc	\\
&	${}^2$DC$^{M(3)}$	&	0.2292	&	-0.0657	&	0.3652	&	nc	&	nc	&	nc	\\
&	$^{2}$DC$^{M,*}$	&	0.2291	&	-0.0560	&	0.2693	&	nc	&	nc	&	nc	\\
	&	$^{2}$DCG$^M$	&	0.2774	&	-0.0442	&	0.3793	&	-0.1555	&	1.8864	&	-47.7646	\\\hline
\end{tabular}
}
    \caption{QR-CCSD (CC) and QR-EOMCCSD (EOM) values for the $\beta_{\text{aniso}}$ component (in a.u.) of the static first hyperpolarizability of hydrogen halides, as well as their difference ($\Delta\beta_{\text{aniso}}$). All calculations correlate eight (valence) electrons virtuals up to 5 a.u., except those marked by `$^*$', in which all electrons and virtuals are correlated; $^\dagger$calculated with Dunning basis sets for the halogens; nc: not calculated; ${}^2$DC$^{M(1)}$, ${}^2$DC$^{M(2)}$ and ${}^2$DC$^{M(3)}$ denote valence electrons correlated and virtual space truncations at 10 a.u., 100 a.u. and no truncation, respectively.}
    \label{tab:hyperpolar-table4}
\end{table}   

\begin{table}[H]
    \centering
    \setlength{\tabcolsep}{2.2mm}
{
    \begin{tabular}{lrrrrrp{0.1in}rrrr}
\hline\hline				
        &              & \multicolumn{4}{c}{$\beta_{||}(\omega,0)$} &&  \multicolumn{4}{c}{$\beta_\text{aniso}(\omega,0)$} \\
        \cline{3-6}
        \cline{8-11}
System	&	$\omega$	&	CC	&	EOM	&	$\Delta$	&	\%	&&	CC	&	EOM	&	$\Delta$ & \%				\\
\hline																				
HF	&	0.01	&	-8.0557	&	-7.5067	&	-0.5490	&	6.81	&&	4.9484	&	5.2266	&	-5.4974	&	-5.62		\\
	&	0.02	&	-8.0733	&	-7.5242	&	-0.5491	&	6.80	&&	4.9535	&	5.2320	&	-5.5026	&	-5.62		\\
	&	0.03	&	-8.1028	&	-7.5536	&	-0.5492	&	6.78	&&	4.9619	&	5.2411	&	-5.5111	&	-5.63		\\
	&	0.05	&	-8.1985	&	-7.6488	&	-0.5496	&	6.70	&&	4.9886	&	5.2696	&	-5.5382	&	-5.63		\\
	&	0.09	&	-8.5492	&	-7.9982	&	-0.5510	&	6.44	&&	5.0776	&	5.3652	&	-5.6286	&	-5.66		\\
	&		&		&		&		&		&&		&		&		&			\\
HCl	&	0.01	&	-8.1047	&	-5.4717	&	-2.6329	&	32.49	&&	10.4421	&	10.4022	&	-13.0750	&	0.38		\\
	&	0.02	&	-8.1362	&	-5.4991	&	-2.6371	&	32.41	&&	10.4639	&	10.4243	&	-13.1009	&	0.38		\\
	&	0.03	&	-8.1891	&	-5.5452	&	-2.6439	&	32.29	&&	10.5001	&	10.4610	&	-13.1441	&	0.37		\\
	&	0.05	&	-8.3626	&	-5.6965	&	-2.6662	&	31.88	&&	10.6151	&	10.5775	&	-13.2813	&	0.35		\\
	&	0.09	&	-9.0230	&	-6.2765	&	-2.7465	&	30.44	&&	10.9995	&	10.9670	&	-13.7460	&	0.30		\\
	&		&		&		&		&		&&		&		&		&			\\
HBr	&	0.01	&	-4.3414	&	0.9411	&	-5.2825	&	121.68	&&	16.1002	&	16.5005	&	-21.3828	&	-2.49		\\
	&	0.02	&	-4.3629	&	0.9310	&	-5.2939	&	121.34	&&	16.1579	&	16.5600	&	-21.4518	&	-2.49		\\
	&	0.03	&	-4.3992	&	0.9138	&	-5.3130	&	120.77	&&	16.2543	&	16.6596	&	-21.5673	&	-2.49		\\
	&	0.05	&	-4.5196	&	0.8553	&	-5.3749	&	118.93	&&	16.5659	&	16.9815	&	-21.9408	&	-2.51		\\
	&	0.09	&	-5.0011	&	0.6012	&	-5.6023	&	112.02	&&	17.6886	&	18.1420	&	-23.2909	&	-2.56		\\
	&		&		&		&		&		&&		&		&		&			\\
HI	&	0.01	&	5.7433	&	14.8309	&	-9.0875	&	-158.23	&&	25.1222	&	25.0273	&	-34.2098	&	0.38		\\
	&	0.02	&	5.7710	&	14.8866	&	-9.1155	&	-157.95	&&	25.2923	&	25.1982	&	-34.4079	&	0.37		\\
	&	0.03	&	5.8175	&	14.9800	&	-9.1626	&	-157.50	&&	25.5800	&	25.4873	&	-34.7425	&	0.36		\\
	&	0.05	&	5.9679	&	15.2841	&	-9.3162	&	-156.10	&&	26.5371	&	26.4494	&	-35.8533	&	0.33		\\
	&	0.09	&	6.5089	&	16.4030	&	-9.8941	&	-152.01	&&	30.4156	&	30.3485	&	-40.3097	&	0.22		\\
\hline																				
\end{tabular}
}
\caption{{Parallel and anisotropy ($\beta_{||}(\omega,0), \beta_\text{aniso}(\omega,0)$) [in atomic units]  components of the electro-optical Pockels effect for the hydrogen halide molecules at different perturbing frequencies $\omega$ [in atomic units], calculated with QR-CCSD (denoted by CC) and QR-EOMCCSD (denoted by EOM) for the $^2$DC$^M$ Hamiltonian upon correlating the $n$s$n$p shells and with virtual threshold at 5 a.u., alongside with the respectice differences between the two approaches ($\Delta \beta_{||} = \beta_{||}^\text{CC} - \beta_{||}^\text{EOM}$ and $\Delta \beta_\text{aniso} = \beta_\text{aniso}^\text{CC} - \beta_\text{aniso}^\text{EOM}$) in absolute and relative (in \%) of the corresponding $\Delta \beta$ terms.} 
}
    \label{tab:pockels-table}
\end{table}

In the following we look more closely at the differences between QR-CC and QR-EOMCC by analyzing the contributions from the different terms to the response functions, found in SI (\autoref{SI-tab:breakdown-qrf-cc-beta} for QR-CC and \autoref{SI-tab:breakdown-qrf-eomcc-beta} for QR-EOMCC respectively), as we go across the series from HF to HI.


For this analysis, it is interesting to also consider the results for the static dipole polarizabilties for the same molecules. {The reason for that is twofold: first, it is well-established that the formal difference between LR-CC and LR-EOMCC is connected to the presence of the $\mathbf{F}t^Y$ term in the LR-CC response function (and therefore the right-hand side ground to excited state transition matrix elements), whereas LR-EOMCC employs a simpler expression--for both methods perturbed amplitudes $t^Y$ are obtained from eq.~\autoref{eq:tPerturbated}, and left-hand side transition matrix elements are formally the same. This means that a large difference in polarizability is indicative of the relative importance of the full $\mathbf{F}t^Y$ term, and consequently of an important difference in the right-hand side ground-to-excited state transition matrix elements.}

{Second, if one considers a sum-over-states expression for the hyperpolarizabilities, we have in the denominator one has electronic state energies (which are identical for CC and EOMCC), while in the numerator appears the product of ground-to-excited state transition matrix elements (from linear response) and state to state transition matrix elements (~\autoref{SI-state-to-state-tm}, given in terms of the contractions $\mathbf{A}^{Y}+ \mathbf{B}t^Y$ found in~\autoref{eq:Otbart}). The consequence of the expression of the numerator is that, if the linear response terms for CC and EOMCC are similar while the full quadratic response function is not, the difference should be coming to the state to state transition matrix elements; or, in other words, in the description of transition moments between excited states.} 

\begin{figure}
    \centering
    \includegraphics[width=1\linewidth]{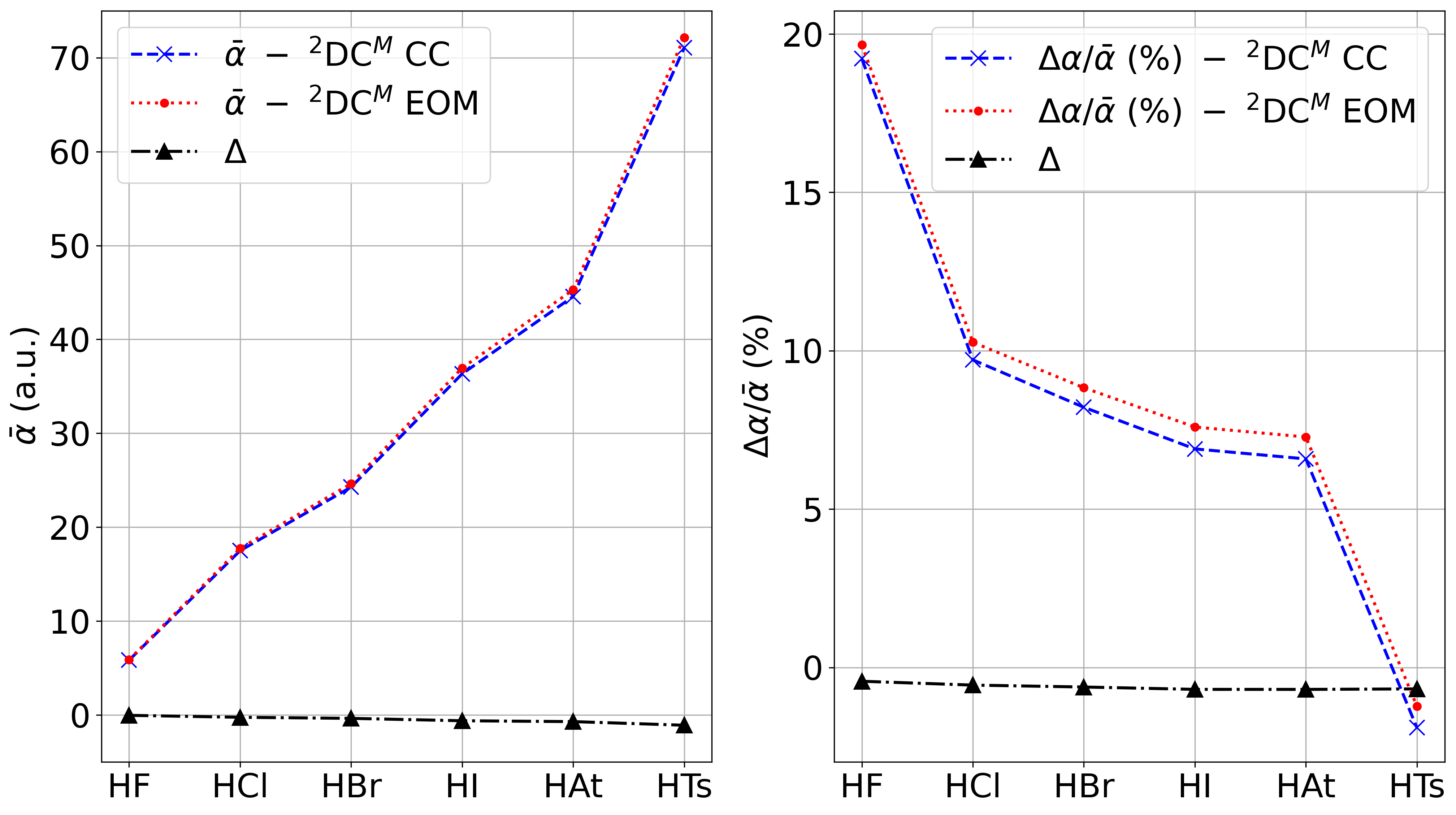}
    \caption{Mean dipole polarizabilities (left, $\bar{\alpha} = 1/3(\alpha_{zz}+2\alpha_{xx})$ in a.u.) and anisotropies as a fraction of mean polarizability (right, $\Delta\alpha / \bar{\alpha}=\left( \alpha_{zz}-\alpha_{xx}\right)/\bar{\alpha}$ in \% ) calculated with $^{2}$DC$^M$ Hamiltonian for the HX series with LR-CC and LR-EOMCC, upon correlating the $n$s$n$p shells (and with virtual threshold at 5 a.u.). The absolute differences between the two approaches ($\Delta$ : LR-CC - LR-EOMCC) are also presented. [Adapted from~\citet{zenodo.17215085}, available under the Creative Commons Attribution 4.0 International license. Copyright X.\ Yuan, L.\  Halbert, L. Visscher, A.\ S.\ P.\ Gomes, 2025].}
    \label{fig:polar-hx-AnisoFracOfTheMean}
\end{figure}

The dipole polarizabilities, shown in~\autoref{fig:polar-hx-AnisoFracOfTheMean} (values for dipole moments and polarizabilities are reported in~\autoref{SI-tab:polar-hx-table}) increase across the series, which is consistent with the increase in size of the halides accompanied by their decreasing electronegativity. The differences between CC and EOMCC do increase across the series for the $\alpha_{zz}$ and $\alpha_{xx}$ components and consequently, for the mean polarizability $\bar{\alpha}$ as well. However, in contrast to the hyperpolarizabilities, in relative terms these differences are roughly the same and of about 2\% for $\alpha_{zz}$ and 1.5\% for $\alpha_{xx}$. 


The difference between CC and EOMCC is more strongly felt for the anisotropy, which is about 3\% for HF, but 8-12\% for HCl to HAt, and reaches about 30\% for HTs, though in absolute terms the differences are still modest. An intriguing observation is that the polarizability anisotropy, while becoming less important for the almost spherical HAt and HTs molecules, changes sign upon going from HAt to HTs. This means that HTs can be more easily polarized in the direction perpendicular to the bond while the reverse is true for the lighter homologues that are more easily polarized along the bond. The sign change occurs with spin-orbit coupling but not without (though in magnitude the spin-orbit and spin-free anisotropies are very close to each other), making further studies necessary to better characterize it.  We note that our CCSD dipole moments compare well with available experimental~\cite{NISTconstants,Hohm2013} or theoretical work (for HTs)~\cite{thierfelder_scalar_2009}, though for reaching benchmark accuracy we would need to account for higher excitations, something which is also beyond the scope of this work. These findings add to the body of data supporting the argument that differences between LR-CC and LR-EOMCC properties are in practice small{, and in particular when correlating the same number of electrons}. 

{As mentioned above, another consequence of the similarity between CC and EOMCC polarizabilities is that it suggests their difference arises from the state to state transition matrix elements. While at this time we have not yet implemented the calculation of excited state properties (and such implementation is beyond the scope of this work), we can nevertheless gain some insight by} considering the individual contributors to the value of the QR-CC functions.

{We see that across the halide series the dominant} contributions typically arise from the terms from eq.~\autoref{eq:Otbart}, involving both the perturbed amplitudes and multipliers ($ \tbXs{~}{~}  \mathbf{A}^{Y} \tZs{}{} $ and $ \tbXs{~}{~}  \mathbf{B} \tYs {}{}  \tZs{}{} $). Furthermore, we observe that these two contributions to the final value in general tend to counteract each other, and that the magnitude of the ``doubles'' contributions increases as the halide becomes heavier. Terms from eq.~\autoref{eq:Ott}, involving only perturbed amplitudes and ground-state multipliers ($\mathbf{F}^{X} \tYs{}{} \tZs{}{} $  and $\mathbf{G} \tXs {}{} \tYs{}{}  \tZs{}{} $) show smaller (but non-negligible) net contributions than those above. They also increase in absolute value as the halide becomes heavier. 

The relative importance of contribution of the terms in eq.~\autoref{eq:Otbart} also changes depending on which response function we look at. For $\langle\langle Z; Z, Z \rangle\rangle_0$, the term $ \tbXs{~}{~}  \mathbf{A}^{Y} \tZs{}{} $ is the one which contributes the most, but for $\langle\langle Z; X, X \rangle\rangle_0$ there are more important cancellations between the ``singles'' and ``doubles'' contributions from it so that the $ \tbXs{~}{~}  \mathbf{B} \tYs {}{}  $ term becomes a more important net contributor, being the dominant one for the HI molecule followed by $\mathbf{F}^{X} \tYs{}{} \tZs{}{} $, whereas $ \tbXs{~}{~}  \mathbf{A}^{Y} \tZs{}{} $ and $\mathbf{G} \tXs {}{} \tYs{}{}  \tZs{}{} $ show similar contributions.

For QR-EOMCC, we see some similarities to the QR-CC picture, in that contributions beyond ``singles'' become increasingly important as the halogen becomes heavier, while the term
$^{\text{EOM}}\eta^X \mathbf{A}^Y \tZs {}{}$ is typically a key contribution and dominates for the lightest elements. The value of the $^{\text{EOM}}\eta^X \mathbf{A}^Y \tZs{}{}$ contribution also appears to follow closely that of $\ \tbXs{}{}  \mathbf{A}^{Y} \tZs{}{} + \tbXs{}{}  \mathbf{B} \tYs{}{} \tZs{}{} $ in QR-CC. The terms $-\tbXs {}{} \tYs{}{} \bar{t} \xi^Z $ and $-\bar{t} \tYs{}{} \tbZs{}{} \xi^X \ $, on the other hand, become increasingly more important for heavier systems, and for $\langle\langle Z; Z, Z \rangle\rangle_0$ of HBr and HI, they offset the other contribution and are largely responsible for the sign changes between QR-CC and QR-EOMCC for these systems.

\subsection{Verdet constant of noble gases (Xe-Og) and hydrogen halides (HF-HI)}

In~\autoref{tab:verdet-table1} we present our results for the calculation of the Verdet constant for Xe, Rn and Og. In line with our prior work, we considered three laser wavelengths (589.0 nm, 694.3 nm, and 1064.0 nm) to compare QR-EOMCC and QR-CC for different Hamiltonians. The Verdet constants are evaluated as quadratic response functions\cite{coriani_coupled_1997,norman_quadratic_2004}:
\begin{equation}
    V(\omega) = \omega\frac{eN\epsilon_{xyz}}{24c_{0}\epsilon_{0}m_{e}}\text{Im}\langle\langle\hat{\mu}_{x};\hat{\mu}_{y},\hat{m}_{z}\rangle\rangle_{\omega,0}
\end{equation}
with $N$ the number density of the gas, $e$ the elementary charge, $m_{e}$ the electron mass, $c_{0}$ the speed of light in vacuo, and $\hat{m}_{z}$ is magnetic dipole moment operator. {For the noble gases, due to atomic} symmetry, we only needed to calculate one unique component and could obtain $V(\omega)$ through the expression 
\begin{equation}
    V(\omega) = \omega\frac{eN}{4c_{0}\epsilon_{0}m_{e}}\text{Im}\langle\langle\hat{Y};\hat{X},\hat{m}_{z}\rangle\rangle_{\omega,0}.
\end{equation}

Comparing first the differences between QR-CC and QR-EOMCC, we observe that these become larger as the wavelength decreases, and increase as the atoms become heavier. However, if we consider the magnitude of these differences, we see that the differences are rather small overall (about 1 to 1.4\%). Furthermore, the differences are very similar for the different wavelengths. 
These additional calculations with respect to those already reported by~\cite{Yuan_Xiang_QR}, allow us to more clearly separate contributions from different relativistic effects.
With respect to the Hamiltonian and as discussed previously, the Verdet constant for Xe is still well described by non-relativistic Hamiltonians~\cite{cadene_circular_2015,Yuan_Xiang_QR}, and scalar relativity is the dominant relativistic effect. Furthermore, we observe no difference between (SF-)X2C and (SF-)$^{2}$DC$^M$ results, and find only rather small differences X2C and $^{2}$DC$^M$. As this is an atomic system, these differences are primarily due to the difference in Hamiltonian used to compute the spin-orbit screening in the mean-field approximation used in the X2C approach. Adding the Gaunt interaction to this mean-field screening acts to slightly decrease the $^{2}$DC$^M$ value, in the direction of the spinfree and X2C (Douglas-Kroll-Hess based screening) result. We also observe that in spite of the fact that QR-CC values are smaller than QR-EOMCC ones, they still overestimate the experimental results.
For Rn relativistic effects are important and spin-orbit coupling effects significantly increase the Verdet constant. We again see some small differences between X2C and $^{2}$DC$^M$. For Og, spin-free results with X2C and $^{2}$DC$^M$ are roughly the same, indicating that the bare nucleus and screened X2C transformation matrices are very similar (this is the only difference between the two spin-free approaches, see \cite{sikkema2009molecular} for a more extensive discussion). With spin-orbit coupling differences between X2C and $^{2}$DC$^M$ are more noticeable at about 3\%. For this superheavy element, the spin-orbit effect is very significant and even dominates over scalar relativity, with $^{2}$DC$^M$ values being roughly three times that of SF-$^{2}$DC$^M$.  While much larger in magnitude, the Gaunt interaction behaves similarly in Og as in Xe, reducing the Verdet constant at all frequencies.

\begin{figure}
    \centering
    \includegraphics[width=1\linewidth]{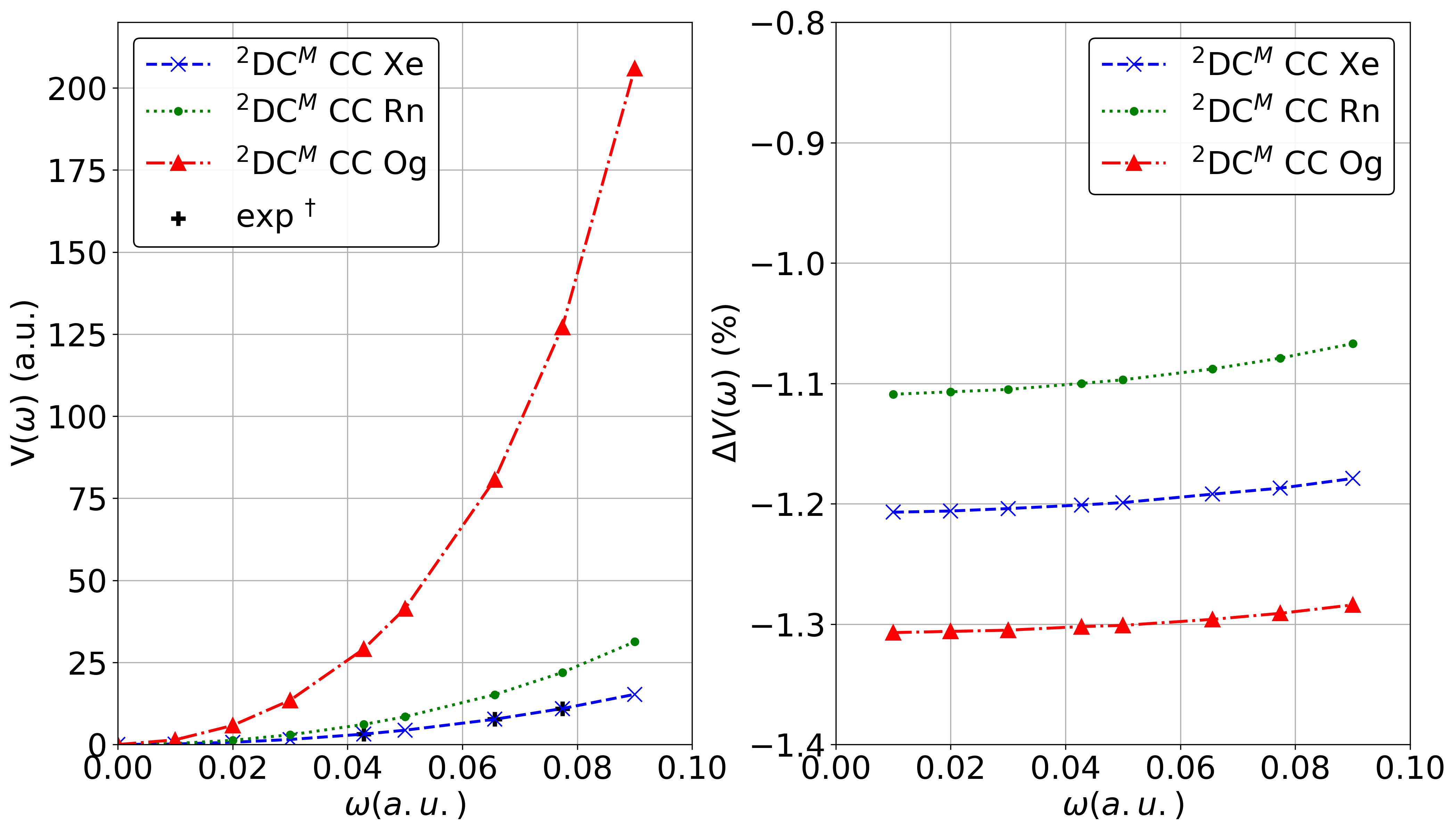}
    \caption{Left: frequency dependence of the QR-CC Verdet constants calculated with $^{2}$DC$^M$ Hamiltonian for Xe, Rn and Og, upon correlating the $n$s$n$p shells (and with virtual threshold at 5 a.u.). ${}^{\dagger} $ Experimental data from~\citet{Ingersoll:56} and~\citet{cadene_circular_2015}. Left: frequency dependence of the difference between QR-CC and QR-EOMCC ($\Delta V = (V_{CC}-V_{EOM})/V_{CC}$, in \%). [Adapted from~\citet{zenodo.17215085}, available under the Creative Commons Attribution 4.0 International license. Copyright X.\ Yuan, L.\  Halbert, L. Visscher, A.\ S.\ P.\ Gomes, 2025].}
    \label{fig:VerdetOmega}
\end{figure}

In~\autoref{fig:VerdetOmega} and~\autoref{SI-tab:verdet-table-scan-nsnp} we present the frequency dependence of the Verdet constant obtained with the $^{2}$DC$^M$ Hamiltonian for additional frequencies than those in~\autoref{tab:verdet-table1}. From these data, we observe that no significant discrepancy arises from the use of  QR-EOMCC instead of QR-CC. Across the range of studied wave lengths, the differences remain on the order of 1.1-1.3\%. 

The magnitude of the Verdet constant for Rn is found to be roughly twice of that for Xe across the range of studied wavelenghts, while the one of Og is much larger still, with its magnitude being 4.5 times larger for the longest to about 6.5 times larger for the shortest wavelenghts. This is in stark contrast to what is observed for the spin-free calculation presented in~\autoref{tab:verdet-table1}, where we see changes of only a factor of roughly 1.5 going from Xe to Rn, and of 1.8 going from Rn to Og. This is consistent with prior theoretical investigations of Og, that predict this element to be much more polarizable due to the extremely large spin-orbit effects in this superheavy element~\cite{Jerabek2019a,Jerabek2019b}. 

While the first excited state is lowered due to relativity, this strongly increased polarizability due to spin-orbit effects manifests itself already far away from this resonance. To document this, we present in~\autoref{SI-tab:excitation-ng} calculations for the lowest excited states of these atoms. These show that while the first excited state is progressively lowered by spin-orbit coupling as the atom gets heavier, the wavelengths probed are still far away from the resonances.

\begin{table}[H]
    \centering
    \setlength{\tabcolsep}{2.2mm}
{
    \begin{tabular}{crccccccccc}
    \hline
& 
    &	\multicolumn{3}{c}{$\lambda$(589 nm)}				
    &	\multicolumn{3}{c}{ $\lambda$(694.3 nm)}
    &	\multicolumn{3}{c}{$\lambda$(1064 nm)} \\
    \cline{3-11}
Atom&		Hamiltonian	&	CC	&	EOM		& $\Delta V$ & 	CC	&	EOM		 & $\Delta V $ &	CC &	EOM	 & $\Delta V$\\
\hline
Xe	&	SF-X2C		&	12.39	&	12.54	&	-0.15	&	8.72	&	8.82	&	-0.11	&	3.60	&	3.64	&	-0.04	\\
	&	SF-$^{2}$DC$^M$		&	12.40	&	12.55	&	-0.15	&	8.72	&	8.83	&	-0.11	&	3.60	&	3.64	&	-0.04	\\
	&	X2C		&	12.95	&	13.10	&	-0.15	&	9.10	&	9.20	&	-0.11	&	3.74	&	3.79	&	-0.05	\\
	&	$^{2}$DC$^M$		&	13.00	&	13.16	&	-0.15	&	9.14	&	9.24	&	-0.11	&	3.76	&	3.81	&	-0.05	\\
	&	$^{2}$DCG$^M$		&	12.97	&	13.12	&	-0.15	&	9.11	&	9.22	&	-0.11	&	3.75	&	3.80	&	-0.05	\\
	&			&		&		&	 	&		&		&		&		&		&		\\
Rn	&	SF-X2C		&	18.79	&	19.01	&	-0.22	&	13.14	&	13.30	&	-0.15	&	5.38	&	5.44	&	-0.06	\\
	&	SF-$^{2}$DC$^M$		&	18.82	&	19.04	&	-0.22	&	13.16	&	13.31	&	-0.15	&	5.38	&	5.45	&	-0.06	\\
	&	X2C		&	25.62	&	25.90	&	-0.28	&	17.73	&	17.92	&	-0.19	&	7.14	&	7.22	&	-0.08	\\
	&	$^{2}$DC$^M$		&	26.07	&	26.35	&	-0.28	&	18.03	&	18.22	&	-0.20	&	7.26	&	7.34	&	-0.08	\\
	&	$^{2}$DCG$^M$		&	25.95	&	26.24	&	-0.28	&	17.95	&	18.15	&	-0.20	&	7.23	&	7.31	&	-0.08	\\
	&			&		&		&	 	&		&		&		&		&		&		\\
Og	&	SF-X2C		&	35.45	&	35.83	&	-0.38	&	24.44	&	24.71	&	-0.27	&	9.81	&	9.91	&	-0.11	\\
	&	SF-$^{2}$DC$^M$		&	35.58	&	35.95	&	-0.38	&	24.53	&	24.79	&	-0.26	&	9.84	&	9.95	&	-0.11	\\
	&	X2C		&	144.66	&	146.52	&	-1.85	&	92.02	&	93.20	&	-1.18	&	33.34	&	33.77	&	-0.43	\\
	&	$^{2}$DC$^M$		&	151.00	&	152.95	&	-1.95	&	95.74	&	96.98	&	-1.24	&	34.55	&	35.00	&	-0.45	\\
	&	$^{2}$DCG$^M$		&	149.96	&	151.92	&	-1.96	&	95.12	&	96.36	&	-1.24	&	34.34	&	34.79	&	-0.45	\\
    \hline
\end{tabular}
}
    \caption{Verdet Constant V($\omega$) [in 10$^{-3}$ rad/(T m)] for Gaseous Xe, Rn and Og at Different Laser Wavelengths, calculated with QR-CCSD (denoted by CC) and QR-EOMCCSD (denoted by EOM) upon correlating the $n$s$n$p shells (and with virtual threshold at 5 a.u.), as well as the difference in Verder constant between the two approaches ($\Delta V = V^\text{CC} - V^\text{EOM}$). For Xe, experimental values at $\lambda$(589 nm) and $\lambda$(1064 nm) are respectively 12.30 10$^{-3}$ rad/(T m)~\cite{Ingersoll:56} and 3.56$\pm$0.10 10$^{-3}$ rad/(T m)\cite{cadene_circular_2015}.}
    \label{tab:verdet-table1}
\end{table}

To further complete the picture, we computed the $^2$DC$^M$ dipole polarizabilities for the noble gases with LR-CC and LR-EOMCC for the same range of frequencies as for the Verdet constant. The results, presented in~\autoref{SI-tab:polar-ng-table}, are in line with the findings for the static polarizabilties for the HX systems and our work~\cite{Yuan_Xiang_LR} in that there is no appreciable difference between the two CC variants for this property. 

Our results are in good agreement with the experimental static dipole polarizabilities for Xe and Rn (27.3 and 35.8 a.u.\ respectively) ~\cite{Hohm1994,CRC2014}, and likewise for the theoretical predictions for Og (57.98 a.u.) of Smits and coworkers~\cite{Smits2023}. Though we did not calculated values for atomic Ts, our results from HTs capture well the trend of decreasing polarizability from Ts to Og (76.3 to 57.98 a.u.)~\cite{Smits2023} predicted in that work. We attribute the observed discrepancies to the lack of higher excitations in CCSD.

To check for the contribution of subvalence electrons, we also considered additional calculations in which we retained the virtual space truncation at 5 a.u.\ but included also the $(n-1)d$ electrons. In our prior QR-EOMCC calculations~\cite{Yuan_Xiang_QR} we found, based on non-relativistic calculations (the difference of calculations with the same correlating space, and by correlating all occupied and virtual orbitals), that the effect of correlating electrons beyond the valence shell for the Xe atom could result in a decrease of the Verdet constant by about -0.08, -0.19 and -0.28 [10$^{-3}$ rad/(T m)], for 1064 nm, 694.3 nm and 589 nm respectively. 

Our current results, shown in~\autoref{SI-tab:verdet-table-scan-n-1dnsnp}, confirm that a decrease in Verdet constant is indeed observed for all atoms, though with the current virtual space truncation for Xe we recover slightly smaller changes (-0.07, -0,18 and -0.26 [10$^{-3}$ rad/(T m)], for 1064 nm, 694.3 nm and 589 nm respectively). This is consistent with our earlier finding, that upon correlating the $(n-1)d$ electrons, the excitation energies for all systems are (slightly) shifted upward. Another change seen upon correlating the $(n-1)d$ electrons is that the difference between QR-CC and QR-EOMCC values increases somewhat, though in relative terms these differences remain close to 1\% overall.

{As a first comparison beyond atomic systems, in~\autoref{tab:verdet-table2} we present the Verdet constants obtained with the $^2$DC$^M$ Hamiltonian for the hydrogen halides up to HI. The values of the unique response functions used to calculate the Verdet constant are given in the SI. whereas  systems.}

{Due to their axial symmetry, for the HX systems we require at least two response function evaluations; in addition to $\langle\langle\hat{Y};\hat{X},\hat{m}_{z}\rangle\rangle_{\omega,0}$ we calculated $\langle\langle\hat{Z};\hat{X},\hat{m}_{y}\rangle\rangle_{\omega,0}$ (and verified that we obtain that it yields the same result as $-\langle\langle\hat{Z};\hat{Y},\hat{m}_{x}\rangle\rangle_{\omega,0}$), so that the Verdet constants are calculated with the expression
\begin{equation}
    V(\omega) = \omega\frac{eN}{24c_{0}\epsilon_{0}m_{e}}\left[2\text{Im}\langle\langle\hat{X};\hat{\mu}_{y},\hat{m}_{z}\rangle\rangle_{\omega,0} + 4\text{Im}\langle\langle\hat{Z};\hat{Y},\hat{m}_{x}\rangle\rangle_{\omega,0}\right].
\end{equation}
}

\begin{table}[H]
    \centering
    \setlength{\tabcolsep}{2.2mm}
{
    \begin{tabular}{lrrrrr}
\hline\hline				
 & & \multicolumn{4}{c}{V($\omega$)}\\
\hline
System	&	$\omega$	&	CC	&	EOM	&	$\Delta$ &	\%	\\
\hline											
HF	&	0.01	&	0.0161	&	0.0162	&	-0.0001	&	-0.41	\\
	&	0.02	&	0.0647	&	0.0650	&	-0.0003	&	-0.41	\\
	&	0.03	&	0.1461	&	0.1467	&	-0.0006	&	-0.41	\\
	&	0.05	&	0.4112	&	0.4128	&	-0.0017	&	-0.40	\\
	&	0.09	&	1.3954	&	1.4009	&	-0.0056	&	-0.40	\\
	&		&		&		&		&		\\
HCl	&	0.01	&	0.0781	&	0.0789	&	-0.0008	&	-0.99	\\
	&	0.02	&	0.3133	&	0.3165	&	-0.0031	&	-0.99	\\
	&	0.03	&	0.7089	&	0.7159	&	-0.0070	&	-0.99	\\
	&	0.05	&	2.0042	&	2.0239	&	-0.0197	&	-0.98	\\
	&	0.09	&	6.9238	&	6.9903	&	-0.0665	&	-0.96	\\
	&		&		&		&		&		\\
HBr	&	0.01	&	0.1356	&	0.1370	&	-0.0014	&	-1.07	\\
	&	0.02	&	0.5445	&	0.5503	&	-0.0058	&	-1.06	\\
	&	0.03	&	1.2335	&	1.2466	&	-0.0131	&	-1.06	\\
	&	0.05	&	3.5036	&	3.5405	&	-0.0369	&	-1.05	\\
	&	0.09	&	12.3195	&	12.4447	&	-0.1252	&	-1.02	\\
	&		&		&		&		&		\\
HI	&	0.01	&	0.2689	&	0.2726	&	-0.0037	&	-1.37	\\
	&	0.02	&	1.0813	&	1.0960	&	-0.0147	&	-1.36	\\
	&	0.03	&	2.4545	&	2.4879	&	-0.0334	&	-1.36	\\
	&	0.05	&	7.0169	&	7.1115	&	-0.0946	&	-1.35	\\
	&	0.09	&	25.3084	&	25.6359	&	-0.3275	&	-1.29	\\	
    \hline
\end{tabular}
}
\caption{Verdet Constant [V($\omega$), in atomic units] for the hydrogen halide molecules at different frequencies $\omega$ [in atomic units], calculated with QR-CCSD (denoted by CC) and QR-EOMCCSD (denoted by EOM) for the $^2$DC$^M$ Hamiltonian upon correlating the $n$s$n$p shells and with virtual threshold at 5 a.u.), as well as the difference in Verdet constant between the two approaches ($\Delta = V^\text{CC} - V^\text{EOM}$) in absolute and relative terms (in \%).}
    \label{tab:verdet-table2}
\end{table} 

{Following the trends observed for the atoms, the Verdet constants increase across the HF to HI series. This is in line with the increase in molecular polarizabilities for the series, and a reflex of the Verdet constant representing the change in dynamic polarizability in the presence of the external static magnetic field. Likewise, we have that QR-CC and QR-EOMCC yield very similar results for a given frequency and molecular system. In relative terms, we see that the differences are on par with those for the noble gases, with the exception of HF which shows even smaller differences. We note that as in the case of SHG and the Pockels effect, we have a small decrease in the difference between QR-CC and QR-EOMCC for larger perturbing frequencies.}

{In view of the arguments put forward in the analysis of the hyperpolarizabilities, and gven the similarity between CC and EOMCC polarizabilities for these systems--and consequently of their ground to excited transition matrix elements--the good agreement for the the Verdet constants suggests the terms connected to the state to state transition moments ($\mathbf{A}^{Y}+ \mathbf{B}t^Y$ in~\autoref{eq:Otbart}) should also be similar}.

{In order to assess this hypothesis, we have analyzed the contributions from different contractions to the response functions in the Verdet constant case as well.  The breakdown of the contributions are found in~\autoref{SI-tab:breakdown-qrf-cc-verdet}  for QR-CC and in~\autoref{SI-tab:breakdown-qrf-eomcc-verdet} for QR-EOMCC. From them, we observe that the contributions arise essentially from the $\mathbf{A}^{Y}$ term across the series, with an increased contribution from the $\mathbf{B}t^Y$ for QR-CC seen for HI. We observe that contributions to these terms are essentially made up of singles for all species, a behavior which is distinct from the hyperpolarizabilities.}

{A comparison between the trends across the halide series and those for the Pockels effect is also interesting, as it highlight the difference between the perturbing operators : for Pockels, the driver for the difference between QR-CC and QR-EOMCC is the $Z$ operator, which drives the response of the polarizability along the molecular axis. For verdet, on the other hand, the magnetic operator induces an action perpendicular to the molecular axis. }

To conclude this section, {we consider that through the analysis of the quadratic response functions, we have been able to identify the mechanisms through which the differences between QR-CC and QR-EOMCC arise for the systems under consideration. We also show that depending on the property under considreration, QR-EOMCC can offer a very similar quality than QR-CC--and as shown previously, while outperforming density functional approximations such as B3LYP~\cite{Yuan_Xiang_QR}--at a substantially less cost.}

\section{Conclusions}
\label{Conclusions}

In this work we have presented an implementation in the GPU-accelerated ExaCorr module of coupled cluster quadratic response (QR-CC) theory which can be used in combination with both (2- and 4-component based) relativistic and non-relativistic Hamiltonians, thus extending our prior implementation quadratic response based on the equation of motion (QR-EOMCC) formalism ~\cite{Yuan_Xiang_QR}. 

We have showcased this implementation by carrying out a detailed comparison between QR-CC and QR-EOMCC in the calculation of both static and dynamic first hyperpolarizabilities ({where we determined dispersion curves for second harmonic generation and the Pockels electro-optical effect}) for the hydrogen halide series (for HF to HTs), as well as the calculation of the Verdet constant for noble gases (Xe, Rn and Og) {and hydrogen halides (HF to HI)}. In addition to the comparison of non-relativistic and X2C Hamiltonians as done previously, we have carried out calculations with the X2C molecular mean field approach, and gauged the contribution of spin-orbit coupled and scalar relativistic effects.

For the first hyperpolarizabilities, we find that for the HF molecule the difference between QR-CC and EOM-CC is relatively small (around 4\%), in line with the studies reported in the literature employing non-relativistic approaches that do not see significant differences between the two approaches. The differences between QR-CC and QR-EOMCC do however increase as one goes down the periodic table, with differences of about 20\% already for HCl and a sign change for $\beta_{zzz}$ in HI. {Our findings also point to the fact that such differences are significantly smaller when basis sets without diffuse functions are employed, and are particularly small (from HF to HBr) when double zeta bases such as cc-pVDZ are used. We have attributed this difference to differences in the treatment of terms related to state to state transition moments between QR-CC and QR-EOMCC, based on the analysis of the dominant contributions to the hyperpolarizabilties, and the very good agreement between CC and EOMCC dipole polarizabilities.} {That said, t}he overall shape of the dispersion curves remains unchanged, indicating that {in qualitative terms the QR-EOMCC model still behaves quite similarly to QR-CC}.

The contribution of spin-orbit coupling to the hyperpolarizability is rather small up to HBr, but becomes significant from HI onwards, also for the static case, {but the fact that we observe similar trends across the HX series for all Hamiltonians considered indicates that, in spite of the non-additivity between relativisic effects and electron correlation, the differences between QR-CC and QR-EOMCC are driven by electron correlation.}

Truncation of the correlating space appears to affect QR-EOMCC more than QR-CC for the $\beta_{zzz}$ component for HBr, but the additional electron correlation recovered does not modify the trends upon going down the series from HF to HBr. {We note, however, that results can only be considered converged with respect to virtual space truncation if thresholds are set to higher energies (around 100~a.u. in our tests). Furthermore, core(-valence) as a non-negligible effect on the hyperpolarizability of HBr, calling for more systematic studies. We also found orbital relaxation to be small but non-negligible for the polarizability of HBr (1.7\%), and it would be of interest to consider such effect for higher-order properties in future work.}  

In contrast to the hyperpolarizabilies, for the Verdet constant the difference between QR-CC and EOM-CC is minor and roughly constant percentage-wise (about 1\%) for all systems studied. Spin-orbit coupling accounts for a relatively small part of the response (about 4\%) in Xe, becomes important for Rn (about 35-40\%) and absolutely dominates for Og. For the noble gases, correlating the $(n-1)$d electrons in addition to the valence reduces the magnitude of the Verdet constants by about 10\%. {The comparison of the Verdet constants to the Pockels effect for the hydrogen halides further underscored our argument that the differences between QR-CC and QR-EOMCC are likely driven by the description of state to state transition moments for the operators representing the external static fields (magnetic dipole for Verdet, electric dipole for Pockels). We intend to revisit this point in future work, once our implementation allows for the calculation of excited state properties that may shed further light on these differences.}

The calculations presented here, albeit restricted to a relatively small set of atomic and molecular systems serve to highlight similarities, but also sometimes important differences, in behavior between the QR-CC and QR-EOMCC approaches across the periodic table and for different molecular properties. {Given the rather favorable computational cost and implementation complexity of QR-EOMCC with respect to QR-CC}, it is important to further characterize differences between the two approaches in the calculation of nonlinear properties for heavier element systems.

\begin{acknowledgement}

This research used resources of the Oak Ridge Leadership Computing Facility, which is a DOE Office of Science User Facility supported under Contract DE-AC05-00OR22725 (allocations CHM160, CHM191 and CHP109). XY, LH and ASPG acknowledge funding from projects CPER WaveTech, Labex CaPPA (Grant No. ANR-11-LABX-0005-01), ANR CompRIXS (Grant Nos. ANR-19-CE29-0019 and DFG JA 2329/6-1), ANR SCREECHES (Grant Nos. ANR-24-CE29-0904) the I-SITE ULNE project OVERSEE and MESONM International Associated Laboratory (LAI) (Grant No. ANR-16-IDEX-0004), as well support from the French national supercomputing facilities (Grant Nos. DARI A0150801859, A0170801859) and the HPC center at the University of Lille.

\end{acknowledgement}

\begin{suppinfo}

The data (input/output) corresponding to the calculations of this paper are available at the Zenodo repository\cite{zenodo.17215085}. 

The Supporting Information is available free of charge on the \href{http://pubs.acs.org}{ACS Publications website} at DOI: \href{}{XXX}. Working equations for CCSD quadratic response theory; QR-EOMCC equations; Remarks on state to state transition strenghts; Comparison to analytic QR-CC results from other codes; Comparison between analytic QR-CC to finite-field calculations with ExaCorr ; QR-CC and QR-EOMCC $\beta_{zxx}$ and $\beta_{zzz}$ results for different Hamiltonians for HF to HTs systems; Breakdown of the contributions of different terms of the QR-CC and QR-EOMCC response functions for the $\beta_{zxx}$ $\beta_{zzz}$ components, as well as those to obtain the Verdet constant, for HF to HI systems;  Basis set convergence study for QR-CC and QR-EOMCC for the $\beta_{zxx}$ $\beta_{zzz}$ components for HF to HBr systems, using the $^2$DC$^M$ Hamiltonian; Dipole moment and polarizabilities for HF to HTs, and Xe to Og systems; Verdet constants for Xe to Og systems; Excitation energies for Xe to Og systems; sample timings for QR-CC and QR-EOMCC response function evaluations.

\end{suppinfo}


\bibliography{SortedBiblio.bib}

@article{aidas2014d,
  title = {The Dalton quantum chemistry program system},
  volume = {4},
  ISSN = {1759-0884},
  url = {http://dx.doi.org/10.1002/wcms.1172},
  DOI = {10.1002/wcms.1172},
  number = {3},
  journal = {WIREs Computational Molecular Science},
  publisher = {Wiley},
  author = {Aidas,  Kestutis and Angeli,  Celestino and Bak,  Keld L. and Bakken,  Vebjørn and Bast,  Radovan and Boman,  Linus and Christiansen,  Ove and Cimiraglia,  Renzo and Coriani,  Sonia and Dahle,  Pål and Dalskov,  Erik K. and Ekstr\"{o}m,  Ulf and Enevoldsen,  Thomas and Eriksen,  Janus J. and Ettenhuber,  Patrick and Fernández,  Berta and Ferrighi,  Lara and Fliegl,  Heike and Frediani,  Luca and Hald,  Kasper and Halkier,  Asger and H\"{a}ttig,  Christof and Heiberg,  Hanne and Helgaker,  Trygve and Hennum,  Alf Christian and Hettema,  Hinne and Hjertenæs,  Eirik and Høst,  Stinne and Høyvik,  Ida‐Marie and Iozzi,  Maria Francesca and Jansík,  Branislav and Jensen,  Hans Jørgen Aa. and Jonsson,  Dan and Jørgensen,  Poul and Kauczor,  Joanna and Kirpekar,  Sheela and Kjærgaard,  Thomas and Klopper,  Wim and Knecht,  Stefan and Kobayashi,  Rika and Koch,  Henrik and Kongsted,  Jacob and Krapp,  Andreas and Kristensen,  Kasper and Ligabue,  Andrea and Lutnæs,  Ola B. and Melo,  Juan I. and Mikkelsen,  Kurt V. and Myhre,  Rolf H. and Neiss,  Christian and Nielsen,  Christian B. and Norman,  Patrick and Olsen,  Jeppe and Olsen,  Jógvan Magnus H. and Osted,  Anders and Packer,  Martin J. and Pawlowski,  Filip and Pedersen,  Thomas B. and Provasi,  Patricio F. and Reine,  Simen and Rinkevicius,  Zilvinas and Ruden,  Torgeir A. and Ruud,  Kenneth and Rybkin,  Vladimir V. and Sałek,  Pawel and Samson,  Claire C. M. and de Merás,  Alfredo Sánchez and Saue,  Trond and Sauer,  Stephan P. A. and Schimmelpfennig,  Bernd and Sneskov,  Kristian and Steindal,  Arnfinn H. and Sylvester‐Hvid,  Kristian O. and Taylor,  Peter R. and Teale,  Andrew M. and Tellgren,  Erik I. and Tew,  David P. and Thorvaldsen,  Andreas J. and Thøgersen,  Lea and Vahtras,  Olav and Watson,  Mark A. and Wilson,  David J. D. and Ziolkowski,  Marcin and Ågren,  Hans},
  year = {2013},
  month = sep,
  pages = {269–284}
}

@article{Almoukhalalati2016,
  doi = {10.1063/1.4959452},
  url = {https://doi.org/10.1063/1.4959452},
  year = {2016},
  month = aug,
  publisher = {{AIP} Publishing},
  volume = {145},
  number = {7},
  pages = {074104},
  author = {Adel Almoukhalalati and Stefan Knecht and Hans J{\o}rgen Aa. Jensen and Kenneth G. Dyall and Trond Saue},
  title = {Electron correlation within the relativistic no-pair approximation},
  journal = {The Journal of Chemical Physics}
}

@article{andersen2022cherry,
  title = {Cherry-Picking Resolvents: Recovering the Valence Contribution in X-ray Two-Photon Absorption within the Core–Valence-Separated Equation-of-Motion Coupled-Cluster Response Theory},
  volume = {18},
  ISSN = {1549-9626},
  url = {http://dx.doi.org/10.1021/acs.jctc.2c00541},
  DOI = {10.1021/acs.jctc.2c00541},
  number = {10},
  journal = {Journal of Chemical Theory and Computation},
  publisher = {American Chemical Society (ACS)},
  author = {Andersen,  Josefine H. and Nanda,  Kaushik D. and Krylov,  Anna I. and Coriani,  Sonia},
  year = {2022},
  month = sep,
  pages = {6189–6202}
}

@article{andersen2022probing,
  title = {Probing Molecular Chirality of Ground and Electronically Excited States in the UV–vis and X-ray Regimes: An EOM-CCSD Study},
  volume = {18},
  ISSN = {1549-9626},
  url = {http://dx.doi.org/10.1021/acs.jctc.1c00937},
  DOI = {10.1021/acs.jctc.1c00937},
  number = {3},
  journal = {Journal of Chemical Theory and Computation},
  publisher = {American Chemical Society (ACS)},
  author = {Andersen,  Josefine H. and Nanda,  Kaushik D. and Krylov,  Anna I. and Coriani,  Sonia},
  year = {2022},
  month = feb,
  pages = {1748–1764}
}

@article{bartlett2007coupled,
  title = {Coupled-cluster theory in quantum chemistry},
  volume = {79},
  ISSN = {1539-0756},
  url = {http://dx.doi.org/10.1103/RevModPhys.79.291},
  DOI = {10.1103/revmodphys.79.291},
  number = {1},
  journal = {Reviews of Modern Physics},
  publisher = {American Physical Society (APS)},
  author = {Bartlett,  Rodney J. and Musiał,  Monika},
  year = {2007},
  month = feb,
  pages = {291–352}
}

@article{bishop_molecular_1990,
  title = {Molecular vibrational and rotational motion in static and dynamic electric fields},
  volume = {62},
  ISSN = {1539-0756},
  url = {http://dx.doi.org/10.1103/RevModPhys.62.343},
  DOI = {10.1103/revmodphys.62.343},
  number = {2},
  journal = {Reviews of Modern Physics},
  publisher = {American Physical Society (APS)},
  author = {Bishop,  David M.},
  year = {1990},
  month = apr,
  pages = {343–374}
}

@book{boyd2020nonlinear,
	title        = {Nonlinear optics},
	author       = {Boyd, Robert W},
  url = {http://dx.doi.org/10.1016/C2015-0-05510-1},
  DOI = {10.1016/c2015-0-05510-1},
  publisher = {Elsevier},
  year = {2020}
  
}

@article{caricato2009difference,
  title = {On the difference between the transition properties calculated with linear response- and equation of motion-CCSD approaches},
  volume = {131},
  ISSN = {1089-7690},
  url = {http://dx.doi.org/10.1063/1.3255990},
  DOI = {10.1063/1.3255990},
  number = {17},
  journal = {The Journal of Chemical Physics},
  publisher = {AIP Publishing},
  author = {Caricato,  Marco and Trucks,  Gary W. and Frisch,  Michael J.},
  year = {2009},
  month = nov 
}

@article{Christiansen1995a,
  title = {The second-order approximate coupled cluster singles and doubles model CC2},
  volume = {243},
  ISSN = {0009-2614},
  url = {http://dx.doi.org/10.1016/0009-2614(95)00841-q},
  DOI = {10.1016/0009-2614(95)00841-q},
  number = {5–6},
  journal = {Chemical Physics Letters},
  publisher = {Elsevier BV},
  author = {Christiansen,  Ove and Koch,  Henrik and Jørgensen,  Poul},
  year = {1995},
  month = sep,
  pages = {409–418}
}

@article{Christiansen1995b,
  title = {Response functions in the CC3 iterative triple excitation model},
  volume = {103},
  ISSN = {1089-7690},
  url = {http://dx.doi.org/10.1063/1.470315},
  DOI = {10.1063/1.470315},
  number = {17},
  journal = {The Journal of Chemical Physics},
  publisher = {AIP Publishing},
  year = {1995},
  month = nov,
  pages = {7429–7441},
  author = {Christiansen,  Ove and Koch,  Henrik and J{\o}rgensen,  Poul}
}

@article{Christiansen1998,
  title = {Response functions from Fourier component variational perturbation theory applied to a time-averaged quasienergy},
	author       = {Ove Christiansen and Poul J{\o}rgensen and Christof H{\"a}ttig},
  volume = {68},
  ISSN = {1097-461X},
  url = {http://dx.doi.org/10.1002/(sici)1097-461x(1998)68:1<1::aid-qua1>3.0.co;2-z},
  DOI = {10.1002/(sici)1097-461x(1998)68:1<1::aid-qua1>3.0.co;2-z},
  number = {1},
  journal = {International Journal of Quantum Chemistry},
  publisher = {Wiley},
  year = {1998},
  pages = {1–52}
}

@article{coriani_molecular_2016,
	author       = {Coriani, Sonia and Paw{\l}owski, Filip and Olsen, Jeppe and J{\o}rgensen, Poul},
  title = {Molecular response properties in equation of motion coupled cluster theory: A time-dependent perspective},
  volume = {144},
  ISSN = {1089-7690},
  url = {http://dx.doi.org/10.1063/1.4939183},
  DOI = {10.1063/1.4939183},
  number = {2},
  journal = {The Journal of Chemical Physics},
  publisher = {AIP Publishing},
  year = {2016},
  month = jan 
}

@article{crawford2007introduction,
  title = {An Introduction to Coupled Cluster Theory for Computational Chemists},
  ISBN = {9780470125915},
  ISSN = {1934-5372},
  url = {http://dx.doi.org/10.1002/9780470125915.ch2},
  DOI = {10.1002/9780470125915.ch2},
  journal = {Reviews in Computational Chemistry},
  publisher = {Wiley},
  author = {Crawford,  T. Daniel and Schaefer,  Henry F.},
  year = {2000},
  month = jan,
  pages = {33–136}
}

@inbook{Crawford2009,
  title = {On the Performance of a Size-Extensive Variant of Equation-of-Motion Coupled Cluster Theory for Optical Rotation in Chiral Molecules},
  ISBN = {9789048125968},
  ISSN = {2215-0129},
  url = {http://dx.doi.org/10.1007/978-90-481-2596-8_10},
  DOI = {10.1007/978-90-481-2596-8_10},
  booktitle = {Advances in the Theory of Atomic and Molecular Systems},
  publisher = {Springer Netherlands},
  author = {Crawford,  T. Daniel and Sekino,  Hideo},
  year = {2009},
  pages = {225–239}
}

@article{cronstrand_multi-photon_2005,
  title = {Multi-Photon Absorption of Molecules},
  ISSN = {0065-3276},
  url = {http://dx.doi.org/10.1016/S0065-3276(05)50001-7},
  DOI = {10.1016/s0065-3276(05)50001-7},
  booktitle = {Response Theory and Molecular Properties (A Tribute to Jan Linderberg and Poul Jørgensen)},
  journal = {Advances in Quantum Chemistry},
  publisher = {Elsevier},
  author = {Cronstrand,  Peter and Luo,  Yi and Ågren,  Hans},
  year = {2005},
  pages = {1–21}
}

@article{dalgaard1980time,
  title = {Time-dependent multiconfigurational Hartree–Fock theory},
  volume = {72},
  ISSN = {1089-7690},
  url = {http://dx.doi.org/10.1063/1.439233},
  DOI = {10.1063/1.439233},
  number = {2},
  journal = {The Journal of Chemical Physics},
  publisher = {AIP Publishing},
  author = {Dalgaard,  Esper},
  year = {1980},
  month = jan,
  pages = {816–823}
}

@misc{Dalton2020p1,
	note         = {{Dalton}, a molecular electronic structure program, Release 2020.1 (2022), see \url{http://daltonprogram.org}}
}

@article{dyall2006relativistic,
  title = {Relativistic Quadruple-Zeta and Revised Triple-Zeta and Double-Zeta Basis Sets for the 4p,  5p,  and 6p Elements},
  volume = {115},
  ISSN = {1432-2234},
  url = {http://dx.doi.org/10.1007/s00214-006-0126-0},
  DOI = {10.1007/s00214-006-0126-0},
  number = {5},
  journal = {Theoretical Chemistry Accounts},
  publisher = {Springer Science and Business Media LLC},
  author = {Dyall,  Kenneth G.},
  year = {2006},
  month = mar,
  pages = {441–447}
}

@article{faber2018resonant,
  title = {Resonant Inelastic X-ray Scattering and Nonesonant X-ray Emission Spectra from Coupled-Cluster (Damped) Response Theory},
  volume = {15},
  ISSN = {1549-9626},
  url = {http://dx.doi.org/10.1021/acs.jctc.8b01020},
  DOI = {10.1021/acs.jctc.8b01020},
  number = {1},
  journal = {Journal of Chemical Theory and Computation},
  publisher = {American Chemical Society (ACS)},
  author = {Faber,  Rasmus and Coriani,  Sonia},
  year = {2018},
  month = nov,
  pages = {520–528}
}

@article{gauss_coupledcluster_1995,
  title = {Coupled-cluster calculations of nuclear magnetic resonance chemical shifts},
  volume = {103},
  ISSN = {1089-7690},
  url = {http://dx.doi.org/10.1063/1.470240},
  DOI = {10.1063/1.470240},
  number = {9},
  journal = {The Journal of Chemical Physics},
  publisher = {AIP Publishing},
  author = {Gauss,  J\"{u}rgen and Stanton,  John F.},
  year = {1995},
  month = sep,
  pages = {3561–3577}
}

@misc{githubRepo2023,
    author = {Lyakh, Dmitry I},
    title = {{TAL-SH: Tensor Algebra Library for Shared Memory Computers}},
    year = {2023},
    howpublished ={\url{github.com/https:/DmitryLyakh/TAL_SH} (accessed 2025-11-11)}
}

@article{hattig_cc2_2000,
  title = {CC2 excitation energy calculations on large molecules using the resolution of the identity approximation},
  volume = {113},
  ISSN = {1089-7690},
  url = {http://dx.doi.org/10.1063/1.1290013},
  DOI = {10.1063/1.1290013},
  number = {13},
  journal = {The Journal of Chemical Physics},
  publisher = {AIP Publishing},
  author = {H\"{a}ttig,  Christof and Weigend,  Florian},
  year = {2000},
  month = oct,
  pages = {5154–5161}
}

@article{he2008multiphoton,
  title = {Multiphoton Absorbing Materials: Molecular Designs, Characterizations, and Applications},
  volume = {108},
  ISSN = {1520-6890},
  url = {http://dx.doi.org/10.1021/cr050054x},
  DOI = {10.1021/cr050054x},
  number = {4},
  journal = {Chemical Reviews},
  publisher = {American Chemical Society (ACS)},
  author = {He,  Guang S. and Tan,  Loon-Seng and Zheng,  Qingdong and Prasad,  Paras N.},
  year = {2008},
  month = mar,
  pages = {1245–1330}
}

@article{helgaker1999ab,
  title = {Ab Initio Methods for the Calculation of NMR Shielding and Indirect Spin-Spin Coupling Constants},
  volume = {99},
  ISSN = {1520-6890},
  url = {http://dx.doi.org/10.1021/cr960017t},
  DOI = {10.1021/cr960017t},
  number = {1},
  journal = {Chemical Reviews},
  publisher = {American Chemical Society (ACS)},
  author = {Helgaker,  Trygve and Jaszuński,  Michał and Ruud,  Kenneth},
  year = {1998},
  month = dec,
  pages = {293–352}
}

@article{helgaker2008quantum,
  title = {The quantum-chemical calculation of NMR indirect spin–spin coupling constants},
  volume = {53},
  ISSN = {0079-6565},
  url = {http://dx.doi.org/10.1016/j.pnmrs.2008.02.002},
  DOI = {10.1016/j.pnmrs.2008.02.002},
  number = {4},
  journal = {Progress in Nuclear Magnetic Resonance Spectroscopy},
  publisher = {Elsevier BV},
  author = {Helgaker,  Trygve and Jaszuński,  Michał and Pecul,  Magdalena},
  year = {2008},
  month = nov,
  pages = {249–268}
}

@article{helgaker_recent_2012,
	author       = {Helgaker, Trygve and Coriani, Sonia and J{\o}rgensen, Poul and Kristensen, Kasper and Olsen, Jeppe and Ruud, Kenneth},
  title = {Recent Advances in Wave Function-Based Methods of Molecular-Property Calculations},
  volume = {112},
  ISSN = {1520-6890},
  url = {http://dx.doi.org/10.1021/cr2002239},
  DOI = {10.1021/cr2002239},
  number = {1},
  journal = {Chemical Reviews},
  publisher = {American Chemical Society (ACS)},
  year = {2012},
  month = jan,
  pages = {543–631}
}

@book{helgaker_molecular_2014,
	title = {Molecular {Electronic}-{Structure} {Theory}},
	isbn = {9780471967552},
	url = {https://www.wiley.com/en-us/Molecular+Electronic+Structure+Theory-p-9780471967552},
	language = {English},
	urldate = {2021-08-27},
	author = {Helgaker, Trygve and J{\o}rgensen, Poul and Olsen, Jeppe},
	year = {2014},
        doi={10.1002/9781119019572},
}

@article{hirao1982generalization,
  title = {A generalization of the Davidson’s method to large nonsymmetric eigenvalue problems},
  volume = {45},
  ISSN = {0021-9991},
  url = {http://dx.doi.org/10.1016/0021-9991(82)90119-X},
  DOI = {10.1016/0021-9991(82)90119-x},
  number = {2},
  journal = {Journal of Computational Physics},
  publisher = {Elsevier BV},
  author = {Hirao,  K. and Nakatsuji,  H.},
  year = {1982},
  month = feb,
  pages = {246–254}
}

@book{huber1979constants,
  title = {Molecular Spectra and Molecular Structure},
  ISBN = {9781475709612},
  url = {http://dx.doi.org/10.1007/978-1-4757-0961-2},
  DOI = {10.1007/978-1-4757-0961-2},
  publisher = {Springer New York, NY},
  author = {Huber,  K. P. and Herzberg,  G.},
  year = {1979}
}

@article{iliavs2007infinite,
	author       = {Ilia{\v{s}}, Miroslav and Saue, Trond},
  title = {An infinite-order two-component relativistic Hamiltonian by a simple one-step transformation},
  volume = {126},
  ISSN = {1089-7690},
  url = {http://dx.doi.org/10.1063/1.2436882},
  DOI = {10.1063/1.2436882},
  number = {6},
  journal = {The Journal of Chemical Physics},
  publisher = {AIP Publishing},
  year = {2007},
  month = feb 
}

@article{kauczor2013communication,
  title = {Communication: A reduced-space algorithm for the solution of the complex linear response equations used in coupled cluster damped response theory},
  volume = {139},
  ISSN = {1089-7690},
  url = {http://dx.doi.org/10.1063/1.4840275},
  DOI = {10.1063/1.4840275},
  number = {21},
  journal = {The Journal of Chemical Physics},
  publisher = {AIP Publishing},
  author = {Kauczor,  Joanna and Norman,  Patrick and Christiansen,  Ove and Coriani,  Sonia},
  year = {2013},
  month = dec 
}

@article{kendall1992electron,
  title = {Electron affinities of the first-row atoms revisited. Systematic basis sets and wave functions},
  volume = {96},
  ISSN = {1089-7690},
  url = {http://dx.doi.org/10.1063/1.462569},
  DOI = {10.1063/1.462569},
  number = {9},
  journal = {The Journal of Chemical Physics},
  publisher = {AIP Publishing},
  author = {Kendall,  Rick A. and Dunning,  Thom H. and Harrison,  Robert J.},
  year = {1992},
  month = may,
  pages = {6796–6806}
}

@article{khani_uv_2019,
  title = {UV Absorption and Magnetic Circular Dichroism Spectra of Purine,  Adenine,  and Guanine: A Coupled Cluster Study in Vacuo and in Aqueous Solution},
  volume = {15},
  ISSN = {1549-9626},
  url = {http://dx.doi.org/10.1021/acs.jctc.8b00930},
  DOI = {10.1021/acs.jctc.8b00930},
  number = {2},
  journal = {Journal of Chemical Theory and Computation},
  publisher = {American Chemical Society (ACS)},
  author = {Khani,  Sarah Karbalaei and Faber,  Rasmus and Santoro,  Fabrizio and H\"{a}ttig,  Christof and Coriani,  Sonia},
  year = {2018},
  month = nov,
  pages = {1242–1254}
}

@article{kobayashi1994calculation,
	title        = {Calculation of frequency-dependent polarizabilities using coupled-cluster response theory},
	author       = {Kobayashi, Rika and Koch, Henrik and J{\o}rgensen, Poul},
  volume = {219},
  ISSN = {0009-2614},
  url = {http://dx.doi.org/10.1016/0009-2614(94)00051-4},
  DOI = {10.1016/0009-2614(94)00051-4},
  number = {1–2},
  journal = {Chemical Physics Letters},
  publisher = {Elsevier BV},
  year = {1994},
  month = mar,
  pages = {30–35}
}

@article{Koch1990a,
  title = {Coupled cluster response functions},
  volume = {93},
  ISSN = {1089-7690},
  url = {http://dx.doi.org/10.1063/1.458814},
  DOI = {10.1063/1.458814},
  number = {5},
  journal = {The Journal of Chemical Physics},
  publisher = {AIP Publishing},
  author = {Koch,  Henrik and J{\o}rgensen,  Poul},
  year = {1990},
  month = sep,
  pages = {3333–3344}
}

@article{Koch1994,
  title = {Calculation of size-intensive transition moments from the coupled cluster singles and doubles linear response function},
  volume = {100},
  ISSN = {1089-7690},
  url = {http://dx.doi.org/10.1063/1.466321},
  DOI = {10.1063/1.466321},
  number = {6},
  journal = {The Journal of Chemical Physics},
  publisher = {AIP Publishing},
  author = {Koch,  Henrik and Kobayashi,  Rika and Sanchez de Merás,  Alfredo and J{\o}rgensen,  Poul},
  year = {1994},
  month = mar,
  pages = {4393–4400}
}

@article{langhoff1972aspects,
  title = {Aspects of Time-Dependent Perturbation Theory},
  volume = {44},
  ISSN = {0034-6861},
  url = {http://dx.doi.org/10.1103/RevModPhys.44.602},
  DOI = {10.1103/revmodphys.44.602},
  number = {3},
  journal = {Reviews of Modern Physics},
  publisher = {American Physical Society (APS)},
  author = {Langhoff,  P. W. and Epstein,  S. T. and Karplus,  M.},
  year = {1972},
  month = jul,
  pages = {602–644}
}

@article{levy1967nonrelativistic,
	author       = {L{\'e}vy-Leblond, Jean-Marc},
  title = {Nonrelativistic particles and wave equations},
  volume = {6},
  ISSN = {1432-0916},
  url = {http://dx.doi.org/10.1007/BF01646020},
  DOI = {10.1007/bf01646020},
  number = {4},
  journal = {Communications in Mathematical Physics},
  publisher = {Springer Science and Business Media LLC},
  year = {1967},
  month = dec,
  pages = {286–311}
}

@article{liu_going_2013,
 title = {Going beyond “no-pair relativistic quantum chemistry”},
  volume = {139},
  ISSN = {1089-7690},
  url = {http://dx.doi.org/10.1063/1.4811795},
  DOI = {10.1063/1.4811795},
  number = {1},
  journal = {The Journal of Chemical Physics},
  publisher = {AIP Publishing},
  author = {Liu,  Wenjian and Lindgren,  Ingvar},
  year = {2013},
  month = jul 
}

@book{liu2017handbook,
	title        = {Handbook of Relativistic Quantum Chemistry},
	author       = {Liu, Wenjian},
	year         = 2017,
  url = {http://dx.doi.org/10.1007/978-3-642-40766-6},
  DOI = {10.1007/978-3-642-40766-6},
	publisher    = {Springer Berlin Heidelberg Berlin, Heidelberg}

}

@article{lyakh_domainspecific_2019,
  title = {Domain‐specific virtual processors as a portable programming and execution model for parallel computational workloads on modern heterogeneous high‐performance computing architectures},
  volume = {119},
  ISSN = {1097-461X},
  url = {http://dx.doi.org/10.1002/qua.25926},
  DOI = {10.1002/qua.25926},
  number = {12},
  journal = {International Journal of Quantum Chemistry},
  publisher = {Wiley},
  author = {Lyakh,  Dmitry I.},
  year = {2019},
  month = mar 
}

@article{matthews2020coupled,
  title = {Coupled-cluster techniques for computational chemistry: The CFOUR program package},
  volume = {152},
  ISSN = {1089-7690},
  url = {http://dx.doi.org/10.1063/5.0004837},
  DOI = {10.1063/5.0004837},
  number = {21},
  journal = {The Journal of Chemical Physics},
  publisher = {AIP Publishing},
  author = {Matthews,  Devin A. and Cheng,  Lan and Harding,  Michael E. and Lipparini,  Filippo and Stopkowicz,  Stella and Jagau,  Thomas-C. and Szalay,  Péter G. and Gauss,  J\"{u}rgen and Stanton,  John F.},
  year = {2020},
  month = jun
}

@article{Nanda2015,
  title = {Two-photon absorption cross sections within equation-of-motion coupled-cluster formalism using resolution-of-the-identity and Cholesky decomposition representations: Theory,  implementation,  and benchmarks},
  volume = {142},
  ISSN = {1089-7690},
  url = {http://dx.doi.org/10.1063/1.4907715},
  DOI = {10.1063/1.4907715},
  number = {6},
  journal = {The Journal of Chemical Physics},
  publisher = {AIP Publishing},
  author = {Nanda,  Kaushik D. and Krylov,  Anna I.},
  year = {2015},
  month = feb 
}

@article{nanda_communication_2018,
  title = {Communication: The pole structure of the dynamical polarizability tensor in equation-of-motion coupled-cluster theory},
  volume = {149},
  ISSN = {1089-7690},
  url = {http://dx.doi.org/10.1063/1.5053727},
  DOI = {10.1063/1.5053727},
  number = {14},
  journal = {The Journal of Chemical Physics},
  publisher = {AIP Publishing},
  author = {Nanda,  Kaushik D. and Krylov,  Anna I. and Gauss,  J\"{u}rgen},
  year = {2018},
  month = oct 
}

@article{nanda_static_2016,
  title = {Static polarizabilities for excited states within the spin-conserving and spin-flipping equation-of-motion coupled-cluster singles and doubles formalism: Theory,  implementation,  and benchmarks},
  volume = {145},
  ISSN = {1089-7690},
  url = {http://dx.doi.org/10.1063/1.4967860},
  DOI = {10.1063/1.4967860},
  number = {20},
  journal = {The Journal of Chemical Physics},
  publisher = {AIP Publishing},
  author = {Nanda,  Kaushik D. and Krylov,  Anna I.},
  year = {2016},
  month = nov 
}

@article{Stanton1993,
  title = {A coupled-cluster based effective Hamiltonian method for dynamic electric polarizabilities},
  volume = {99},
  ISSN = {1089-7690},
  url = {http://dx.doi.org/10.1063/1.466019},
  DOI = {10.1063/1.466019},
  number = {7},
  journal = {The Journal of Chemical Physics},
  publisher = {AIP Publishing},
  author = {Stanton,  John F. and Bartlett,  Rodney J.},
  year = {1993},
  month = oct,
  pages = {5178–5183}
}

@incollection{NISTconstants,
  doi = {10.18434/T4T59X},
  url = {http://www.nist.gov/pml/data/msd-di/index.cfm},
  author = {Lovas,  Frank},
  language = {en},
  title = {Diatomic Spectral Database,  NIST Standard Reference Database 114},
  publisher = {National Institute of Standards and Technology},
  year = {2002},
  copyright = {License Information for NIST data}
}

@misc{NISTAtomic,
  doi = {10.18434/T4W30F},
  url = {http://www.nist.gov/pml/data/asd.cfm},
  author = {Kramida,  Alexander and Ralchenko,  Yuri},
  language = {en},
  title = {NIST Atomic Spectra Database,  NIST Standard Reference Database 78},
  publisher = {National Institute of Standards and Technology},
  year = {1999},
  copyright = {License Information for NIST data}
}

@article{norman_perspective_2011,
  title = {A perspective on nonresonant and resonant electronic response theory for time-dependent molecular properties},
  volume = {13},
  ISSN = {1463-9084},
  url = {http://dx.doi.org/10.1039/c1cp21951k},
  DOI = {10.1039/c1cp21951k},
  number = {46},
  journal = {Physical Chemistry Chemical Physics},
  publisher = {Royal Society of Chemistry (RSC)},
  author = {Norman,  Patrick},
  year = {2011},
  pages = {20519}
}

@book{norman_principles_2018,
  title = {Principles and Practices of Molecular Properties: Theory,  Modeling and Simulations},
  ISBN = {9781118794821},
  url = {http://dx.doi.org/10.1002/9781118794821},
  DOI = {10.1002/9781118794821},
  publisher = {Wiley},
  author = {Norman,  Patrick and Ruud,  Kenneth and Saue,  Trond},
  year = {2018},
  month = feb 
}

@article{olsen1988solution,
	author       = {Olsen, Jeppe and Jensen, Hans J{\o}rgen Aa and J{\o}rgensen, Poul},
  title = {Solution of the large matrix equations which occur in response theory},
  volume = {74},
  ISSN = {0021-9991},
  url = {http://dx.doi.org/10.1016/0021-9991(88)90081-2},
  DOI = {10.1016/0021-9991(88)90081-2},
  number = {2},
  journal = {Journal of Computational Physics},
  publisher = {Elsevier BV},
  year = {1988},
  month = feb,
  pages = {265–282}
}

@article{olsen_linear_1985,
	author       = {Olsen, Jeppe and J{\o}rgensen, Poul},
  title = {Linear and nonlinear response functions for an exact state and for an MCSCF state},
  volume = {82},
  ISSN = {1089-7690},
  url = {http://dx.doi.org/10.1063/1.448223},
  DOI = {10.1063/1.448223},
  number = {7},
  journal = {The Journal of Chemical Physics},
  publisher = {AIP Publishing},
  year = {1985},
  month = apr,
  pages = {3235–3264}
}

@book{papadopoulos_non-linear_2006,
	title        = {Non-linear optical properties of matter: from molecules to condensed phases},
	year         = 2006,
  url = {http://dx.doi.org/10.1007/1-4020-4850-5},
  DOI = {10.1007/1-4020-4850-5},
	publisher    = {Springer},
	number       = 1,
	isbn         = {978-1-4020-4849-4 978-1-4020-4850-0},
	language     = {en},
	editor       = {Papadopoulos, Manthos G. and Sadlej, Andrzej Jerzy and Leszczynski, Jerzy},
}

@article{pawlowski_molecular_2015,
  title = {Molecular response properties from a Hermitian eigenvalue equation for a time-periodic Hamiltonian},
  volume = {142},
  ISSN = {1089-7690},
  url = {http://dx.doi.org/10.1063/1.4913364},
  DOI = {10.1063/1.4913364},
  number = {11},
  journal = {The Journal of Chemical Physics},
  publisher = {AIP Publishing},
  author = {Pawłowski,  Filip and Olsen,  Jeppe and Jørgensen,  Poul},
  year = {2015},
  month = mar 
}

@article{Koch1990b,
  title = {Coupled cluster energy derivatives. Analytic Hessian for the closed-shell coupled cluster singles and doubles wave function: Theory and applications},
  volume = {92},
  ISSN = {1089-7690},
  url = {http://dx.doi.org/10.1063/1.457710},
  DOI = {10.1063/1.457710},
  number = {8},
  journal = {The Journal of Chemical Physics},
  publisher = {AIP Publishing},
  author = {Koch,  Henrik and Jensen,  Hans J{\o}rgen Aa. and J{\o}rgensen,  Poul and Helgaker,  Trygve and Scuseria,  Gustavo E. and Schaefer,  Henry F.},
  year = {1990},
  month = apr,
  pages = {4924–4940}
}

@inbook{Pedersen2015,
  title = {Introduction to Response Theory},
  ISBN = {9789400761698},
  url = {http://dx.doi.org/10.1007/978-94-007-6169-8_5-2},
  DOI = {10.1007/978-94-007-6169-8_5-2},
  booktitle = {Handbook of Computational Chemistry},
  publisher = {Springer Netherlands},
  author = {Pedersen,  Thomas Bondo},
  year = {2015},
  pages = {1–26}
}

@article{pototschnig_implementation_2021,
  title = {Implementation of Relativistic Coupled Cluster Theory for Massively Parallel GPU-Accelerated Computing Architectures},
  volume = {17},
  ISSN = {1549-9626},
  url = {http://dx.doi.org/10.1021/acs.jctc.1c00260},
  DOI = {10.1021/acs.jctc.1c00260},
  number = {9},
  journal = {Journal of Chemical Theory and Computation},
  publisher = {American Chemical Society (ACS)},
  author = {Pototschnig,  Johann V. and Papadopoulos,  Anastasios and Lyakh,  Dmitry I. and Repisky,  Michal and Halbert,  Loïc and Severo Pereira Gomes,  André and Jensen,  Hans Jørgen Aa and Visscher,  Lucas},
  year = {2021},
  month = aug,
  pages = {5509–5529}
}

@article{rice1991calculation,
  title = {The calculation of frequency-dependent polarizabilities as pseudo-energy derivatives},
  volume = {94},
  ISSN = {1089-7690},
  url = {http://dx.doi.org/10.1063/1.460558},
  DOI = {10.1063/1.460558},
  number = {7},
  journal = {The Journal of Chemical Physics},
  publisher = {AIP Publishing},
  author = {Rice,  Julia E. and Handy,  Nicholas C.},
  year = {1991},
  month = apr,
  pages = {4959–4971}
}

@article{salek2005linear,
	title        = {Linear response at the 4-component relativistic density-functional level: application to the frequency-dependent dipole polarizability of Hg, AuH and PtH$_2$},
	author       = {Sa{\l}ek, Pawel and Helgaker, Trygve and Saue, Trond},
  volume = {311},
  ISSN = {0301-0104},
  url = {http://dx.doi.org/10.1016/j.chemphys.2004.10.011},
  DOI = {10.1016/j.chemphys.2004.10.011},
  number = {1–2},
  journal = {Chemical Physics},
  publisher = {Elsevier BV},
  year = {2005},
  month = apr,
  pages = {187–201}
}

@article{sasagane1993higher,
  title = {Higher-order response theory based on the quasienergy derivatives: The derivation of the frequency-dependent polarizabilities and hyperpolarizabilities},
  volume = {99},
  ISSN = {1089-7690},
  url = {http://dx.doi.org/10.1063/1.466123},
  DOI = {10.1063/1.466123},
  number = {5},
  journal = {The Journal of Chemical Physics},
  publisher = {AIP Publishing},
  author = {Sasagane,  Kotoku and Aiga,  Fumihiko and Itoh,  Reikichi},
  year = {1993},
  month = sep,
  pages = {3738–3778}
}

@article{saue2002four,
  title = {Four‐component relativistic Kohn–Sham theory},
  volume = {23},
  ISSN = {1096-987X},
  url = {http://dx.doi.org/10.1002/jcc.10066},
  DOI = {10.1002/jcc.10066},
  number = {8},
  journal = {Journal of Computational Chemistry},
  publisher = {Wiley},
  author = {Saue,  Trond and Helgaker,  Trygve},
  year = {2002},
  month = apr,
  pages = {814–823}
}

@inbook{Saue2002_PostDirac,
  title = {Chapter 7 Post Dirac-Hartree-Fock methods—properties},
  ISSN = {1380-7323},
  url = {http://dx.doi.org/10.1016/s1380-7323(02)80033-4},
  DOI = {10.1016/s1380-7323(02)80033-4},
  booktitle = {Relativistic Electronic Structure Theory},
  publisher = {Elsevier},
  author = {Saue,  Trond},
  year = {2002},
  pages = {332–400}
}

@article{saue2020dirac,
  title = {The DIRAC code for relativistic molecular calculations},
  volume = {152},
  ISSN = {1089-7690},
  url = {http://dx.doi.org/10.1063/5.0004844},
  DOI = {10.1063/5.0004844},
  number = {20},
  journal = {The Journal of Chemical Physics},
  publisher = {AIP Publishing},
  author = {Saue,  Trond and Bast,  Radovan and Gomes,  André Severo Pereira and Jensen,  Hans Jørgen Aa. and Visscher,  Lucas and Aucar,  Ignacio Agustín and Di Remigio,  Roberto and Dyall,  Kenneth G. and Eliav,  Ephraim and Fasshauer,  Elke and Fleig,  Timo and Halbert,  Loïc and Hedegård,  Erik Donovan and Helmich-Paris,  Benjamin and Iliaš,  Miroslav and Jacob,  Christoph R. and Knecht,  Stefan and Laerdahl,  Jon K. and Vidal,  Marta L. and Nayak,  Malaya K. and Olejniczak,  Małgorzata and Olsen,  Jógvan Magnus Haugaard and Pernpointner,  Markus and Senjean,  Bruno and Shee,  Avijit and Sunaga,  Ayaki and van Stralen,  Joost N. P.},
  year = {2020},
  month = may 
}

@article{saue_relativistic_2011,
  title = {Relativistic Hamiltonians for Chemistry: A Primer},
  volume = {12},
  ISSN = {1439-7641},
  url = {http://dx.doi.org/10.1002/cphc.201100682},
  DOI = {10.1002/cphc.201100682},
  number = {17},
  journal = {ChemPhysChem},
  publisher = {Wiley},
  author = {Saue,  Trond},
  year = {2011},
  month = nov,
  pages = {3077–3094}
}

@article{Schimmelpfennig.Gropen.1998f8, 
  title = {On the combination of ECP-based CI calculations with all-electron spin-orbit mean-field integrals},
  volume = {286},
  ISSN = {0009-2614},
  url = {http://dx.doi.org/10.1016/S0009-2614(98)00121-3},
  DOI = {10.1016/s0009-2614(98)00121-3},
  number = {3–4},
  journal = {Chemical Physics Letters},
  publisher = {Elsevier BV},
  author = {Schimmelpfennig,  Bernd and Maron,  Laurent and Wahlgren,  Ulf and Teichteil,  Christian and Fagerli,  Hilde and Gropen,  Odd},
  year = {1998},
  month = apr,
  pages = {267–271}
}

@article{Sekino1994,
  title = {Nuclear coupling constants obtained by the equation-of-motion coupled cluster theory},
  volume = {225},
  ISSN = {0009-2614},
  url = {http://dx.doi.org/10.1016/0009-2614(94)87116-7},
  DOI = {10.1016/0009-2614(94)87116-7},
  number = {4–6},
  journal = {Chemical Physics Letters},
  publisher = {Elsevier BV},
  author = {Sekino,  Hideo and Bartlett,  Rodney J.},
  year = {1994},
  month = aug,
  pages = {486–493}
}

@incollection{Sekino1999,
  doi = {10.1016/s0065-3276(08)60459-1},
  url = {https://doi.org/10.1016/s0065-3276(08)60459-1},
  year = {1999},
  publisher = {Elsevier},
  pages = {149--173},
  author = {Hideo Sekino and Rodney J. Bartlett},
  title = {On the Extensivity Problem in Coupled-Cluster Property Evaluation},
  booktitle = {Advances in Quantum Chemistry}
}

@book{shavitt2009many,
	title        = {Many-Body Methods in Chemistry and Physics: MBPT and Coupled-Cluster Theory},
	author       = {Shavitt, I. and Bartlett, R.J.},
	year         = 2009,
	publisher    = {Cambridge University Press},
	series       = {Cambridge Molecular Science},
	isbn         = 9780521818322,
	url          = {https://books.google.gl/books?id=gU1eHAAACAAJ},
	lccn         = 2009517972
}

@article{shee2018equation,
	author       = {Shee, Avijit and Saue, Trond and Visscher, Lucas and  Gomes, Andr{\'e} Severo Pereira},
  title = {Equation-of-motion coupled-cluster theory based on the 4-component Dirac–Coulomb(–Gaunt) Hamiltonian. Energies for single electron detachment,  attachment,  and electronically excited states},
  volume = {149},
  ISSN = {1089-7690},
  url = {http://dx.doi.org/10.1063/1.5053846},
  DOI = {10.1063/1.5053846},
  number = {17},
  journal = {The Journal of Chemical Physics},
  publisher = {AIP Publishing},
  year = {2018},
  month = nov 
}

@article{shelton_measurements_1994,
  title = {Measurements and calculations of the hyperpolarizabilities of atoms and small molecules in the gas phase},
  volume = {94},
  ISSN = {1520-6890},
  url = {http://dx.doi.org/10.1021/cr00025a001},
  DOI = {10.1021/cr00025a001},
  number = {1},
  journal = {Chemical Reviews},
  publisher = {American Chemical Society (ACS)},
  author = {Shelton,  David P. and Rice,  Julia E.},
  year = {1994},
  month = jan,
  pages = {3–29}
}

@article{sikkema2009molecular,
	author       = {Sikkema, Jetze and Visscher, Lucas and Saue, Trond and Ilia{\v{s}}, Miroslav},
  title = {The molecular mean-field approach for correlated relativistic calculations},
  volume = {131},
  ISSN = {1089-7690},
  url = {http://dx.doi.org/10.1063/1.3239505},
  DOI = {10.1063/1.3239505},
  number = {12},
  journal = {The Journal of Chemical Physics},
  publisher = {AIP Publishing},
  year = {2009},
  month = sep 
}

@article{vaara2007theory,
  title = {Theory and computation of nuclear magnetic resonance parameters},
  volume = {9},
  ISSN = {1463-9084},
  url = {http://dx.doi.org/10.1039/B706135H},
  DOI = {10.1039/b706135h},
  number = {40},
  journal = {Physical Chemistry Chemical Physics},
  publisher = {Royal Society of Chemistry (RSC)},
  author = {Vaara,  Juha},
  year = {2007},
  pages = {5399}
}

@article{visscher1996formulation,
  title = {Formulation and implementation of a relativistic unrestricted coupled-cluster method including noniterative connected triples},
  volume = {105},
  ISSN = {1089-7690},
  url = {http://dx.doi.org/10.1063/1.472655},
  DOI = {10.1063/1.472655},
  number = {19},
  journal = {The Journal of Chemical Physics},
  publisher = {AIP Publishing},
  author = {Visscher,  Lucas and Lee,  Timothy J. and Dyall,  Kenneth G.},
  year = {1996},
  month = nov,
  pages = {8769–8776}
}

@article{visscher1997dirac,
	title        = {Dirac--Fock atomic electronic structure calculations using different nuclear charge distributions},
  volume = {67},
  ISSN = {0092-640X},
  url = {http://dx.doi.org/10.1006/adnd.1997.0751},
  DOI = {10.1006/adnd.1997.0751},
  number = {2},
  journal = {Atomic Data and Nuclear Data Tables},
  publisher = {Elsevier BV},
  author = {Visscher,  L. and Dyall,  K.G.},
  year = {1997},
  month = nov,
  pages = {207–224}
}

@article{visscher2000approximate,
  title = {Approximate relativistic electronic structure methods based on the quaternion modified Dirac equation},
  volume = {113},
  ISSN = {1089-7690},
  url = {http://dx.doi.org/10.1063/1.1288371},
  DOI = {10.1063/1.1288371},
  number = {10},
  journal = {The Journal of Chemical Physics},
  publisher = {AIP Publishing},
  author = {Visscher,  Lucas and Saue,  Trond},
  year = {2000},
  month = sep,
  pages = {3996–4002}
}

@article{visscher_4-component_1997,
  title = {The 4-component random phase approximation method applied to the calculation of frequency-dependent dipole polarizabilities},
  volume = {274},
  ISSN = {0009-2614},
  url = {http://dx.doi.org/10.1016/S0009-2614(97)00675-1},
  DOI = {10.1016/s0009-2614(97)00675-1},
  number = {1–3},
  journal = {Chemical Physics Letters},
  publisher = {Elsevier BV},
  author = {Visscher,  Lucas and Saue,  Trond and Oddershede,  Jens},
  year = {1997},
  month = aug,
  pages = {181–188}
}

@article{wilson1999gaussian,
  title = {Gaussian basis sets for use in correlated molecular calculations. IX. The atoms gallium through krypton},
  volume = {110},
  ISSN = {1089-7690},
  url = {http://dx.doi.org/10.1063/1.478678},
  DOI = {10.1063/1.478678},
  number = {16},
  journal = {The Journal of Chemical Physics},
  publisher = {AIP Publishing},
  author = {Wilson,  Angela K. and Woon,  David E. and Peterson,  Kirk A. and Dunning,  Thom H.},
  year = {1999},
  month = apr,
  pages = {7667–7676}
}

@article{woon1993gaussian,
  title = {Gaussian basis sets for use in correlated molecular calculations. III. The atoms aluminum through argon},
  volume = {98},
  ISSN = {1089-7690},
  url = {http://dx.doi.org/10.1063/1.464303},
  DOI = {10.1063/1.464303},
  number = {2},
  journal = {The Journal of Chemical Physics},
  publisher = {AIP Publishing},
  author = {Woon,  David E. and Dunning,  Thom H.},
  year = {1993},
  month = jan,
  pages = {1358–1371}
}

@article{Yuan_Xiang_LR,
	title = {Formulation and {Implementation} of {Frequency}-{Dependent} {Linear} {Response} {Properties} with {Relativistic} {Coupled} {Cluster} {Theory} for {GPU}-{Accelerated} {Computer} {Architectures}},
	volume = {20},
	copyright = {https://doi.org/10.15223/policy-029},
	issn = {1549-9618, 1549-9626},
	url = {https://pubs.acs.org/doi/10.1021/acs.jctc.3c00812},
	doi = {10.1021/acs.jctc.3c00812},
	language = {en},
	number = {2},
	urldate = {2025-05-14},
	journal = {Journal of Chemical Theory and Computation},
	author = {Yuan, Xiang and Halbert, Loïc and Pototschnig, Johann Valentin and Papadopoulos, Anastasios and Coriani, Sonia and Visscher, Lucas and Gomes, André Severo Pereira},
	month = jan,
	year = {2024},
	pages = {677--694},
}

@article{Yuan_Xiang_QR,
	title = {Frequency-{Dependent} {Quadratic} {Response} {Properties} and {Two}-{Photon} {Absorption} from {Relativistic} {Equation}-of-{Motion} {Coupled} {Cluster} {Theory}},
	volume = {19},
	copyright = {https://doi.org/10.15223/policy-029},
	issn = {1549-9618, 1549-9626},
	url = {https://pubs.acs.org/doi/10.1021/acs.jctc.3c01011},
	doi = {10.1021/acs.jctc.3c01011},
	language = {en},
	number = {24},
	urldate = {2025-05-14},
	journal = {Journal of Chemical Theory and Computation},
	author = {Yuan, Xiang and Halbert, Loïc and Visscher, Lucas and Pereira Gomes, André Severo},
	month = dec,
	year = {2023},
	pages = {9248--9259},
}

@Misc{DIRAC25,
      note = "{DIRAC}, a relativistic ab initio electronic structure program,
      Release {DIRAC25} (2025), written by T.~Saue, L.~Visscher, H.~J.~{\relax Aa}.~Jensen, R.~Bast and A.~S.~P.~Gomes, with contributions from I.~A.~Aucar, V.~Bakken, J.~Brandejs, C.~Chibueze, J.~Creutzberg,  K.~G.~Dyall, S.~Dubillard, U.~Ekstr{\"o}m, E.~Eliav, T.~Enevoldsen, E.~Fa{\ss}hauer, T.~Fleig, O.~Fossgaard, K.~G.~Gaul, L.~Halbert, E.~D.~Hedeg{\aa}rd, T.~Helgaker, B.~Helmich--Paris, J.~Henriksson, M.~van~Horn, M.~Ilia{\v{s}}, Ch.~R.~Jacob, S.~Knecht, S.~Komorovsk{\'y}, O.~Kullie, J.~K.~L{\ae}rdahl, C.~V.~Larsen, Y.~S.~Lee, N.~H.~List, H.~S.~Nataraj, M.~K.~Nayak, P.~Norman, A.~Nyvang, G.~Olejniczak, J.~Olsen, J.~M.~H.~Olsen, A.~Papadopoulos, Y.~C.~Park, J.~K.~Pedersen,  M.~Pernpointner, J.~V.~Pototschnig, R.~di~Remigio, M.~Repisky, C. M. R. Rocha, K.~Ruud, P.~Sa{\l}ek, B.~Schimmelpfennig, B.~Senjean, A.~Shee, J.~Sikkema, A.~Sunaga, A.~J.~Thorvaldsen, J.~Thyssen, J.~van~Stralen, M.~L.~Vidal, S.~Villaume, O.~Visser, T.~Winther, S.~Yamamoto and X.~Yuan (available at \url{https://doi.org/10.5281/zenodo.14833106}, see also  \url{https://www.diracprogram.org})"
}

@article{Jerabek2019a,
  title = {Solid Oganesson via a Many-Body Interaction Expansion Based on Relativistic Coupled-Cluster Theory and from Plane-Wave Relativistic Density Functional Theory},
  volume = {123},
  ISSN = {1520-5215},
  url = {http://dx.doi.org/10.1021/acs.jpca.9b01947},
  DOI = {10.1021/acs.jpca.9b01947},
  number = {19},
  journal = {The Journal of Physical Chemistry A},
  publisher = {American Chemical Society (ACS)},
  author = {Jerabek,  Paul and Smits,  Odile R. and Mewes,  Jan-Michael and Peterson,  Kirk A. and Schwerdtfeger,  Peter},
  year = {2019},
  month = apr,
  pages = {4201–4211}
}

@article{Jerabek2019b,
  title = {Correction to “Solid Oganesson via a Many-Body Interaction Expansion Based on Relativistic Coupled-Cluster Theory and from Plane-Wave Relativistic Density Functional Theory”},
  volume = {123},
  ISSN = {1520-5215},
  url = {http://dx.doi.org/10.1021/acs.jpca.9b08424},
  DOI = {10.1021/acs.jpca.9b08424},
  number = {38},
  journal = {The Journal of Physical Chemistry A},
  publisher = {American Chemical Society (ACS)},
  author = {Jerabek,  Paul and Smits,  Odile R. and Mewes,  Jan-Michael and Peterson,  Kirk A. and Schwerdtfeger,  Peter},
  year = {2019},
  month = sep,
  pages = {8333–8333}
}

@article{Smits2023,
  title = {Pushing the limits of the periodic table — A review on atomic relativistic electronic structure theory and calculations for the superheavy elements},
  volume = {1035},
  ISSN = {0370-1573},
  url = {http://dx.doi.org/10.1016/j.physrep.2023.09.004},
  DOI = {10.1016/j.physrep.2023.09.004},
  journal = {Physics Reports},
  publisher = {Elsevier BV},
  author = {Smits,  O.R. and Indelicato,  P. and Nazarewicz,  W. and Piibeleht,  M. and Schwerdtfeger,  P.},
  year = {2023},
  month = sep,
  pages = {1–57}
}

@article{Smits2023b,
  title = {The quest for superheavy elements and the limit of the periodic table},
  volume = {6},
  ISSN = {2522-5820},
  url = {http://dx.doi.org/10.1038/s42254-023-00668-y},
  DOI = {10.1038/s42254-023-00668-y},
  number = {2},
  journal = {Nature Reviews Physics},
  publisher = {Springer Science and Business Media LLC},
  author = {Smits,  Odile R. and D\"{u}llmann,  Christoph E. and Indelicato,  Paul and Nazarewicz,  Witold and Schwerdtfeger,  Peter},
  year = {2023},
  month = dec,
  pages = {86–98}
}

@article{cadene_circular_2015,
	title = {Circular and linear magnetic birefringences in xenon at $\lambda$ = 1064 nm},
  volume = {142},
  ISSN = {1089-7690},
  url = {http://dx.doi.org/10.1063/1.4916049},
  DOI = {10.1063/1.4916049},
  number = {12},
  journal = {The Journal of Chemical Physics},
  publisher = {AIP Publishing},
  author = {Cadène,  Agathe and Fouché,  Mathilde and Rivère,  Alice and Battesti,  Rémy and Coriani,  Sonia and Rizzo,  Antonio and Rizzo,  Carlo},
  year = {2015},
  month = mar 
}

@article{Ingersoll:56,
  title = {Faraday Effect in Gases and Vapors II*},
  volume = {46},
  ISSN = {0030-3941},
  url = {http://dx.doi.org/10.1364/JOSA.46.000538},
  DOI = {10.1364/josa.46.000538},
  number = {7},
  journal = {Journal of the Optical Society of America},
  publisher = {Optica Publishing Group},
  author = {Ingersoll,  L. R. and Liebenberg,  D. H.},
  year = {1956},
  month = jul,
  pages = {538}
}

@article{coriani_coupled_1997,
  title = {Coupled cluster calculations of Verdet constants},
  volume = {281},
  ISSN = {0009-2614},
  url = {http://dx.doi.org/10.1016/S0009-2614(97)01286-4},
  DOI = {10.1016/s0009-2614(97)01286-4},
  number = {4–6},
  journal = {Chemical Physics Letters},
  publisher = {Elsevier BV},
  author = {Coriani,  Sonia and H\"{a}ttig,  Christof and Jørgensen,  Poul and Halkier,  Asger and Rizzo,  Antonio},
  year = {1997},
  month = dec,
  pages = {445–451}
}

@article{norman_quadratic_2004,
  title = {Quadratic response functions in the time-dependent four-component Hartree-Fock approximation},
  volume = {121},
  ISSN = {1089-7690},
  url = {http://dx.doi.org/10.1063/1.1785774},
  DOI = {10.1063/1.1785774},
  number = {13},
  journal = {The Journal of Chemical Physics},
  publisher = {AIP Publishing},
  author = {Norman,  Patrick and Jensen,  Hans Jørgen Aa.},
  year = {2004},
  month = oct,
  pages = {6145–6154}
}

@article{pereira_gomes_influence_2004,
  title = {The influence of core correlation on the spectroscopic constants of HAt},
  volume = {399},
  ISSN = {0009-2614},
  url = {http://dx.doi.org/10.1016/j.cplett.2004.09.132},
  DOI = {10.1016/j.cplett.2004.09.132},
  number = {1–3},
  journal = {Chemical Physics Letters},
  publisher = {Elsevier BV},
  author = {Pereira Gomes,  André Severo and Visscher,  Lucas},
  year = {2004},
  month = nov,
  pages = {1–6}
}

@article{deFarias2017,
  title = {Estimation of some physical properties for tennessine and tennessine hydride (TsH)},
  volume = {667},
  ISSN = {0009-2614},
  url = {http://dx.doi.org/10.1016/j.cplett.2016.11.023},
  DOI = {10.1016/j.cplett.2016.11.023},
  journal = {Chemical Physics Letters},
  publisher = {Elsevier BV},
  author = {de Farias,  Robson Fernandes},
  year = {2017},
  month = jan,
  pages = {1–3}
}

@article{irek2025,
  title = {Excited-State Absorption: Reference Oscillator Strengths,  Wave Function,  and TDDFT Benchmarks},
  volume = {21},
  ISSN = {1549-9626},
  url = {http://dx.doi.org/10.1021/acs.jctc.5c00159},
  DOI = {10.1021/acs.jctc.5c00159},
  number = {9},
  journal = {Journal of Chemical Theory and Computation},
  publisher = {American Chemical Society (ACS)},
  author = {Širůček,  Jakub and Le Guennic,  Boris and Damour,  Yann and Loos,  Pierre-Fran\c{c}ois and Jacquemin,  Denis},
  year = {2025},
  month = apr,
  pages = {4688–4703}
}

@article{Rozyczko1997,
  title = {Frequency dependent equation-of-motion coupled cluster hyperpolarizabilities: Resolution of the discrepancy between theory and experiment for HF?},
  volume = {107},
  ISSN = {1089-7690},
  url = {http://dx.doi.org/10.1063/1.474225},
  DOI = {10.1063/1.474225},
  number = {24},
  journal = {The Journal of Chemical Physics},
  publisher = {AIP Publishing},
  author = {Rozyczko,  Piotr and Bartlett,  Rodney J.},
  year = {1997},
  month = dec,
  pages = {10823–10826}
}

@article{Sekino1995,
  title = {Frequency-dependent hyperpolarizabilities in the coupled-cluster method: the Kerr effect for molecules},
  volume = {234},
  ISSN = {0009-2614},
  url = {http://dx.doi.org/10.1016/0009-2614(95)00007-Q},
  DOI = {10.1016/0009-2614(95)00007-q},
  number = {1–3},
  journal = {Chemical Physics Letters},
  publisher = {Elsevier BV},
  author = {Sekino,  Hideo and Bartlett,  Rodney J.},
  year = {1995},
  month = mar,
  pages = {87–93}
}

@article{Hofener2012,
  title = {Molecular properties via a subsystem density functional theory formulation: A common framework for electronic embedding},
  volume = {136},
  ISSN = {1089-7690},
  url = {http://dx.doi.org/10.1063/1.3675845},
  DOI = {10.1063/1.3675845},
  number = {4},
  journal = {The Journal of Chemical Physics},
  publisher = {AIP Publishing},
  author = {H\"{o}fener,  Sebastian and Gomes,  André Severo Pereira and Visscher,  Lucas},
  year = {2012},
  month = jan 
}

@article{Hofener2013,
  title = {Solvatochromic shifts from coupled-cluster theory embedded in density functional theory},
  volume = {139},
  ISSN = {1089-7690},
  url = {http://dx.doi.org/10.1063/1.4820488},
  DOI = {10.1063/1.4820488},
  number = {10},
  journal = {The Journal of Chemical Physics},
  publisher = {AIP Publishing},
  author = {H\"{o}fener,  Sebastian and Gomes,  André Severo Pereira and Visscher,  Lucas},
  year = {2013},
  month = sep 
}

@article{Hofener2012b,
  title = {Calculation of electronic excitations using wave-function in wave-function frozen-density embedding},
  volume = {137},
  ISSN = {1089-7690},
  url = {http://dx.doi.org/10.1063/1.4767981},
  DOI = {10.1063/1.4767981},
  number = {20},
  journal = {The Journal of Chemical Physics},
  publisher = {AIP Publishing},
  author = {H\"{o}fener,  Sebastian and Visscher,  Lucas},
  year = {2012},
  month = nov 
}

@article{Hofener2016,
  title = {Wave Function Frozen-Density Embedding: Coupled Excitations},
  volume = {12},
  ISSN = {1549-9626},
  url = {http://dx.doi.org/10.1021/acs.jctc.5b00821},
  DOI = {10.1021/acs.jctc.5b00821},
  number = {2},
  journal = {Journal of Chemical Theory and Computation},
  publisher = {American Chemical Society (ACS)},
  author = {H\"{o}fener,  Sebastian and Visscher,  Lucas},
  year = {2016},
  month = jan,
  pages = {549–557}
}

@article{Niemeyer2025,
  title = {Response properties from frozen-density embedding approximate second-order coupled-cluster theory},
  volume = {162},
  ISSN = {1089-7690},
  url = {http://dx.doi.org/10.1063/5.0260850},
  DOI = {10.1063/5.0260850},
  number = {17},
  journal = {The Journal of Chemical Physics},
  publisher = {AIP Publishing},
  author = {Niemeyer,  Niklas and Neugebauer,  Johannes},
  year = {2025},
  month = may 
}

@article{Myhre2014,
  title = {Multi-level coupled cluster theory},
  volume = {141},
  ISSN = {1089-7690},
  url = {http://dx.doi.org/10.1063/1.4903195},
  DOI = {10.1063/1.4903195},
  number = {22},
  journal = {The Journal of Chemical Physics},
  publisher = {AIP Publishing},
  author = {Myhre,  Rolf H. and Sánchez de Merás,  Alfredo M. J. and Koch,  Henrik},
  year = {2014},
  month = dec 
}

@article{Myhre2016,
  title = {The multilevel CC3 coupled cluster model},
  volume = {145},
  ISSN = {1089-7690},
  url = {http://dx.doi.org/10.1063/1.4959373},
  DOI = {10.1063/1.4959373},
  number = {4},
  journal = {The Journal of Chemical Physics},
  publisher = {AIP Publishing},
  author = {Myhre,  Rolf H. and Koch,  Henrik},
  year = {2016},
  month = jul 
}

@article{Goletto2021,
  title = {Combining multilevel Hartree–Fock and multilevel coupled cluster approaches with molecular mechanics: a study of electronic excitations in solutions},
  volume = {23},
  ISSN = {1463-9084},
  url = {http://dx.doi.org/10.1039/D0CP06359B},
  DOI = {10.1039/d0cp06359b},
  number = {7},
  journal = {Physical Chemistry Chemical Physics},
  publisher = {Royal Society of Chemistry (RSC)},
  author = {Goletto,  Linda and Giovannini,  Tommaso and Folkestad,  Sarai D. and Koch,  Henrik},
  year = {2021},
  pages = {4413–4425}
}

@article{Hettema1992,
  title = {Quadratic response functions for a multiconfigurational self-consistent field wave function},
  volume = {97},
  ISSN = {1089-7690},
  url = {http://dx.doi.org/10.1063/1.463245},
  DOI = {10.1063/1.463245},
  number = {2},
  journal = {The Journal of Chemical Physics},
  publisher = {AIP Publishing},
  author = {Hettema,  Hinne and Jensen,  Hans J{\o}rgen Aa. and J{\o}rgensen,  Poul and Olsen,  Jeppe},
  year = {1992},
  month = jul,
  pages = {1174–1190}
}

@inbook{Casida1996,
  title = {Time-Dependent Density Functional Response Theory of Molecular Systems: Theory,  Computational Methods,  and Functionals},
  ISSN = {1380-7323},
  url = {http://dx.doi.org/10.1016/S1380-7323(96)80093-8},
  DOI = {10.1016/s1380-7323(96)80093-8},
  booktitle = {Recent Developments and Applications of Modern Density Functional Theory},
  publisher = {Elsevier},
  author = {Casida,  Mark E.},
  year = {1996},
  pages = {391–439}
}

@inbook{Norman2006,
  title = {Microscopic Theory of Nonlinear Optics},
  ISBN = {9781402048500},
  ISSN = {2542-4483},
  url = {http://dx.doi.org/10.1007/1-4020-4850-5_1},
  DOI = {10.1007/1-4020-4850-5_1},
  booktitle = {Non-Linear Optical Properties of Matter},
  publisher = {Springer Netherlands},
  author = {Norman,  Patrick and Ruud,  Kenneth},
  year = {2006},
  pages = {1–49}
}

@article{thierfelder_scalar_2009,
  title = {Scalar relativistic and spin-orbit effects in closed-shell superheavy-element monohydrides},
  volume = {80},
  ISSN = {1094-1622},
  url = {http://dx.doi.org/10.1103/PhysRevA.80.022501},
  DOI = {10.1103/physreva.80.022501},
  number = {2},
  journal = {Physical Review A},
  publisher = {American Physical Society (APS)},
  author = {Thierfelder,  Christian and Schwerdtfeger,  Peter and Koers,  Anton and Borschevsky,  Anastasia and Fricke,  Burkhard},
  year = {2009},
  month = aug 
}

@article{Hohm2013,
  title = {Experimental static dipole–dipole polarizabilities of molecules},
  volume = {1054–1055},
  ISSN = {0022-2860},
  url = {http://dx.doi.org/10.1016/j.molstruc.2013.10.003},
  DOI = {10.1016/j.molstruc.2013.10.003},
  journal = {Journal of Molecular Structure},
  publisher = {Elsevier BV},
  author = {Hohm,  U.},
  year = {2013},
  month = dec,
  pages = {282–292}
}

@article{Hohm1994,
  title = {Temperature dependence of the dipole polarizability of xenon (1S0) due to dynamic non-resonant Stark effect caused by black-body radiation},
  volume = {189},
  ISSN = {0301-0104},
  url = {http://dx.doi.org/10.1016/0301-0104(94)00296-7},
  DOI = {10.1016/0301-0104(94)00296-7},
  number = {3},
  journal = {Chemical Physics},
  publisher = {Elsevier BV},
  author = {Hohm,  Uwe and Tr\"{u}mper,  Ulf},
  year = {1994},
  month = dec,
  pages = {443–449}
}

@book{CRC2014,
  title = {CRC Handbook of Chemistry and Physics},
  ISBN = {9780429170195},
  url = {http://dx.doi.org/10.1201/b17118},
  DOI = {10.1201/b17118},
  publisher = {CRC Press},
  year = {2014},
  month = jun 
}

@inbook{Jaszuski2015,
  title = {Molecular Electric,  Magnetic,  and Optical Properties},
  ISBN = {9789400761698},
  url = {http://dx.doi.org/10.1007/978-94-007-6169-8_11-2},
  DOI = {10.1007/978-94-007-6169-8_11-2},
  booktitle = {Handbook of Computational Chemistry},
  publisher = {Springer Netherlands},
  author = {Jaszuński,  Michał and Rizzo,  Antonio and Ruud,  Kenneth},
  year = {2015},
  pages = {1–97}
}

@article{Gauss1998,
  title = {Triple excitation effects in coupled-cluster calculations of frequency-dependent hyperpolarizabilities},
  volume = {296},
  ISSN = {0009-2614},
  url = {http://dx.doi.org/10.1016/S0009-2614(98)01013-6},
  DOI = {10.1016/s0009-2614(98)01013-6},
  number = {1–2},
  journal = {Chemical Physics Letters},
  publisher = {Elsevier BV},
  author = {Gauss,  J\"{u}rgen and Christiansen,  Ove and Stanton,  John F.},
  year = {1998},
  month = oct,
  pages = {117–124}
}

@article{Stanton1993b,
  title = {Many-body methods for excited state potential energy surfaces. I. General theory of energy gradients for the equation-of-motion coupled-cluster method},
  volume = {99},
  ISSN = {1089-7690},
  url = {http://dx.doi.org/10.1063/1.465552},
  DOI = {10.1063/1.465552},
  number = {11},
  journal = {The Journal of Chemical Physics},
  publisher = {AIP Publishing},
  author = {Stanton,  John F.},
  year = {1993},
  month = dec,
  pages = {8840–8847}
}

@article{Christiansen1996,
  title = {Excitation energies of {$H_2O$},  {$N_2$} and {$C_2$} in full configuration interaction and coupled cluster theory},
  volume = {256},
  ISSN = {0009-2614},
  url = {http://dx.doi.org/10.1016/0009-2614(96)00394-6},
  DOI = {10.1016/0009-2614(96)00394-6},
  number = {1–2},
  journal = {Chemical Physics Letters},
  publisher = {Elsevier BV},
  author = {Christiansen,  Ove and Koch,  Henrik and Jørgensen,  Poul and Olsen,  Jeppe},
  year = {1996},
  month = jun,
  pages = {185–194}
}

@article{Hattig1997,
  title = {Frequency-dependent first hyperpolarizabilities using coupled cluster quadratic response theory},
  volume = {269},
  ISSN = {0009-2614},
  url = {http://dx.doi.org/10.1016/S0009-2614(97)00311-4},
  DOI = {10.1016/s0009-2614(97)00311-4},
  number = {5–6},
  journal = {Chemical Physics Letters},
  publisher = {Elsevier BV},
  author = {H\"{a}ttig,  Christof and Christiansen,  Ove and Koch,  Henrik and Jørgensen,  Poul},
  year = {1997},
  month = may,
  pages = {428–434}
}

@article{Stanton1994,
  title = {Separability properties of reduced and effective density matrices in the equation-of-motion coupled cluster method},
  volume = {101},
  ISSN = {1089-7690},
  url = {http://dx.doi.org/10.1063/1.468021},
  DOI = {10.1063/1.468021},
  number = {10},
  journal = {The Journal of Chemical Physics},
  publisher = {AIP Publishing},
  author = {Stanton,  John F.},
  year = {1994},
  month = nov,
  pages = {8928–8937}
}

@article{bast:JCP2009,
  title = {Role of noncollinear magnetization for the first-order electric-dipole hyperpolarizability at the four-component Kohn–Sham density functional theory level},
  volume = {130},
  ISSN = {1089-7690},
  url = {http://dx.doi.org/10.1063/1.3054302},
  DOI = {10.1063/1.3054302},
  number = {2},
  journal = {The Journal of Chemical Physics},
  publisher = {AIP Publishing},
  author = {Bast,  Radovan and Saue,  Trond and Henriksson,  Johan and Norman,  Patrick},
  year = {2009},
  month = jan 
}

@article{Bast_ijqc2009,
  title = {Relativistic adiabatic time‐dependent density functional theory using hybrid functionals and noncollinear spin magnetization},
  volume = {109},
  ISSN = {1097-461X},
  url = {http://dx.doi.org/10.1002/qua.22065},
  DOI = {10.1002/qua.22065},
  number = {10},
  journal = {International Journal of Quantum Chemistry},
  publisher = {Wiley},
  author = {Bast,  Radovan and Jensen,  Hans Jørgen Aa. and Saue,  Trond},
  year = {2009},
  month = jan,
  pages = {2091–2112}
}

@article{Villaume_JCP2010,
  title = {Linear complex polarization propagator in a four-component Kohn–Sham framework},
  volume = {133},
  ISSN = {1089-7690},
  url = {http://dx.doi.org/10.1063/1.3461163},
  DOI = {10.1063/1.3461163},
  number = {6},
  journal = {The Journal of Chemical Physics},
  publisher = {AIP Publishing},
  author = {Villaume,  Sebastien and Saue,  Trond and Norman,  Patrick},
  year = {2010},
  month = aug 
}

@article{Henriksson:2005,
  title = {Two-photon absorption in the relativistic four-component Hartree–Fock approximation},
  volume = {122},
  ISSN = {1089-7690},
  url = {http://dx.doi.org/10.1063/1.1869469},
  DOI = {10.1063/1.1869469},
  number = {11},
  journal = {The Journal of Chemical Physics},
  publisher = {AIP Publishing},
  author = {Henriksson,  Johan and Norman,  Patrick and Jensen,  Hans Jørgen Aa.},
  year = {2005},
  month = mar 
}

@article{Henriksson_JCP2006,
  title = {On the evaluation of quadratic response functions at the four-component Hartree-Fock level: Nonlinear polarization and two-photon absorption in bromo- and iodobenzene},
  volume = {124},
  ISSN = {1089-7690},
  url = {http://dx.doi.org/10.1063/1.2204604},
  DOI = {10.1063/1.2204604},
  number = {21},
  journal = {The Journal of Chemical Physics},
  publisher = {AIP Publishing},
  author = {Henriksson,  Johan and Ekstr\"{o}m,  Ulf and Norman,  Patrick},
  year = {2006},
  month = jun 
}

@article{Saue:Jensen:JCP2003,
  title = {Linear response at the 4-component relativistic level: Application to the frequency-dependent dipole polarizabilities of the coinage metal dimers},
  volume = {118},
  ISSN = {1089-7690},
  url = {http://dx.doi.org/10.1063/1.1522407},
  DOI = {10.1063/1.1522407},
  number = {2},
  journal = {The Journal of Chemical Physics},
  publisher = {AIP Publishing},
  author = {Saue,  T. and Jensen,  H. J. Aa.},
  year = {2003},
  month = jan,
  pages = {522–536}
}

@article{Tellgren_JCP2007,
  title = {First-order excited state properties in the four-component Hartree-Fock approximation: The excited state electric dipole moments in CsAg and CsAu},
  volume = {126},
  ISSN = {1089-7690},
  url = {http://dx.doi.org/10.1063/1.2436877},
  DOI = {10.1063/1.2436877},
  number = {6},
  journal = {The Journal of Chemical Physics},
  publisher = {AIP Publishing},
  author = {Tellgren,  Erik and Henriksson,  Johan and Norman,  Patrick},
  year = {2007},
  month = feb 
}

@article{Bast2009b,
  title = {Atomic orbital-based cubic response theory for one-,  two-,  and four-component relativistic self-consistent field models},
  volume = {356},
  ISSN = {0301-0104},
  url = {http://dx.doi.org/10.1016/j.chemphys.2008.10.033},
  DOI = {10.1016/j.chemphys.2008.10.033},
  number = {1–3},
  journal = {Chemical Physics},
  publisher = {Elsevier BV},
  author = {Bast,  Radovan and Thorvaldsen,  Andreas J. and Ringholm,  Magnus and Ruud,  Kenneth},
  year = {2009},
  month = feb,
  pages = {177–186}
}

@article{Schwerdtfeger2020,
  title = {The periodic table and the physics that drives it},
  volume = {4},
  ISSN = {2397-3358},
  url = {http://dx.doi.org/10.1038/s41570-020-0195-y},
  DOI = {10.1038/s41570-020-0195-y},
  number = {7},
  journal = {Nature Reviews Chemistry},
  publisher = {Springer Science and Business Media LLC},
  author = {Schwerdtfeger,  Peter and Smits,  Odile R. and Pyykk\"{o},  Pekka},
  year = {2020},
  month = jun,
  pages = {359–380}
}

@article{Pyykko1988a,
  title = {Relativistic effects in structural chemistry},
  volume = {88},
  ISSN = {1520-6890},
  url = {http://dx.doi.org/10.1021/cr00085a006},
  DOI = {10.1021/cr00085a006},
  number = {3},
  journal = {Chemical Reviews},
  publisher = {American Chemical Society (ACS)},
  author = {Pyykk\"{o},  Pekka},
  year = {1988},
  month = may,
  pages = {563–594}
}

@article{Pyykk2012,
  title = {Relativistic Effects in Chemistry: More Common Than You Thought},
  volume = {63},
  ISSN = {1545-1593},
  url = {http://dx.doi.org/10.1146/annurev-physchem-032511-143755},
  DOI = {10.1146/annurev-physchem-032511-143755},
  number = {1},
  journal = {Annual Review of Physical Chemistry},
  publisher = {Annual Reviews},
  author = {Pyykk\"{o},  Pekka},
  year = {2012},
  month = may,
  pages = {45–64}
}

@article{Autschbach2012a,
  title = {Perspective: Relativistic effects},
  volume = {136},
  ISSN = {1089-7690},
  url = {http://dx.doi.org/10.1063/1.3702628},
  DOI = {10.1063/1.3702628},
  number = {15},
  journal = {The Journal of Chemical Physics},
  publisher = {AIP Publishing},
  author = {Autschbach,  Jochen},
  year = {2012},
  month = apr 
}

@article{Halbert2021,
  doi = {10.1021/acs.jctc.0c01203},
  url = {https://doi.org/10.1021/acs.jctc.0c01203},
  year = {2021},
  month = may,
  publisher = {American Chemical Society ({ACS})},
  volume = {17},
  number = {6},
  pages = {3583--3598},
  author = {Loïc Halbert and Marta L. Vidal and Avijit Shee and Sonia Coriani and Andr{\'{e}} Severo Pereira Gomes},
  title = {Relativistic {EOM}-{CCSD} for Core-Excited and Core-Ionized State Energies Based on the Four-Component Dirac{\textendash}Coulomb(-Gaunt) Hamiltonian},
  journal = {Journal of Chemical Theory and Computation}
}

@article{Halbert2023,
  title = {The performance of approximate equation of motion coupled cluster for valence and core states of heavy element systems},
  volume = {123},
  ISSN = {1362-3028},
  url = {http://dx.doi.org/10.1080/00268976.2023.2246592},
  DOI = {10.1080/00268976.2023.2246592},
  number = {5–6},
  journal = {Molecular Physics},
  publisher = {Informa UK Limited},
  author = {Halbert,  Loic and Gomes,  André Severo Pereira},
  year = {2023},
  month = aug 
}

@misc{Banerjee2025,
  doi = {10.48550/ARXIV.2506.09008},
  author = {Banerjee,  Samragni and Li,  Run R. and Cooper,  Brandon C. and Zhang,  Tianyuan and Valeev,  Edward F. and Li,  Xiaosong and DePrince,  A. Eugene},
  title = {Relativistic Core-Valence-Separated Molecular Mean-Field Exact-Two-Component Equation-of-Motion Coupled Cluster Theory: Applications to L-edge X-ray Absorption Spectroscopy},
  publisher = {arXiv},
  year = {2025},
  copyright = {arXiv.org perpetual,  non-exclusive license},
  howpublished =  {\url{https://arxiv.org/abs/2506.09008} (accessed 2025-11-11)}, 
}

@article{Li2025,
  title = {Relativistic two-component double ionization potential equation-of-motion coupled cluster with the Dirac–Coulomb–Breit Hamiltonian},
  volume = {163},
  ISSN = {1089-7690},
  url = {http://dx.doi.org/10.1063/5.0278675},
  DOI = {10.1063/5.0278675},
  number = {10},
  journal = {The Journal of Chemical Physics},
  publisher = {AIP Publishing},
  author = {Li,  Run R. and Yuwono,  Stephen H. and Liebenthal,  Marcus D. and Zhang,  Tianyuan and Li,  Xiaosong and DePrince,  A. Eugene},
  year = {2025},
  month = sep 
}

@article{Yuwono2025,
  title = {Two-component relativistic equation-of-motion coupled cluster for electron ionization},
  volume = {162},
  ISSN = {1089-7690},
  url = {http://dx.doi.org/10.1063/5.0248535},
  DOI = {10.1063/5.0248535},
  number = {8},
  journal = {The Journal of Chemical Physics},
  publisher = {AIP Publishing},
  author = {Yuwono,  Stephen H. and Li,  Run R. and Zhang,  Tianyuan and Li,  Xiaosong and DePrince,  A. Eugene},
  year = {2025},
  month = feb 
}

@article{Zhang2024,
  title = {Dirac–Coulomb–Breit Molecular Mean-Field Exact-Two-Component Relativistic Equation-of-Motion Coupled-Cluster Theory},
  volume = {128},
  ISSN = {1520-5215},
  url = {http://dx.doi.org/10.1021/acs.jpca.3c08167},
  DOI = {10.1021/acs.jpca.3c08167},
  number = {17},
  journal = {The Journal of Physical Chemistry A},
  publisher = {American Chemical Society (ACS)},
  author = {Zhang,  Tianyuan and Banerjee,  Samragni and Koulias,  Lauren N. and Valeev,  Edward F. and DePrince,  A. Eugene and Li,  Xiaosong},
  year = {2024},
  month = apr,
  pages = {3408–3418}
}

@article{Koulias2019,
  title = {Relativistic Real-Time Time-Dependent Equation-of-Motion Coupled-Cluster},
  volume = {15},
  ISSN = {1549-9626},
  url = {http://dx.doi.org/10.1021/acs.jctc.9b00729},
  DOI = {10.1021/acs.jctc.9b00729},
  number = {12},
  journal = {Journal of Chemical Theory and Computation},
  publisher = {American Chemical Society (ACS)},
  author = {Koulias,  Lauren N. and Williams-Young,  David B. and Nascimento,  Daniel R. and DePrince,  A. Eugene and Li,  Xiaosong},
  year = {2019},
  month = oct,
  pages = {6617–6624}
}

@article{Chakraborty2025,
  title = {Spin-Free Exact Two-Component Linear Response Coupled Cluster Theory for the Estimation of Frequency-Dependent Second-Order Properties},
  volume = {129},
  ISSN = {1520-5215},
  url = {http://dx.doi.org/10.1021/acs.jpca.4c03584},
  DOI = {10.1021/acs.jpca.4c03584},
  number = {14},
  journal = {The Journal of Physical Chemistry A},
  publisher = {American Chemical Society (ACS)},
  author = {Chakraborty,  Sudipta and Mukhopadhyay,  Tamoghna and Dutta,  Achintya Kumar},
  year = {2025},
  month = mar,
  pages = {3315–3330}
}

@article{Mukhopadhyay2025,
  title = {Analytic calculation of transition dipole moment using four-component relativistic equation-of-motion coupled-cluster expectation value approach},
  volume = {162},
  ISSN = {1089-7690},
  url = {http://dx.doi.org/10.1063/5.0229955},
  DOI = {10.1063/5.0229955},
  number = {5},
  journal = {The Journal of Chemical Physics},
  publisher = {AIP Publishing},
  author = {Mukhopadhyay,  Tamoghna and Chakraborty,  Sudipta and Chamoli,  Somesh and Nayak,  Malaya K. and Dutta,  Achintya Kumar},
  year = {2025},
  month = feb 
}

@article{Chamoli2025,
  title = {Frozen Natural Spinors for Cholesky Decomposition-Based Two-Component Relativistic Coupled Cluster Method},
  volume = {21},
  ISSN = {1549-9626},
  url = {http://dx.doi.org/10.1021/acs.jctc.5c00199},
  DOI = {10.1021/acs.jctc.5c00199},
  number = {9},
  journal = {Journal of Chemical Theory and Computation},
  publisher = {American Chemical Society (ACS)},
  author = {Chamoli,  Somesh and Wang,  Xubo and Zhang,  Chaoqun and Nayak,  Malaya K. and Dutta,  Achintya Kumar},
  year = {2025},
  month = apr,
  pages = {4532–4542}
}

@article{Liu2021,
  title = {Relativistic coupled‐cluster and equation‐of‐motion coupled‐cluster methods},
  volume = {11},
  ISSN = {1759-0884},
  url = {http://dx.doi.org/10.1002/wcms.1536},
  DOI = {10.1002/wcms.1536},
  number = {6},
  journal = {WIREs Computational Molecular Science},
  publisher = {Wiley},
  author = {Liu,  Junzi and Cheng,  Lan},
  year = {2021},
  month = may 
}

@article{Brandejs2025,
  title = {Generating Coupled Cluster Code for Modern Distributed-Memory Tensor Software},
  volume = {21},
  ISSN = {1549-9626},
  url = {http://dx.doi.org/10.1021/acs.jctc.5c00219},
  DOI = {10.1021/acs.jctc.5c00219},
  number = {15},
  journal = {Journal of Chemical Theory and Computation},
  publisher = {American Chemical Society (ACS)},
  author = {Brandejs,  Jan and Pototschnig,  Johann and Saue,  Trond},
  year = {2025},
  month = jul,
  pages = {7320–7334}
}

@article{Yuwono2024,
  title = {Relativistic Coupled Cluster with Completely Renormalized and Perturbative Triples Corrections},
  volume = {128},
  ISSN = {1520-5215},
  url = {http://dx.doi.org/10.1021/acs.jpca.4c02583},
  DOI = {10.1021/acs.jpca.4c02583},
  number = {31},
  journal = {The Journal of Physical Chemistry A},
  publisher = {American Chemical Society (ACS)},
  author = {Yuwono,  Stephen H. and Li,  Run R. and Zhang,  Tianyuan and Surjuse,  Kshitijkumar A. and Valeev,  Edward F. and Li,  Xiaosong and Eugene DePrince,  A.},
  year = {2024},
  month = jul,
  pages = {6521–6539}
}

@article{Zhang2024b,
  title = {Cholesky Decomposition-Based Implementation of Relativistic Two-Component Coupled-Cluster Methods for Medium-Sized Molecules},
  volume = {20},
  ISSN = {1549-9626},
  url = {http://dx.doi.org/10.1021/acs.jctc.3c01236},
  DOI = {10.1021/acs.jctc.3c01236},
  number = {2},
  journal = {Journal of Chemical Theory and Computation},
  publisher = {American Chemical Society (ACS)},
  author = {Zhang,  Chaoqun and Lipparini,  Filippo and Stopkowicz,  Stella and Gauss,  J\"{u}rgen and Cheng,  Lan},
  year = {2024},
  month = jan,
  pages = {787–798}
}

@article{Uhlov2024,
  title = {Cholesky Decomposition in Spin-Free Dirac–Coulomb Coupled-Cluster Calculations},
  volume = {128},
  ISSN = {1520-5215},
  url = {http://dx.doi.org/10.1021/acs.jpca.4c04353},
  DOI = {10.1021/acs.jpca.4c04353},
  number = {38},
  journal = {The Journal of Physical Chemistry A},
  publisher = {American Chemical Society (ACS)},
  author = {Uhlířová,  Tereza and Cianchino,  Davide and Nottoli,  Tommaso and Lipparini,  Filippo and Gauss,  J\"{u}rgen},
  year = {2024},
  month = sep,
  pages = {8292–8303}
}

@article{Karol2016a,
  title = {Discovery of the elements with atomic numbers Z=113,  115 and 117 (IUPAC Technical Report)},
  volume = {88},
  ISSN = {0033-4545},
  url = {http://dx.doi.org/10.1515/pac-2015-0502},
  DOI = {10.1515/pac-2015-0502},
  number = {1–2},
  journal = {Pure and Applied Chemistry},
  publisher = {Walter de Gruyter GmbH},
  author = {Karol,  Paul J. and Barber,  Robert C. and Sherrill,  Bradley M. and Vardaci,  Emanuele and Yamazaki,  Toshimitsu},
  year = {2016},
  month = jan,
  pages = {139–153}
}

@article{Karol2016b,
  title = {Discovery of the element with atomic number Z = 118 completing the 7th row of the periodic table (IUPAC Technical Report)},
  volume = {88},
  ISSN = {0033-4545},
  url = {http://dx.doi.org/10.1515/pac-2015-0501},
  DOI = {10.1515/pac-2015-0501},
  number = {1–2},
  journal = {Pure and Applied Chemistry},
  publisher = {Walter de Gruyter GmbH},
  author = {Karol,  Paul J. and Barber,  Robert C. and Sherrill,  Bradley M. and Vardaci,  Emanuele and Yamazaki,  Toshimitsu},
  year = {2016},
  month = jan,
  pages = {155–160}
}

@article{piecuch_molecular_1996,
	title = {Molecular quadrupole moment functions of {$HF$} and {$N_2$}. {I}. \textit{{Ab}} \textit{initio} linear-response coupled-cluster results},
	volume = {104},
	issn = {0021-9606, 1089-7690},
	url = {https://pubs.aip.org/jcp/article/104/12/4699/293460/Molecular-quadrupole-moment-functions-of-HF-and-N2},
	doi = {10.1063/1.471164},
	language = {en},
	number = {12},
	urldate = {2025-09-03},
	journal = {The Journal of Chemical Physics},
	author = {Piecuch, Piotr and Kondo, Anne E. and Špirko, Vladimír and Paldus, Josef},
	month = mar,
	year = {1996},
	pages = {4699--4715},
}

@article{spirko_molecular_1996,
	title = {Molecular quadrupole moment functions of {HF} and {$N_2$}. {II}. {Rovibrational} effects},
	volume = {104},
	issn = {0021-9606, 1089-7690},
	url = {https://pubs.aip.org/jcp/article/104/12/4716/293367/Molecular-quadrupole-moment-functions-of-HF-and-N2},
	doi = {10.1063/1.471165},
	language = {en},
	number = {12},
	urldate = {2025-09-03},
	journal = {The Journal of Chemical Physics},
	author = {Špirko, Vladimír and Piecuch, Piotr and Kondo, Anne E. and Paldus, Josef},
	month = mar,
	year = {1996},
	pages = {4716--4727},
}

@article{kondo_orthogonally_1995,
	title = {Orthogonally spin-adapted single-reference coupled-cluster formalism: {Linear} response calculation of static properties},
	volume = {102},
	issn = {0021-9606, 1089-7690},
	shorttitle = {Orthogonally spin-adapted single-reference coupled-cluster formalism},
	url = {https://pubs.aip.org/jcp/article/102/16/6511/178870/Orthogonally-spin-adapted-single-reference-coupled},
	doi = {10.1063/1.469365},
	language = {en},
	number = {16},
	urldate = {2025-09-03},
	journal = {The Journal of Chemical Physics},
	author = {Kondo, A. E. and Piecuch, P. and Paldus, J.},
	month = apr,
	year = {1995},
	pages = {6511--6524},
}

@article{piecuch_property_1995,
	title = {Property {Evaluation} {Using} the {Two}-{Reference} {State}-{Universal} {Coupled}-{Cluster} {Method}},
	volume = {99},
	issn = {0022-3654, 1541-5740},
	url = {https://pubs.acs.org/doi/abs/10.1021/j100042a006},
	doi = {10.1021/j100042a006},
	language = {en},
	number = {42},
	urldate = {2025-09-03},
	journal = {The Journal of Physical Chemistry},
	author = {Piecuch, Piotr and Paldus, Josef},
	month = oct,
	year = {1995},
	pages = {15354--15368},
}

@article{piecuch_convergence_1997,
	title = {The convergence of energy expansions for molecules in electrostatic fields: {A} linear‐response coupled‐cluster study},
	volume = {21},
	copyright = {https://www.springernature.com/gp/researchers/text-and-data-mining},
	issn = {0259-9791, 1572-8897},
	shorttitle = {The convergence of energy expansions for molecules in electrostatic fields},
	url = {https://link.springer.com/10.1023/A:1019110116658},
	doi = {10.1023/A:1019110116658},
	language = {en},
	number = {1},
	urldate = {2025-09-03},
	journal = {Journal of Mathematical Chemistry},
	author = {Piecuch, Piotr and Paldus, Josef},
	month = jun,
	year = {1997},
	pages = {51--70},
}

@article{Monkhorst2009,
  title = {Calculation of properties with the coupled-cluster method},
  volume = {12},
  ISSN = {1097-461X},
  url = {http://dx.doi.org/10.1002/qua.560120850},
  DOI = {10.1002/qua.560120850},
  number = {S11},
  journal = {International Journal of Quantum Chemistry},
  publisher = {Wiley},
  author = {Monkhorst,  Hendrik J.},
  year = {2009},
  month = jun,
  pages = {421–432}
}

@article{dalgaard_aspects_1983,
	title = {Some aspects of the time-dependent coupled-cluster approach to dynamic response functions},
	volume = {28},
	copyright = {http://link.aps.org/licenses/aps-default-license},
	issn = {0556-2791},
	url = {https://link.aps.org/doi/10.1103/PhysRevA.28.1217},
	doi = {10.1103/PhysRevA.28.1217},
	language = {en},
	number = {3},
	urldate = {2025-09-12},
	journal = {Physical Review A},
	author = {Dalgaard, Esper and Monkhorst, Hendrik J.},
	month = sep,
	year = {1983},
	pages = {1217--1222},
}

@article{Fernandez1998,
  title = {MCSCF polarizability and hyperpolarizabilities of HCl and HBr},
  volume = {288},
  ISSN = {0009-2614},
  url = {http://dx.doi.org/10.1016/S0009-2614(98)00355-8},
  DOI = {10.1016/s0009-2614(98)00355-8},
  number = {5–6},
  journal = {Chemical Physics Letters},
  publisher = {Elsevier BV},
  author = {Fernández,  Berta and Coriani,  Sonia and Rizzo,  Antonio},
  year = {1998},
  month = may,
  pages = {677–688}
}

@misc{fdcc,
  title={Finite Difference Coefficients Calculator},
  author={Taylor, Cameron R.},
  year={2016},
  howpublished="\url{https://web.media.mit.edu/~crtaylor/calculator.html}. (accessed 2025-11-11)"
}

@article{Pathak:2016ck,
  title = {Relativistic equation-of-motion coupled-cluster method for the electron attachment problem},
  volume = {1076},
  ISSN = {2210-271X},
  url = {http://dx.doi.org/10.1016/j.comptc.2015.12.015},
  DOI = {10.1016/j.comptc.2015.12.015},
  journal = {Computational and Theoretical Chemistry},
  publisher = {Elsevier BV},
  author = {Pathak,  Himadri and Sasmal,  Sudip and Nayak,  Malaya K. and Vaval,  Nayana and Pal,  Sourav},
  year = {2016},
  month = jan,
  pages = {94–100}
}

@article{Pathak:2015he,
  title = {A relativistic equation-of-motion coupled-cluster investigation of the trends of single and double ionization potentials in the He and Be isoelectronic systems},
  volume = {48},
  ISSN = {1361-6455},
  url = {http://dx.doi.org/10.1088/0953-4075/48/11/115009},
  DOI = {10.1088/0953-4075/48/11/115009},
  number = {11},
  journal = {Journal of Physics B: Atomic,  Molecular and Optical Physics},
  publisher = {IOP Publishing},
  author = {Pathak,  Himadri and Sahoo,  B K and Sengupta,  Turbasu and Das,  B P and Vaval,  Nayana and Pal,  Sourav},
  year = {2015},
  month = may,
  pages = {115009}
}

@article{Pathak:2014gv,
  title = {Relativistic equation-of-motion coupled-cluster method for the ionization problem: Application to molecules},
  volume = {90},
  ISSN = {1094-1622},
  url = {http://dx.doi.org/10.1103/PhysRevA.90.062501},
  DOI = {10.1103/physreva.90.062501},
  number = {6},
  journal = {Physical Review A},
  publisher = {American Physical Society (APS)},
  author = {Pathak,  Himadri and Sasmal,  Sudip and Nayak,  Malaya K. and Vaval,  Nayana and Pal,  Sourav},
  year = {2014},
  month = dec 
}

@article{Pathak:2014dm,
  title = {Relativistic equation-of-motion coupled-cluster method: Application to closed-shell atomic systems},
  volume = {89},
  ISSN = {1094-1622},
  url = {http://dx.doi.org/10.1103/PhysRevA.89.042510},
  DOI = {10.1103/physreva.89.042510},
  number = {4},
  journal = {Physical Review A},
  publisher = {American Physical Society (APS)},
  author = {Pathak,  Himadri and Sahoo,  B. K. and Das,  B. P. and Vaval,  Nayana and Pal,  Sourav},
  year = {2014},
  month = apr 
}

@article{Pathak:2016dk,
  title = {Relativistic equation-of-motion coupled-cluster method using open-shell reference wavefunction: Application to ionization potential},
  volume = {145},
  ISSN = {1089-7690},
  url = {http://dx.doi.org/10.1063/1.4960954},
  DOI = {10.1063/1.4960954},
  number = {7},
  journal = {The Journal of Chemical Physics},
  publisher = {AIP Publishing},
  author = {Pathak,  Himadri and Sasmal,  Sudip and Nayak,  Malaya K. and Vaval,  Nayana and Pal,  Sourav},
  year = {2016},
  month = aug 
}

@article{Klein:2008hx,
  title = {Perturbative calculation of spin-orbit splittings using the equation-of-motion ionization-potential coupled-cluster ansatz},
  volume = {129},
  ISSN = {1089-7690},
  url = {http://dx.doi.org/10.1063/1.3013199},
  DOI = {10.1063/1.3013199},
  number = {19},
  journal = {The Journal of Chemical Physics},
  publisher = {AIP Publishing},
  author = {Klein,  Kerstin and Gauss,  J\"{u}rgen},
  year = {2008},
  month = nov 
}

@article{Yang:2012gd,
  title = {Equation of motion coupled cluster method for electron attached states with spin–orbit coupling},
  volume = {531},
  ISSN = {0009-2614},
  url = {http://dx.doi.org/10.1016/j.cplett.2012.02.014},
  DOI = {10.1016/j.cplett.2012.02.014},
  journal = {Chemical Physics Letters},
  publisher = {Elsevier BV},
  author = {Yang,  Dong-Dong and Wang,  Fan and Guo,  Jingwei},
  year = {2012},
  month = apr,
  pages = {236–241}
}

@article{Wang:2015jw,
  title = {Equation-of-motion coupled-cluster method for doubly ionized states with spin-orbit coupling},
  volume = {142},
  ISSN = {1089-7690},
  url = {http://dx.doi.org/10.1063/1.4917041},
  DOI = {10.1063/1.4917041},
  number = {14},
  journal = {The Journal of Chemical Physics},
  publisher = {AIP Publishing},
  author = {Wang,  Zhifan and Hu,  Shu and Wang,  Fan and Guo,  Jingwei},
  year = {2015},
  month = apr 
}

@article{Epifanovsky:2015hsa,
  title = {Spin-orbit couplings within the equation-of-motion coupled-cluster framework: Theory,  implementation,  and benchmark calculations},
  volume = {143},
  ISSN = {1089-7690},
  url = {http://dx.doi.org/10.1063/1.4927785},
  DOI = {10.1063/1.4927785},
  number = {6},
  journal = {The Journal of Chemical Physics},
  publisher = {AIP Publishing},
  author = {Epifanovsky,  Evgeny and Klein,  Kerstin and Stopkowicz,  Stella and Gauss,  J\"{u}rgen and Krylov,  Anna I.},
  year = {2015},
  month = aug 
}

@article{Cao:2016fx,
  title = {Spin-orbit coupling with approximate equation-of-motion coupled-cluster method for ionization potential and electron attachment},
  volume = {145},
  ISSN = {1089-7690},
  url = {http://dx.doi.org/10.1063/1.4964859},
  DOI = {10.1063/1.4964859},
  number = {15},
  journal = {The Journal of Chemical Physics},
  publisher = {AIP Publishing},
  author = {Cao,  Zhanli and Wang,  Fan and Yang,  Mingli},
  year = {2016},
  month = oct 
}

@article{Cao:2017id,
  title = {Combining the spin-separated exact two-component relativistic Hamiltonian with the equation-of-motion coupled-cluster method for the treatment of spin–orbit splittings of light and heavy elements},
  volume = {19},
  ISSN = {1463-9084},
  url = {http://dx.doi.org/10.1039/C6CP07588F},
  DOI = {10.1039/c6cp07588f},
  number = {5},
  journal = {Physical Chemistry Chemical Physics},
  publisher = {Royal Society of Chemistry (RSC)},
  author = {Cao,  Zhanli and Li,  Zhendong and Wang,  Fan and Liu,  Wenjian},
  year = {2017},
  pages = {3713–3721}
}

@article{Zhang:2017jj,
  title = {Excitation Energies of UO22+,  NUO+,  and NUN Based on Equation-of-Motion Coupled-Cluster Theory with Spin–Orbit Coupling},
  volume = {121},
  ISSN = {1520-5215},
  url = {http://dx.doi.org/10.1021/acs.jpca.7b02985},
  DOI = {10.1021/acs.jpca.7b02985},
  number = {20},
  journal = {The Journal of Physical Chemistry A},
  publisher = {American Chemical Society (ACS)},
  author = {Zhang,  Shuo and Wang,  Fan},
  year = {2017},
  month = may,
  pages = {3966–3975}
}

@article{Akinaga:2017hv,
  title = {Two-Component Relativistic Equation-of-Motion Coupled-Cluster Methods for Excitation Energies and Ionization Potentials of Atoms and Molecules},
  volume = {121},
  ISSN = {1520-5215},
  url = {http://dx.doi.org/10.1021/acs.jpca.6b10921},
  DOI = {10.1021/acs.jpca.6b10921},
  number = {4},
  journal = {The Journal of Physical Chemistry A},
  publisher = {American Chemical Society (ACS)},
  author = {Akinaga,  Yoshinobu and Nakajima,  Takahito},
  year = {2017},
  month = jan,
  pages = {827–835}
}

@inbook{Wang:2016hx,
  title = {Relativistic Equation-of-Motion Coupled-Cluster Theory (EOM-CC)},
  ISBN = {9783642416118},
  url = {http://dx.doi.org/10.1007/978-3-642-41611-8_33-1},
  DOI = {10.1007/978-3-642-41611-8_33-1},
  booktitle = {Handbook of Relativistic Quantum Chemistry},
  publisher = {Springer Berlin Heidelberg},
  author = {Wang,  Fan},
  year = {2015},
  pages = {1–27}
}

@article{basis-Dyall-TCA2012-131-3962,
  title = {Relativistic double-zeta,  triple-zeta,  and quadruple-zeta basis sets for the 7p elements,  with atomic and molecular applications},
  volume = {131},
  ISSN = {1432-2234},
  url = {http://dx.doi.org/10.1007/s00214-012-1172-4},
  DOI = {10.1007/s00214-012-1172-4},
  number = {3},
  journal = {Theoretical Chemistry Accounts},
  publisher = {Springer Science and Business Media LLC},
  author = {Dyall,  Kenneth G.},
  year = {2012},
  month = mar 
}

@misc{zenodo.17215085,
  doi = {10.5281/ZENODO.17215085},
  url = {https://zenodo.org/doi/10.5281/zenodo.17215085},
  author = {Yuan,  Xiang and Halbert,  Loïc and Visscher,  Lucas and Gomes,  Andre Severo Pereira},
  title = {Dataset: A Comparison of Relativistic Coupled Cluster and Equation of Motion Coupled Cluster Quadratic Response Theory},
  publisher = {Zenodo},
  year = {2025},
  copyright = {Creative Commons Attribution 4.0 International}
}

\end{document}


\maketitle
\clearpage

\section{Working equations for CCSD quadratic response}

In the quadratic response function~\cite{Christiansen1998},
\begin{align}
	\langle\langle X; Y, Z\rangle\rangle_{\omega_Z; \omega_X, \omega_Y} 
	&= \frac{1}{2} C^{\pm \omega} P^{'}\left( X\left(\omega_{X}\right), Y\left(\omega_{Y}\right), Z\left(\omega_{Z}\right)\right) \nonumber \\
	& \times\left\{ 
\left[\frac{1}{2} \mathbf{F}^{X}+\frac{1}{6} \mathbf{G} \tXs {}{} \left(\omega_{X}\right)\right] \tYs{}{} \left(\omega_{Y}\right) +\tbXs{~}{~}\left(\omega_X\right)\left[\mathbf{A}^{Y}+\frac{1}{2} \mathbf{B} \tYs {}{}\left(\omega_{Y}\right)\right]\right\} \tZs{}{} \left(\omega_{Z}\right)
\label{qrf-cc-eq}
\end{align}
we have that, with the exception of $\mathbf{G} \tXs{}{} \tYs{}{} \tZs{}{}$, all other terms 
can be obtained by adapting the working equations reported previously~\cite{Yuan_Xiang_LR} and the respective computer implementations.  Employing Einstein notation, for $\mathbf{G} \tXs{}{} \tYs{}{} \tZs{}{}$ the programmable expressions for are:
\allowdisplaybreaks
\begin{align}
	\left(G\tXs{}{}\tYs{}{}\right)^{m}_{e} = &- \tbs ic   \left( \V kmae  \right)   . \tXs ai .\tYs ck- \tbs ka \left(  \V imce \right).   \tXs ai . \tYs ck - \tbs ke  \left( \V mica  \right) . \tXs ai . \tYs ck   \nonumber \\
	&- \tbs ie  \left( \V mkac \right) . \tXs ai. \tYs ck  - \tbs mc   \left( \V kiea  \right)   . \tXs ai . \tYs ck  - \tbs ma \left(  \V ikec  \right) . \tXs ai . \tYs ck   \nonumber \\
	& + \tbd nkae \left( \V imnc   \right) . \tXs ai . \tYs ck - \tbd mkfa \left( \V fiec  \right) . \tXs ai . \tYs ck  - \tbd imfc \left(  \V fkae  \right) . \tXs ai  .\tYs ck  \nonumber \\
	&+  \tbd niec \left( \V mkna \right) . \tXs ai .\tYs ck -  \tbd ikfe \left( \V fmac  \right) .\tXs ai . \tYs ck + \tbd nmac \left( \V ikne  \right). \tXs ai . \tYs ck   \nonumber \\
	&+  \tbd koae \left( \V imcg \ts go  \right). \tXs ai .\tYs ck +   \tbd imcf \left( \V koae \ts go \right) . \tXs ai . \tYs ck  \nonumber \\
	&+  \tbd kieg \left( \V moca  \ts go \right).  \tXs ai . \tYs ck +  \tbd moca \left( \V kieg  \ts go \right).  \tXs ai .\tYs ck   \nonumber \\
	& +  \tbd ioce \left( \V kmag \ts go  \right). \tXs ai .\tYs ck     +   \tbd kmag \left( \V ioce \ts go \right) . \tXs ai . \tYs ck     \nonumber \\
	& -  \undemi \tbd jibc \left( \V kmae  \right) . \tXd abij . \tYs ck  - \undemi \tbd jkba \left( \V imce   \right) . \tXd abij . \tYs ck    -\undemi \tbd ijeb \left( \V mkac   \right) . \tXd abij . \tYs ck   \nonumber \\
	&    - \undemi  \tbd jmba \left( \V ikec  \right) . \tXd abij . \tYs ck   +\unquart \tbd ijce \left(  \V kmab \right) . \tXd abij  . \tYs ck  + \unquart \tbd kmab \left( \V ijce  \right) . \tXd abij . \tYs ck   \nonumber \\
	& - \tbd jkbe \left(\V imac  \right) . \tXd abij . \tYs ck - \tbd jmbc \left( \V ikae   \right) .  \tXd abij  . \tYs ck  \nonumber \\
	&-  \undemi  \tbd klad \left( \V imce \right) .\tXs ai .\tYd cdkl   - \undemi \tbd ilcd \left( \V kmae   \right) . \tXs ai .\tYd cdkl   -  \undemi  \tbd kled \left( \V mica   \right) . \tXs ai .\tYd cdkl \nonumber \\
	& - \undemi   \tbd mlcd \left( \V kiea \right) . \tXs ai .\tYd cdkl    \nonumber + \unquart  \tbd lkae \left( \V imdc    \right) . \tXs ai .\tYd cdkl   + \unquart \tbd micd \left( \V lkae  \right) . \tXs ai .\tYd cdkl  \nonumber \\
	& - \tbd iled \left( \V kmca  \right) . \tXs ai . \tYd cdkl - \tbd mlad \left( \V kice \right) .  \tXs ai . \tYd cdkl  \nonumber\\
	~\nonumber\\
	\left(G\tXs{}{}\tYs{}{}\right)^{mn}_{ef} = & -  \Pmm ef \tbd mnec \left( \V kifa  \right) .\tXs ai  . \tYs ck - \Pmm mn \tbd mkef \left( \V nica \right) . \tXs ai  . \tYs ck   \nonumber \\
	& -\Pmm ef  \tbd mnea \left( \V ikfc   \right). \tXs ai   . \tYs ck   - \Pmm mn \tbd mief \left( \V nkac \right) . \tXs ai  . \tYs ck   \nonumber \\
	& +  \tbd kief \left( \V mnca  \right) . \tXs ai  . \tYs ck    + \tbd mnca \left( \V kief   \right) .  \tXs ai    . \tYs ck    \nonumber \\
	&- \Pmm mn \Pmm ef \tbd nkfa \left( \V miec  \right) .\tXs ai  . \tYs ck    - \Pmm mn \Pmm ef \tbd mkec \left( \V nkfa  \right) . \tXs ai  . \tYs ck 
    \label{eq:EquationGtt}
\end{align}
where $a,b,c,.. $  indicate {particle lines}, $ i,j,k,... $ {hole lines}, and $ p,q,r,s,... $ general indexes~\cite{crawford2007introduction}, and $P_{pq}$ indicates a permutation operator, with : $P_{-pq} f\left(\dots pq \dots\right)= f\left(\dots pq \dots\right) - f\left(\dots qp \dots\right)$. Furthermore, $\tbs pq, \tbd pqrs$ denote the single and double ground-state Lagrange multipliers, and $\V pqrs$ an antisymmetrized two-electron integral.

\section{QR-EOMCC equations}

Within the EOM-CC approximation, the quadratic response function is expressed below\cite{pawlowski_molecular_2015,coriani_molecular_2016,faber2018resonant}:

\begin{equation}
    \begin{split}
        {}^{\mathrm{EOM}}\langle\langle X;Y,Z\rangle\rangle_{\omega_{Y},\omega_{Z}} =& \frac{1}{2}C^{\pm \omega}P^{X,Y,Z}\\
        &[-\ {}^{\mathrm{EOM}} \tbXs{}{} (\omega_{X}) \tYs{}{}(\omega_{Y})\bar{t} \xi^{Z}\\
        &+\ {}^{\mathrm{EOM}} \tbXs {}{}(\omega_{X}) {}^{\mathrm{EOM}}\mathbf{A}^{Y} \tZs{}{}(\omega_{Z})\\
        &-\bar{t} \tYs{}{}(\omega_{Y}) {}^{\mathrm{EOM}} \tbZs{}{}(\omega_{Z})\xi^{X}]
        \label{eq:qreomcc-function}
    \end{split}
\end{equation}

where $\bar{t}$ indicates the zeroth-order multipliers and ${}^{\mathrm{EOM}}\mathbf{A}^{X}$ is the EOM-CC property Jacobian matrix  
\begin{equation}
    {}^{\mathrm{EOM}}\mathbf{A}_{\mu\nu}^{Y} = \bra{\mu}\Big[\bar{Y},\ket{\nu}\bra{{\mathrm{HF}}}\Big]\ket{{\mathrm{HF}}}
\end{equation}
\begin{equation}
    \bar{Y} = e^{-\hat{T}}\hat{Y}e^{\hat{T}}
\end{equation}

The EOM-CC right-hand side first order response equation is the same as the one presented in the manuscript, whereas the left-hand side first order response equation the $\mathbf{F}t^Y$ term is replaced by $\bar{t}_{D}\xi_{S}^{Y}+(\bar{t}\xi^{Y})\bar{t}$, yielding
\begin{equation}
     {}^{\mathrm{EOM}}\tbYs{}{}(\bar{\mathbf{H}}+\omega_{Y}\mathbf{I})=-\eta^{X}-\bar{t}_{D}\xi_{S}^{Y}+(\bar{t}\xi^{Y})\bar{t}.
     \label{rsp-lhs-eom}
\end{equation}
The subscripts "S" and "D" denote singles and doubles blocks, respectively.

\section{Remarks on state to state transition strengths}

While in the manuscript we do no report the implementation of excited state properties, as a support to the discussion we note that in the QR-CC formalism (see eq.\ 5.64 of ~\citet{Christiansen1998}), the state to state transition strengths ($^{cc}T^X_{if}$) are obtained as 
\begin{equation}
^{cc}T^X_{if} =L^i [A^X + Bt^X(\omega_i - \omega_f)] R^f,
\label{state-to-state-tm}
\end{equation}
where $L^i (R^f)$ represents the $i$-th left ($f$-th right) eigenvector(s) of $\bar{H}$, and $\omega_i, \omega_f$ are the frequencies corresponding to excitations from the ground state to states $i,f$ respectively.

The expression for $^{cc}T^X_{if}$ depends therefore on the $[A^X +Bt^X]$ quantity, which appears in the quadratic response expression (see the ``Theory and Implementation'' section). 

As discussed in the text, we have identified these terms as leading contributions to the response functions based on the results reported on~\autoref{tab:breakdown-qrf-cc-beta} and~\autoref{tab:breakdown-qrf-eomcc-beta} for hyperpolarizabilties, and note that equivalent terms in QR-EOMCC drive its difference to QR-CC. We observe a similar trend for the Verdet constants reported in~\autoref{tab:breakdown-qrf-cc-verdet} and~\autoref{tab:breakdown-qrf-eomcc-verdet}, though there the QR functions are essentially driven by the value of $A^X$.

\clearpage

\section{Validation of the QR-CC implementation\label{sec-validation}}

\subsection{Comparison to analytic results from other codes\label{sec:compare-analytic}}

In this subsection we report the comparison of analytic results for the Exacorr, Dalton and CFOUR codes. The Dalton and CFOUR calculations were carried out with the latest public releases. In all calculations we employed a point charge nucleus, contracted basis functions, and correlated all electrons.

In the case of Dalton, we have requested more than one set of perturbing frequencies (including zero for the static case) in order to verify both the dynamic and static cases. Below we only show the static case since for the dynamic case showed equivalent results. The results for HF are found in~\autoref{tab:verification-analytic-hf}, for HCl in~\autoref{tab:verification-analytic-hcl} and for HBr in~\autoref{tab:verification-analytic-hbr}.

We have verified that our QR-EOMCC results (and in particular the difference calculated to QR-CC) are consistent with those reported by~\citet{coriani_molecular_2016} (analytic $\beta_{zzz}$(QR-CC) = 13.2699 a.u., $\beta_{zzz}$(QR-EOMCC) = 13.2858 a.u., $\Delta = -0.0159$ a.u.; finite field $\beta_{zzz}$(QR-EOMCC) = 13.2858, that is, a difference of -3.1E-5 to the analyic result).
\begin{table}[H]
    \centering
    \setlength{\tabcolsep}{2.2mm}
{
    \begin{tabular}{lcccc}
    \hline
    \hline
Basis 	&	code	&	$\langle\langle Z; Z\rangle\rangle_0$ 
                    &	$\langle\langle Z; Z, Z\rangle\rangle_0$
                    &	$\langle\langle Z; X, X\rangle\rangle_0$ \\									
\hline
cc-pVDZ	&	a	&	-4.2044	&	13.2656	&	2.2715	\\
	&	b	&	-4.2044	&	13.2656	&	2.2715	\\
	&	c	&	-4.2044	&		&		\\
	&	d	&	-4.1621	&		&		\\
	&		&		&		&		\\
	&	$\Delta$b	&	-5.20E-07	&	3.65E-06	&	3.24E-07	\\
	&	$\Delta$c	&	-2.00E-08	&		&		\\
	&	$\Delta$d	&	-4.23E-02	&		&		\\
	&		&		&		&		\\
aug-cc-pVDZ	&	a	&	-6.2805	&	12.1295	&	1.7035	\\
	&	b	&	-6.2805	&	12.1295	&	1.7035	\\
	&	c	&	-6.2805	&		&		\\
	&	d	&	-6.1115	&		&		\\
	&		&		&		&		\\
	&	$\Delta$b	&	4.95E-06	&	-4.87E-05	&	1.13E-06	\\
	&	$\Delta$c	&	5.00E-08	&		&		\\
	&	$\Delta$d	&	-1.69E-01	&		&		\\
	&		&		&		&		\\
cc-pVTZ	&	a	&	-5.2703	&	12.7704	&	3.7566	\\
	&	b	&	-5.2703	&	12.7704	&	3.7566	\\
	&	c	&	-5.2703	&		&		\\
	&	d	&	-5.1882	&		&		\\
	&		&		&		&		\\
	&	$\Delta$b	&	-2.28E-06	&	1.33E-05	&	-1.92E-07	\\
	&	$\Delta$c	&	2.00E-08	&		&		\\
	&	$\Delta$d	&	-8.21E-02	&		&		\\
	&		&		&		&		\\
aug-cc-pVTZ	&	a	&	-6.3506	&	11.5603	&	2.3841	\\
	&	b	&	-6.3506	&	11.5604	&	2.3841	\\
	&	c	&	-6.3506	&		&		\\
	&	d	&	-6.1972	&		&		\\
	&		&		&		&		\\
	&	$\Delta$b	&	2.39E-06	&	-2.41E-05	&	-1.16E-08	\\
	&	$\Delta$c	&	-1.00E-08	&		&		\\
	&	$\Delta$d	&	-1.53E-01	&		&		\\
    \hline
\end{tabular}
}
    \caption[Comparison of analytic QR-CC results between codes for HF]{Values for $\langle\langle Z; Z\rangle\rangle_0$, $\langle\langle Z; Z, Z\rangle\rangle_0$ and $\langle\langle Z; X, X\rangle\rangle_0$ (in a.u.) obtained for the HF molecule with ExaCorr (a); Dalton ((b); and CFOUR (c: orbital-unrelaxed and d: orbital-relaxed) and the respective differences with respect to ExaCorr ($\Delta x = a - x$). All calculations carried out with the non-relativistic Hamiltonian, contracted Dunning basis sets and correlating all electrons and virtuals. nc: not calculated.}
    \label{tab:verification-analytic-hf}
\end{table}      									
									
\begin{table}[H]
    \centering
    \setlength{\tabcolsep}{2.2mm}
{
    \begin{tabular}{lcccc}
    \hline
    \hline
Basis 	&	code	&	$\langle\langle Z; Z\rangle\rangle_0$ 
                    &	$\langle\langle Z; Z, Z\rangle\rangle_0$
                    &	$\langle\langle Z; X, X\rangle\rangle_0$ \\									
\hline
cc-pVDZ	&	a	&	-12.8125	&	41.5655	&	2.2939	\\
	&	b	&	-12.8125	&	41.5654	&	2.2939	\\
	&	c	&	-12.8125	&		&		\\
	&	d	&	-12.7772	&		&		\\
	&		&		&		&		\\
	&	$\Delta$b	&	-3.53E-06	&	2.89E-05	&	4.89E-06	\\
	&	$\Delta$c	&	-3.00E-08	&		&		\\
	&	$\Delta$d	&	-3.53E-02	&		&		\\
	&		&		&		&		\\
aug-cc-pVDZ	&	a	&	-17.659	&	28.3104	&	1.2542	\\
	&	b	&	-17.659	&	28.3103	&	1.2542	\\
	&	c	&	-17.659	&		&		\\
	&	d	&	-17.4393	&		&		\\
	&		&		&		&		\\
	&	$\Delta$b	&	-4.43E-06	&	8.03E-05	&	4.68E-05	\\
	&	$\Delta$c	&	-3.00E-08	&		&		\\
	&	$\Delta$d	&	-2.20E-01	&		&		\\
	&		&		&		&		\\
cc-pVTZ	&	a	&	-14.9256	&	30.1416	&	3.4218	\\
	&	b	&	-14.9256	&	30.1416	&	3.4218	\\
	&	c	&	-14.9256	&		&		\\
	&	d	&	-14.8006	&		&		\\
	&		&		&		&		\\
	&	$\Delta$b	&	-1.09E-06	&	1.57E-05	&	-1.42E-06	\\
	&	$\Delta$c	&	1.00E-08	&		&		\\
	&	$\Delta$d	&	-1.25E-01	&		&		\\
	&		&		&		&		\\
aug-cc-pVTZ	&	a	&	-18.3596	&	19.5906	&	-0.1705	\\
	&	b	&	-18.3596	&	19.5906	&	-0.1705	\\
	&	c	&	-18.3596	&		&		\\
	&	d	&	-18.0987	&		&		\\
	&		&		&		&		\\
	&	$\Delta$b	&	1.03E-06	&	-1.87E-05	&	1.82E-05	\\
	&	$\Delta$c	&	7.30E-07	&		&		\\
	&	$\Delta$d	&	-2.61E-01	&		&		\\
    \hline
\end{tabular}
}
    \caption[Comparison of analytic QR-CC results between codes for HCl]{Values for $\langle\langle Z; Z\rangle\rangle_0$, $\langle\langle Z; Z, Z\rangle\rangle_0$ and $\langle\langle Z; X, X\rangle\rangle_0$ (in a.u.) obtained for the HCl molecule with ExaCorr (a); Dalton ((b); and CFOUR (c: orbital-unrelaxed and d: orbital-relaxed) and the respective differences with respect to ExaCorr ($\Delta x = a - x$). All calculations carried out with the non-relativistic Hamiltonian, contracted Dunning basis sets and correlating all electrons and virtuals. nc: not calculated.}
    \label{tab:verification-analytic-hcl}
\end{table}      									
									
\begin{table}[H]
    \centering
    \setlength{\tabcolsep}{2.2mm}
{
    \begin{tabular}{lcccc}
    \hline
    \hline
Basis 	&	code	&	$\langle\langle Z; Z\rangle\rangle_0$ 
                    &	$\langle\langle Z; Z, Z\rangle\rangle_0$
                    &	$\langle\langle Z; X, X\rangle\rangle_0$ \\									
\hline
cc-pVDZ	&	a	&	-18.1219	&	53.5305	&	0.3463	\\
	&	b	&	-18.1219	&	53.5305	&	0.3463	\\
	&	c	&	-18.1219	&		&		\\
	&	d	&	-18.0522	&		&		\\
	&		&		&		&		\\
	&	$\Delta$b	&	-2.93E-06	&	1.35E-05	&	-1.69E-06	\\
	&	$\Delta$c	&	-3.00E-08	&		&		\\
	&	$\Delta$d	&	-6.97E-02	&		&		\\
	&		&		&		&		\\
aug-cc-pVDZ	&	a	&	-23.8818	&	35.7128	&	0.4634	\\
	&	b	&	-23.8818	&	35.7127	&	0.4634	\\
	&	c	&	-23.8818	&		&		\\
	&	d	&	-23.5683	&		&		\\
	&		&		&		&		\\
	&	$\Delta$b	&	-1.80E-06	&	2.99E-05	&	7.65E-06	\\
	&	$\Delta$c	&	0.00E+00	&		&		\\
	&	$\Delta$d	&	-3.14E-01	&		&		\\
	&		&		&		&		\\
cc-pVTZ	&	a	&	-20.9216	&	32.919	&	-0.2587	\\
	&	b	&	-20.9216	&	32.919	&	-0.2587	\\
	&	c	&	-20.9216	&		&		\\
	&	d	&	-20.7024	&		&		\\
	&		&		&		&		\\
	&	$\Delta$b	&	-1.87E-06	&	-2.99E-05	&	-3.57E-06	\\
	&	$\Delta$c	&	3.00E-08	&		&		\\
	&	$\Delta$d	&	-2.19E-01	&		&		\\
	&		&		&		&		\\
aug-cc-pVTZ	&	a	&	-25.2354	&	-2.8646	&		\\
	&	b	&	-25.2354	&	18.7169	&	-2.8646	\\
	&	c	&	-25.2354	&		&		\\
	&	d	&	-24.8174	&		&		\\
	&		&		&		&		\\
	&	$\Delta$b	&	-4.62E-06	&	-2.16E+01	&	2.86E+00	\\
	&	$\Delta$c	&	-2.00E-08	&		&		\\
	&	$\Delta$d	&	-4.18E-01	&		&		\\
    \hline
\end{tabular}
}
    \caption[Comparison of analytic QR-CC results between codes for HBr]{Values for $\langle\langle Z; Z\rangle\rangle_0$, $\langle\langle Z; Z, Z\rangle\rangle_0$ and $\langle\langle Z; X, X\rangle\rangle_0$ (in a.u.) obtained for the HBr molecule with  ExaCorr (a); Dalton ((b); and CFOUR (c: orbital-unrelaxed and d: orbital-relaxed) and the respective differences with respect to ExaCorr ($\Delta x = a - x$). All calculations carried out with the non-relativistic Hamiltonian, contracted Dunning basis sets and correlating all electrons and virtuals. nc: not calculated.}
    \label{tab:verification-analytic-hbr}
\end{table}

\subsection{Comparison to finite field calculations with ExaCorr\label{sec:compare-numeric}}

In this section we report the values for the $\langle\langle Z; Z\rangle\rangle_0$ and $\langle\langle Z; Z, Z\rangle\rangle_0$ response functions obtained via numerical differentiation of the CCSD ground-state energies. The starting point of this procedure is the calculation of perturbed CCSD energies, and for that we have obtained sets energies perturbed by multiples of base perturbation strength (5E-5, 5E-4 and 1E-3).

We have followed two different procedures: in the first, we carried out least-square fits for a subset of the calculated points around the unperturbed energy for a polynomial expression, and from the fitted expression obtained the required second or third derivative. In the second procedure, after choosing a suitable step size $h$ we selected a suitable subset of points and calculated the derivative according to the expressions below. These derivative formulas were obtained with the FDCC code\cite{fdcc}, employing as inputs the derivative order and the following stencils: (-1,0,1) for the 3-point formula: ; (-2,-1,0,1,2) for the 5-point formulas; (-3,-2,-1,0,1,2,3) for the 7-point formula: (-4,-3,-2,-1,0,1,2,3,4) for the 9-point formula; and (-5,-4,-3,-2,-1,0,1,2,3,4,5) for the 11-point formula. We note that in this case, for second derivatives we explored the following 3- and 5-point formulas:
\begin{align}
    \begin{aligned}\alpha^{3}=&\frac{\partial^{(2)}f}{\partial x^{(2)}} \approx \frac{f\left(x-h\right)-2.f\left(x+0h\right)+f\left(x+h\right)}{h^2} \\
    \alpha^{5}=&\frac{\partial^{(2)}f}{\partial x^{(2)}} \approx \frac{-f\left(x-2h\right)+16f\left(x-h\right)-30.f\left(x+0h\right)+16f\left(x+h\right) +\left(x+2h\right)}{12h^2},
    \end{aligned}
    \label{eq:AlphaPowerN} 
\end{align}
whereas for third derivatives we explored the following 5-, 7-, 9- and 11-point formulas:
\begin{align}
    \begin{aligned}
\beta^{5}=\frac{\partial^{(3)}f}{\partial x^{(3)}} \approx &\frac{-f\left(x-2h\right)+2f\left(x-h\right)+0.f\left(x+0h\right)-f\left(x+h\right) +\left(x+2h\right)}{2h^3}\\
    \beta^{7}=\frac{\partial^{(3)}f}{\partial x^{(3)}} \approx &\frac{1}{8h^3} \left(f\left(x-3h\right)-8f\left(x-2h\right)+13f\left(x-h\right)+0.f\left(x+0h\right) \right. \\ 
    &\left. -13f\left(x+h\right) +8\left(x+2h\right)-\left(x+3h\right) \right)\\
\beta^{9}=\frac{\partial^{(3)}f}{\partial x^{(3)}} \approx &  \frac{1}{240h^3} \left( -7 f\left(x-4h\right) +72 f\left(x-3h\right) \right.  \\
& \left. -338f\left(x-2h\right) +488 \left(x-1h\right) + 0 f\left(x-0h\right)  -488 f\left(x+1h\right) \right.  \\ 
& \left. +338 f\left(x+2h\right)-72f\left(x+3h\right) +7 f\left(x+4h\right)   \right)\\
\beta^{11}\equiv\frac{\partial^{(3)}f}{\partial x^{(3)}} \approx &  \frac{1}{30240h^3} \left( 205 f\left(x-5h\right) -2522 f\left(x-4h\right) +14607 f\left(x-3h\right)  \right.  \\
& \left. -52428f\left(x-2h\right) +70098 \left(x-1h\right) + 0 f\left(x-0h\right)  -70098 f\left(x+1h\right) \right.  \\ 
& \left. +52428 f\left(x+2h\right)-14607f\left(x+3h\right) +2522 f\left(x+4h\right) -205f\left(x+5h\right)  \right)
    \end{aligned}
\label{eq:BetaPowerN}
\end{align}
In in the following we will employ the shorthand notation $\alpha^n, \beta^n$ to refer to a particular formula. 

The results reported in~\autoref{tab:compare-nrdz-hf-hbr} correspond to calculations employing the Levy-Leblond Hamiltonian, contracted cc-pVDZ basis set, a point charge nucleus, and correlated all electrons. These results are therefore to be compared to those in~\autoref{tab:verification-analytic-hf}, for HCl in~\autoref{tab:verification-analytic-hcl} and for HBr in~\autoref{tab:verification-analytic-hbr}.
From this comparison, we observe that depending on the numerical procedure (and in particular the step size), we were able to reproduce the analytic results to about 1E-4 or better, thereby further verifying the correctness of the code. We note that the determination of hyperpolarizabilities was significantly more sensitive to the numerical procedure than the polarizabilities.   


\npdecimalsign{.}
\nprounddigits{4}
\npthousandthpartsep{}

\begin{table}[]
\begin{tabular}{lcrrrrrr}
\hline\hline
& 
& \multicolumn{2}{c}{HF}                  
& \multicolumn{2}{c}{HCl}                  
& \multicolumn{2}{c}{HBr}                  \\
\cline{3-8}
&$h$
& $\langle\langle Z; Z\rangle\rangle$  
& $\langle\langle Z; Z\, Z\rangle\rangle$ 
& $\langle\langle Z; Z\rangle\rangle$    
& $\langle\langle Z; Z\, Z\rangle\rangle$ 
& $\langle\langle Z; Z\rangle\rangle$    
& $\langle\langle Z; Z\, Z\rangle\rangle$  \\
\hline
\hline
analytic		&		&	-4.2044	&	13.2656	&	-12.8125	&	41.5655	&	-18.1219	&	53.5305	\\
\hline
fit	(b)	&		&	-4.2044	&	13.2617	&	-12.8125	&	41.5651	&	-18.1218	&	53.4527	\\
fit	(c)	&		&	-4.2044	&	13.2667	&	-12.8125	&	41.5733	&	-18.1219	&	53.5375	\\
fit	(d)	&		&	-4.2044	&	13.2666	&	-12.8125	&	41.5728	&	-18.1218	&	53.5370	\\
\hline															
$\alpha^3$		&	5.00E-05	&	-4.2044	&		&	-12.8128	&		&		&		\\
$\alpha^5$,	$\beta^7$	&		&	-4.2044	&	12.0650	&	-12.8129	&	24.9543	&		&		\\
$\beta^9$		&		&		&	12.3388	&		&	19.4632	&		&		\\
$\beta^{11}$		&		&		&	12.3585	&		&	16.0302	&		&		\\
\hline																
$\alpha^3$		&	5.00E-04	&	-4.2044	&		&	-12.8125	&		&	-4.1136	&		\\
$\alpha^5$,	$\beta^7$	&		&	-4.2044	&	13.2691	&	-12.8125	&	41.5623	&	-4.1136	&	53.5147	\\
$\beta^9$		&		&		&	13.2697	&		&	41.5616	&		&	428.1982	\\
$\beta^{11}$		&		&		&	13.2699	&		&	41.5611	&		&	53.4980	\\
\hline																
$\alpha^3$		&	1.00E-03	&	-4.2044	&		&	-12.8125	&		&	-18.1219	&		\\
$\alpha^5$,	$\beta^7$	&		&	-4.2044	&	13.2661	&	-12.8125	&	41.5663	&	-18.1219	&	53.5356	\\
$\beta^9$		&		&		&	13.2663	&		&	41.5666	&		&	53.5359	\\
$\beta^{11}$		&		&		&	13.2664	&		&	41.5669	&		&	53.5372	\\
\hline																
$\alpha^3$		&	5.00E-03	&	-4.2046	&		&	-12.8134	&		&	-18.1230	&		\\
$\alpha^5$,	$\beta^7$	&		&	-4.2044	&	13.2656	&	-12.8125	&	41.5655	&	-18.1219	&	53.5307	\\
\hline																
\multicolumn{8}{c}{$\Delta$}\\																
\hline																
fit	(b)	&		&	2.35E-07	&	-3.85E-03	&	3.20E-06	&	-3.57E-04	&	2.16E-05	&	-7.79E-02	\\
fit	(c)	&		&	-1.12E-08	&	1.05E-03	&	4.79E-08	&	7.85E-03	&	-5.32E-07	&	6.95E-03	\\
fit	(d)	&		&	1.96E-05	&	1.00E-03	&	-1.15E-05	&	7.35E-03	&	6.39E-05	&	6.46E-03	\\
\hline																
$\alpha^3$		&	5.00E-05	&	1.89E-05	&		&	-3.36E-04	&		&	1.81E+01	&		\\
$\alpha^5$,	$\beta^7$	&		&	6.20E-05	&	-1.20E+00	&	-4.22E-04	&	-1.66E+01	&	1.81E+01	&	-5.35E+01	\\
$\beta^9$		&		&		&	-9.27E-01	&		&	-2.21E+01	&		&	-5.35E+01	\\
$\beta^{11}$		&		&		&	-9.07E-01	&		&	-2.55E+01	&		&	-5.35E+01	\\
\hline																
$\alpha^3$		&	5.00E-04	&	-3.89E-07	&		&	-7.62E-06	&		&	1.40E+01	&		\\
$\alpha^5$,	$\beta^7$	&		&	2.04E-06	&	3.49E-03	&	1.66E-06	&	-3.14E-03	&	1.40E+01	&	-1.59E-02	\\
$\beta^9$		&		&		&	4.08E-03	&		&	-3.83E-03	&		&	3.75E+02	\\
$\beta^{11}$		&		&		&	4.27E-03	&		&	-4.38E-03	&		&	-3.25E-02	\\
\hline																
$\alpha^3$		&	1.00E-03	&	-7.38E-06	&		&	-3.45E-05	&		&	-3.63E-05	&		\\
$\alpha^5$,	$\beta^7$	&		&	-3.55E-07	&	5.19E-04	&	9.57E-07	&	8.70E-04	&	9.31E-06	&	5.05E-03	\\
$\beta^9$		&		&		&	7.39E-04	&		&	1.18E-03	&		&	5.39E-03	\\
$\beta^{11}$		&		&		&	8.49E-04	&		&	1.43E-03	&		&	6.67E-03	\\
\hline																
$\alpha^3$		&	5.00E-03	&	-1.77E-04	&		&	-8.80E-04	&		&	-1.09E-03	&		\\
$\alpha^5$,	$\beta^7$	&		&	5.72E-08	&	3.19E-06	&	4.44E-08	&	6.64E-05	&	-7.39E-07	&	1.57E-04	\\
\hline
\hline
\end{tabular}
\caption{Linear and quadratic response functions for calculations employing the LL Hamiltonian, contracted cc-pVDZ basis sets and correlating all electrons. Here, $h$ the step of the numerical derivative. (b,c,d) are calculated by the fitting procedure (with polynomials of order 4, 5 and 6), on 39 points, with h=5E-5 and h=5E-4 respectively for (b) and (c)  and with h=1E-3 on 19 points for (d). For $\alpha^n$ and $\beta^n$, $n$ is the number of points use for the derivative equation (see equations \ref{eq:AlphaPowerN} and \ref{eq:BetaPowerN}). $\Delta$ indicates the differences to the analytic results\label{tab:compare-nrdz-hf-hbr}.}
\end{table}

\section{First hyperpolarizabilities of HX (X = F-Ts)}


\begin{table}[H]
    \centering
    \setlength{\tabcolsep}{2.2mm}
{
    \begin{tabular}{clrrrrrr}
    \hline
Method&	Hamiltonian	&	HF	&	HCl	&	HBr	&	HI	&	HAt	&	HTs	\\
    \hline
CC	&	LL	&	-1.6922	&	-0.6742	&	1.0636	&	3.7952	&	7.0220	&	25.5116	\\
	&	LL$^\dagger$	&	-1.7528	&	-0.9107	&	0.9085	&	nc	&	nc	&	nc	\\
	&	SF-X2C	&	-1.6937	&	-0.6120	&	1.7100	&	6.1264	&	20.1247	&	96.2015	\\
	&	SF-$^{2}$DC$^M$	&	-1.6937	&	-0.6122	&	1.7086	&	6.1240	&	20.1191	&	96.3081	\\
	&	X2C	&	-1.6939	&	-0.6125	&	1.7702	&	6.8921	&	29.8295	&	278.7012	\\
	&	X2C$^\dagger$	&	-1.7547	&	-0.8486	&	1.6143	&	nc	&	nc	&	nc	\\
	&	$^{2}$DC$^M$	&	-1.6939	&	-0.6111	&	1.7714	&	6.9246	&	30.2199	&	288.7374	\\
    & ${}^2$DC$^{M(1)}$ &   -1.6279 &   -0.5947 &   1.7746  &    nc & nc & nc \\
    & ${}^2$DC$^{M(2)}$ &   -1.6192 &   -0.5825 &   1.7963  &    nc & nc & nc \\
    & ${}^2$DC$^{M(3)}$ &   -1.6192 &   -0.5825 &   1.7963  &    nc & nc & nc \\
	& $^{2}$DC$^{M,*}$	&	-1.6121	&	-0.5613	&	1.7282	&	nc	&	nc	&	nc	\\
	&	$^{2}$DCG$^M$	&	-1.6936	&	-0.6083	&	1.7840	&	6.9473	&	30.2659	&	288.1002	\\
	&		&		&		&		&		&		&		\\
EOM	&	LL	&	-1.4545	&	0.1867	&	2.8480	&	6.7308	&	11.1312	&	31.0130	\\
	&	LL$^\dagger$	&	-1.5276	&	-0.0858	&	2.6604	&	nc	&	nc	&	nc	\\
	&	SF-X2C	&	-1.4550	&	0.2573	&	3.5540	&	9.2200	&	24.1946	&	93.8942	\\
	&	SF-$^{2}$DC$^M$	&	-1.4551	&	0.2570	&	3.5530	&	9.2194	&	24.1958	&	93.9776	\\
	&	X2C	&	-1.4552	&	0.2566	&	3.6077	&	9.8901	&	32.2297	&	245.9024	\\
	&	X2C$^\dagger$	&	-1.5286	&	-0.0157	&	3.4213	&	nc	&	nc	&	nc	\\
	&	$^{2}$DC$^M$	&	-1.4553	&	0.2581	&	3.6110	&	9.9316	&	32.6382	&	255.1885	\\
    & ${}^2$DC$^{M(1)}$ &   -1.4562 &   0.2271  &   3.5614  & nc & nc & nc \\
    & ${}^2$DC$^{M(2)}$ &   -1.4633 &   0.2433  &   3.5831  & nc & nc & nc \\
    & ${}^2$DC$^{M(3)}$ &   -1.4633 &   0.2433  &   3.5829  & nc & nc & nc \\
	&  $^{2}$DC$^{M,*}$	&	-1.4632	&	0.2319	&	3.2191	& nc	&	nc	&	nc	\\
	&	$^{2}$DCG$^M$	&	-1.4555	&	0.2582	&	3.6116	&	9.9241	&	32.6553	&	254.9151	\\
	&		&		&		&		&		&		&		\\
$\Delta\beta_{zxx}$	&	LL	&	-0.2377	&	-0.8609	&	-1.7844	&	-2.9356	&	-4.1092	&	-5.5015	\\
	&	LL$^\dagger$	&	-0.2253	&	-0.8249	&	-1.7519	&	nc	&	nc	&	nc	\\
	&	SF-X2C	&	-0.2386	&	-0.8693	&	-1.8440	&	-3.0935	&	-4.0698	&	2.3073	\\
	&	SF-$^{2}$DC$^M$	&	-0.2386	&	-0.8692	&	-1.8445	&	-3.0954	&	-4.0768	&	2.3304	\\
	&	X2C	&	-0.2386	&	-0.8691	&	-1.8375	&	-2.9980	&	-2.4002	&	32.7988	\\
	&	X2C$^\dagger$	&	-0.2261	&	-0.8329	&	-1.8070	&	nc	&	nc	&	nc	\\
	&	$^{2}$DC$^M$	&	-0.2386	&	-0.8692	&	-1.8395	&	-3.0070	&	-2.4183	&	33.5489	\\
    & ${}^2$DC$^{M(1)}$ &   -0.1718 &   -0.8218 &    -1.7817 & nc & nc & nc \\
    & ${}^2$DC$^{M(2)}$ &   -0.1559 &   -0.8259 &    -1.7868 & nc & nc & nc \\
    & ${}^2$DC$^{M(3)}$ &   -0.1559 &   -0.8258 &    -1.7866 & nc & nc & nc \\
	&$^{2}$DC$^{M,*}$	&	-0.1489	&	-0.7932	&	-1.4909	&	nc	&	nc	&	nc	\\
	&	$^{2}$DCG$^M$	&	-0.2381	&	-0.8665	&	-1.8276	&	-2.9769	&	-2.3894	&	33.1851	\\    \hline
\end{tabular}
}
    \caption{QR-CCSD (CC) and QR-EOMCCSD (EOM)  values for the $\beta_{zxx}$ component (in a.u.) of the first hyperpolarizability of hydrogen halides, as well as their difference ($\Delta\beta_{zxx}$). All calculations correlate eight (valence) electrons virtuals up to 5 a.u., except those marked by `$^*$'', in which all electrons and virtuals are correlated; $^\dagger$calculated with Dunning basis sets for the halogens; nc: not calculated; ${}^2$DC$^{M(1)}$, ${}^2$DC$^{M(2)}$ and ${}^2$DC$^{M(3)}$ denote valence electrons correlated and virtual space truncations at 10 a.u., 100 a.u. and no truncation, respectively.}
    \label{tab:hyperpolar-table1}
\end{table}      

\begin{table}[H]
    \centering
    \setlength{\tabcolsep}{2.2mm}
{
    \begin{tabular}{clrrrrrr}
    \hline
Method&	Hamiltonian	&	HF	&	HCl	&	HBr	&	HI	&	HAt	&	HTs	\\
    \hline
CC	&	LL	&	-10.0225	&	-12.3576	&	-11.3099	&	-4.9920	&	2.5122	&	26.4008	\\
	&	LL$^\dagger$	&	-9.8894	&	-12.8290	&	-11.6437	&	nc	&	nc	&	nc	\\
	&	SF-X2C	&	-10.0265	&	-12.2467	&	-10.2951	&	-1.7249	&	19.1581	&	97.0937	\\
	&	SF-$^{2}$DC$^M$	&	-10.0265	&	-12.2457	&	-10.2927	&	-1.7207	&	19.1714	&	97.2916	\\
	&	X2C	&	-10.0351	&	-12.2800	&	-10.7858	&	-4.1339	&	6.2096	&	236.5265	\\
	&	X2C$^\dagger$	&	-9.8918	&	-12.7406	&	-11.0795	&	nc	&	nc	&	nc	\\
	&	$^{2}$DC$^M$	&	-10.0285	&	-12.2682	&	-10.7667	&	-4.2923	&	5.8862	&	247.7845	\\
    & ${}^2$DC$^{M(1)}$ & -10.0325 & -12.3174 & -10.8755 & nc & nc & nc \\
    & ${}^2$DC$^{M(2)}$ & -9.9828 &-12.2854  & -10.8324 & nc & nc & nc \\
    & ${}^2$DC$^{M(3)}$ & -9.9829 & -12.2856  & -10.8327  & nc & nc & nc \\
	&	$^{2}$DC$^{M,*}$	&	-9.9662	&	-12.3004	&	-11.3303	&	nc	&	nc	&	nc	\\
	&	$^{2}$DCG$^M$	&	-10.0285	&	-12.2698	&	-10.7700	&	-4.2988	&	5.8167	&	246.5360	\\
	&		&		&		&		&		&		&		\\
EOM	&	LL	&	-9.5863	&	-9.7123	&	-6.1185	&	5.1071	&	16.9523	&	49.5616	\\
	&	LL$^\dagger$	&	-9.4973	&	-10.2439	&	-6.4784	&	nc	&	nc	&	nc	\\
	&	SF-X2C	&	-9.5887	&	-9.5979	&	-5.1513	&	7.7038	&	29.0053	&	85.6894	\\
	&	SF-$^{2}$DC$^M$	&	-9.5887	&	-9.5976	&	-5.1508	&	7.7011	&	28.9897	&	85.6045	\\
	&	X2C	&	-9.5972	&	-9.6318	&	-5.6680	&	4.9753	&	11.4539	&	184.6568	\\
	&	X2C$^\dagger$	&	-9.4983	&	-10.1522	&	-5.9790	&	nc	&	nc	&	nc	\\
	&	$^{2}$DC$^M$	&	-9.5908	&	-9.6206	&	-5.6478	&	4.8241	&	11.1031	&	194.7558	\\
    &  ${}^2$DC$^{M(1)}$ & -9.7647 & -9.7781 & -5.8843 & nc & nc & nc \\
    &  ${}^2$DC$^{M(2)}$ & -9.7443 & -9.7420 & -5.8373 & nc & nc  & nc \\
    &  ${}^2$DC$^{M(3)}$ & -9.7442 & -9.7423 & -5.8381 & nc & nc & nc \\
	&  $^{2}$DC$^{M,*}$  & -9.7486 & -9.8647 & -7.1271 &	nc	&	nc	&	nc	\\
	&	$^{2}$DCG$^M$    & -9.5917 & -9.6261 & -5.6664 &	4.7873	&	11.0986	&	194.7453	\\
	&		&		&		&		&		&		&		\\
$\Delta\beta_{zzz}$	&	LL	&	-0.4362	&	-2.6452	&	-5.1914	&	-10.0991	&	-14.4401	&	-23.1608	\\
	&	LL$^\dagger$	&	-0.3920	&	-2.5851	&	-5.1653	&	nc	&	nc	&	nc	\\
	&	SF-X2C	&	-0.4378	&	-2.6488	&	-5.1438	&	-9.4287	&	-9.8472	&	11.4044	\\
	&	SF-$^{2}$DC$^M$	&	-0.4377	&	-2.6481	&	-5.1419	&	-9.4218	&	-9.8182	&	11.6871	\\
	&	X2C	&	-0.4379	&	-2.6482	&	-5.1179	&	-9.1092	&	-5.2443	&	51.8697	\\
	&	X2C$^\dagger$	&	-0.3935	&	-2.5884	&	-5.1005	&	nc	&	nc	&	nc	\\
	&	$^{2}$DC$^M$	&	-0.4377	&	-2.6476	&	-5.1189	&	-9.1163	&	-5.2169	&	53.0287	\\
    &  ${}^2$DC$^{M(1)}$ & -0.2678 & -2.5774 & -4.9712 & nc & nc &  nc\\
    &  ${}^2$DC$^{M(2)}$ & -0.2385 & -2.5434 & -4.9951 & nc & nc & nc \\
    &  ${}^2$DC$^{M(3)}$ & -0.2387 & -2.5433 &  -4.9946 & nc & nc & nc \\
	&	$^{2}$DC$^{M,*}$	&	-0.2176	&	-2.4356	&	-4.2032	&	nc	&	nc	&	nc	\\
	&	$^{2}$DCG$^M$	&	-0.4368	&	-2.6437	&	-5.1036	&	-9.0861	&	-5.2818	&	51.7907	\\ \hline
\end{tabular}
}
    \caption{QR-CCSD (CC) and QR-EOMCCSD (EOM) values for the $\beta_{zzz}$ component (in a.u.) of the first hyperpolarizability of hydrogen halides, as well as their difference ($\Delta\beta_{zzz}$). All calculations correlate eight (valence) electrons virtuals up to 5 a.u., except those marked by `$^*$'', in which all electrons and virtuals are correlated; $^\dagger$calculated with Dunning basis sets for the halogens; nc: not calculated; ${}^2$DC$^{M(1)}$, ${}^2$DC$^{M(2)}$ and ${}^2$DC$^{M(3)}$ denote valence electrons correlated and virtual space truncations at 10 a.u., 100 a.u. and no truncation, respectively.}
    \label{tab:hyperpolar-table2}
\end{table}

\begin{table}[H]
    \centering
    \setlength{\tabcolsep}{2.2mm}
{
    \begin{tabular}{c|c|ccc|ccc}
    \hline
\hline															
&		&	\multicolumn{3}{c}{$\langle\langle Z; Z, Z\rangle\rangle_0$}		&	\multicolumn{3}{c}{$\langle\langle Z; X, X\rangle\rangle_0$}	\\
\hline				
System	& Term & total		& singles		& doubles		& total		& singles		& doubles	\\
\hline
HF	&	$\mathbf{F}^{X} \tYs{}{} \left(\omega_{Y}\right)  \tZs{}{} \left(\omega_{Z}\right)$	
    &	-0.5139	&	-0.2901	&	-0.2238	&	-0.0748	&	-0.0458	&	-0.0290	\\
	&	$\mathbf{G} \tXs {}{} \left(\omega_{X}\right)  \tYs{}{} \left(\omega_{Y}\right)  \tZs{}{} \left(\omega_{Z}\right)$
    &	-0.0513	&	-0.0698	&	0.0185	&	0.0011	&	-0.0017	&	0.0028	\\
	&	$\tbXs{}{}\left(\omega_X\right) \mathbf{B} \tYs{}{} \left(\omega_{Y}\right) \tZs{}{} \left(\omega_{Z}\right)$
    &	-3.1113	&	-3.0875	&	-0.0238	&	-0.9085	&	-0.9718	&	0.0634	\\
	&	$\tbXs{}{}\left(\omega_X\right) \mathbf{A}^{Y} \tZs{}{} \left(\omega_{Z}\right)$
    &	13.7050	&	15.3507	&	-1.6457	&	2.6761	&	3.2016	&	-0.5255	\\
	&	sum	&	10.0285	&	11.9034	&	-1.8749	&	1.6939	&	2.1823	&	-0.4884	\\
	&		&		&		&		&		&		&		\\
HCl	&	$\mathbf{F}^{X} \tYs{}{} \left(\omega_{Y}\right)  \tZs{}{} \left(\omega_{Z}\right)$	
    &	-1.2492	&	-0.7032	&	-0.5460	&	-0.3202	&	-0.2411	&	-0.0791	\\
	&	$\mathbf{G} \tXs {}{} \left(\omega_{X}\right)  \tYs{}{} \left(\omega_{Y}\right)  \tZs{}{} \left(\omega_{Z}\right)$
    &	-0.0845	&	-0.1599	&	0.0754	&	0.0442	&	0.0353	&	0.0088	\\
	&	$\tbXs{}{}\left(\omega_X\right) \mathbf{B} \tYs{}{}\left(\omega_{Y}\right) \tZs{}{}  \left(\omega_{Z}\right)$
    &	-6.3625	&	-6.7142	&	0.3517	&	-2.6790	&	-2.9266	&	0.2476	\\
	&	$\tbXs{}{}\left(\omega_X\right) \mathbf{A}^{Y} \tZs{}{} \left(\omega_{Z}\right)$
    &	19.9643	&	25.0924	&	-5.1281	&	3.5662	&	5.4898	&	-1.9235	\\
	&	sum	&	12.2682	&	17.5151	&	-5.2469	&	0.6111	&	2.3573	&	-1.7462	\\
	&		&		&		&		&		&		&		\\
HBr	&	$\mathbf{F}^{X} \tYs{}{} \left(\omega_{Y}\right)  \tZs{}{} \left(\omega_{Z}\right)$
    &	-1.9007	&	-1.0708	&	-0.8299	&	-0.5289	&	-0.4014	&	-0.1275	\\
	&	$\mathbf{G} \tXs {}{} \left(\omega_{X}\right)  \tYs{}{} \left(\omega_{Y}\right)  \tZs{}{} \left(\omega_{Z}\right)$
    &	-0.0834	&	-0.2054	&	0.1219	&	0.0779	&	0.0622	&	0.0157	\\
	&	$\tbXs{}{}\left(\omega_X\right) \mathbf{B} \tYs{}{} \left(\omega_{Y}\right) \tZs{}{} \left(\omega_{Z}\right)$
    &	-7.7564	&	-7.9359	&	0.1794	&	-4.1734	&	-4.5494	&	0.3759	\\
	&	$\tbXs{~}{~}\left(\omega_X\right) \mathbf{A}^{Y} \tZs{}{} \left(\omega_{Z}\right)$
    &	20.5073	&	26.9694	&	-6.4621	&	2.8530	&	6.1112	&	-3.2582	\\
	&	sum	&	10.7667	&	17.7573	&	-6.9906	&	-1.7714	&	1.2225	&	-2.9940	\\
	&		&		&		&		&		&		&		\\
HI	&	$\mathbf{F}^{X} \tYs{}{} \left(\omega_{Y}\right)  \tZs{}{} \left(\omega_{Z}\right)$
    &	-1.9820	&	-1.1233	&	-0.8587	&	-0.9588	&	-0.7776	&	-0.1812	\\
	&	$\mathbf{G} \tXs {}{} \left(\omega_{X}\right)  \tYs{}{} \left(\omega_{Y}\right)  \tZs{}{} \left(\omega_{Z}\right)$
    &	-0.0214	&	-0.1844	&	0.1630	&	 0.1575	&	 0.1370	&	 0.0205	\\
	&	$\tbXs{}{}\left(\omega_X\right) \mathbf{B} \tYs{}{} \left(\omega_{Y}\right) \tZs{}{} \left(\omega_{Z}\right)$
    &	-7.0690	&	-6.8321	&	-0.2369	&	-6.3643	&	-6.8736	&	 0.5093	\\
	&	$\tbXs{}{}\left(\omega_X\right) \mathbf{A}^{Y} \tZs{}{} \left(\omega_{Z}\right)$
    &	13.3647	&	19.4919	&	-6.1272	&	0.2410	&	 5.1993	&	-4.9583	\\
	&	sum	&	4.2923	&	11.3520	&	-7.0597	&	-6.9246	&	-2.3150	&	-4.6096	\\
\hline															
\end{tabular}
}
    \caption{Breakdown of the contributions (in a.u.) from each of the terms in eq.~\ref{qrf-cc-eq} to the static QR-CCSD response functions $\langle\langle X; Y, Z\rangle\rangle_0$, (after all permutations) for the calculations with the $^{2}$DC$^M$ Hamiltonian and correlating eight (valence) electrons virtuals up to 5 a.u.~The column ``singles'' (``doubles'') denote the contributions from the contraction between the singles(doubles) block of the terms in square brackets and the $\tZs{}{}$ tensor. The column ``total'' correspond to the sum of the ``singles'' and ``doubles'' contributions.}
    \label{tab:breakdown-qrf-cc-beta}
\end{table}

\begin{table}[H]
    \centering
    \setlength{\tabcolsep}{2.2mm}
{
    \begin{tabular}{c|c|rrrr|rrrr}
    \hline
\hline															
&		&	\multicolumn{4}{c}{$\langle\langle Z; Z, Z\rangle\rangle_0$}		&	\multicolumn{4}{c}{$\langle\langle Z; X, X\rangle\rangle_0$}	\\
\hline				
System	& Term & total		& singles		& doubles	& sd+ds 	& total		& singles		& doubles	& sd+sd \\
\hline
	HF	& $-\tbXs{}{} \tYs{}{} \bar{t} \xi^Z $	&	-0.7616	&	-0.2959	&	-0.0249	&	-0.4407	&	-0.2216	&	-0.0860	&	-0.0074	&	-0.1282	\\
		&	${}^\text{EOM}\eta^X \mathbf{A}^Y \tZs{}{}$			&	11.0952	&	13.1353	&	-2.0401	&	0.0000	&	1.8815	&	2.4852	&	-0.6037	&	0.0000	\\
		& $-\bar{t} \tYs{}{} \tbZs{}{} \xi^X$		&	-0.7429	&	-0.2827	&	-0.0218	&	-0.4384	&	-0.2046	&	-0.0780	&	-0.0058	&	-0.1208	\\
		&	sum			&	9.5908	&	12.5567	&	-2.0868	&	-0.8791	&	1.4553	&	2.3212	&	-0.6168	&	-0.2490	\\
		&				&		    &		    &		    &		    &		    &		    &	        &			\\
	HCl	&	$-\tbXs{}{} \tYs{}{} \bar{t}^0 \xi^Z $			&	-2.3768	&	-1.3300	&	-0.0495	&	-0.9962	&	-0.7690	&	-0.4307	&	-0.0160	&	-0.3223	\\
		&	${}^\text{EOM}\eta^X \mathbf{A}^Y \tZs{}{}$			&	14.6500	&	21.2000	&	-6.5375	&	0.0000	&	1.3102	&	3.4600	&	-2.1498	&	0.0000	\\
		&	$-\bar{t} \tYs{}{} \tbZs{}{} \xi^X$			&	-2.6526	&	-1.3700	&	-0.0607	&	-1.2181	&	-0.7992	&	-0.4150	&	-0.0174	&	-0.3668	\\
		&	sum			&	9.6206	&	18.5000	&	-6.6477	&	-2.2143	&	-0.2581	&	2.6144	&	-2.1833	&	-0.6892	\\
		&				&		    &		    &		    &		    &		    &		    &	        &			\\
	HBr	&	$-\tbXs{}{} \tYs{}{} \bar{t} \xi^Z $			&	-4.0194	&	-2.1189	&	-0.0887	&	-1.8118	&	-1.3421	&	-0.7077	&	-0.0295	&	-0.6049	\\
		&	${}^\text{EOM}\eta^X \mathbf{A}^Y \tZs{}{}$				&	13.9372	&	22.4470	&	-8.5098	&	0.0000	&	-0.9642	&	2.6929	&	-3.4997	&	0.0000	\\
		&	$-\bar{t} \tYs{}{} \tbZs{}{} \xi^X$			&	-4.2700	&	-2.1914	&	-0.0980	&	-1.9806	&	-1.3046	&	-0.6717	&	-0.0281	&	-0.6048	\\
		&	sum			&	5.6478	&	18.1367	&	-8.6965	&	-3.7923	&	-3.6110	&	1.3135	&	-3.5573	&	-1.2097	\\
		&				&		    &		    &		    &		    &		    &		    &	        &			\\
	HI	&	$-\tbXs{}{} \tYs{}{} \bar{t} \xi^Z $		    &	-6.6787	&	-3.5975	&	-0.1503	&	-2.9309	&	-2.2608	&	-1.2200	&	-0.0492	&	-0.9916	\\
		&	${}^\text{EOM}\eta^X \mathbf{A}^Y \tZs{}{}$	    &	 8.4536	&	17.1299	&	-8.6763	&	0.0000	&	-5.6300	&	-0.1012	&	-5.5288	&	-0.0000	\\
		&	$-\bar{t} \tYs{}{} \tbZs{}{} \xi^X$			&	-6.5990	&	-3.8031	&	-0.1373	&	-2.6586	&	-2.0407	&	-1.1815	&	-0.0390	&	-0.8203	\\
		&	sum			                                   &	-4.8241	&	 9.7293	&	-8.9639	&	-5.5895	&	-9.9316	&	-2.5026	&	-5.6171	&	-1.8119	\\\hline															
\end{tabular}
}
    \caption{Breakdown of the contributions (in a.u.) from each of the terms in eq.~\ref{eq:qreomcc-function} to the static QR-EOMCCSD response functions $\langle\langle X; Y, Z\rangle\rangle_0$, (after all permutations) for the calculations with the $^{2}$DC$^M$ Hamiltonian and correlating eight (valence) electrons virtuals up to 5 a.u.~The column ``singles'' (``doubles'') denote the contributions from the contraction between the singles(doubles) block of the [$^\text{EOM}\eta^X \mathbf{A}^Y$] and $\tZs{}{}$  tensors. The column ``total'' correspond to the sum of the ``singles'' and ``doubles'' contributions, plus the "sd+ds" column, which corresponds to the cross-terms arising from the 4-tensor products ($[-\tbXs{}{} \tYs{}{}][\bar{t} \xi^Z]$ and $-[\bar{t} \tYs {}{}][\tbZs {}{} \xi^X]$).}
    \label{tab:breakdown-qrf-eomcc-beta}
\end{table}

\clearpage

\begin{table}[H]
    \centering
    \setlength{\tabcolsep}{2.2mm}
{
    \begin{tabular}{c|c|ccc|ccc}
    \hline
\hline															
& 
& \multicolumn{3}{c}{Im$\langle\langle Y; X, \hat{\mu}_z\rangle\rangle_{\omega,0}$}		
& \multicolumn{3}{c}{Im$\langle\langle Z; Y, \hat{\mu}_x\rangle\rangle_{\omega,0}$}	\\
\hline				
System	& Term & total		& singles		& doubles		& total		& singles		& doubles	\\
\hline
HF	&	$\mathbf{F}^{X}	\tYs{}{}	\left(\omega	{Y}\right)	\tZs{}{}	\left(\omega	{Z}\right)$						&	0.0000	&	0.0000	&	0.0000	&	0.0000	&	0.0000	&	0.0000	\\
	&	$\mathbf{G}	\tXs	{}{}	\left(\omega	{X}\right)	\tYs{}{}	\left(\omega	{Y}\right)	\tZs{}{}	\left(\omega	{Z}\right)$		&	0.0000	&	0.0000	&	0.0000	&	0.0000	&	0.0000	&	0.0000	\\
	&	$\tbXs{}{}\left(\omega	X\right)	\mathbf{B}	\tYs{}{}	\left(\omega	{Y}\right)	\tZs{}{}	\left(\omega	{Z}\right)$				&	0.0000	&	0.0000	&	0.0000	&	-0.0031	&	-0.0031	&	0.0000	\\
	&	$\tbXs{}{}\left(\omega	X\right)	\mathbf{A}^{Y}	\tZs{}{}	\left(\omega	{Z}\right)$							&	0.2752	&	0.2584	&	0.0168	&	0.2689	&	0.2555	&	0.0134	\\
	&	sum												&	0.2752	&	0.2584	&	0.0168	&	0.2657	&	0.2524	&	0.0133	\\
	&													&		&		&		&		&		&		\\
HCl	&	$\mathbf{F}^{X}	\tYs{}{}	\left(\omega	{Y}\right)	\tZs{}{}	\left(\omega	{Z}\right)$						&	0.0000	&	0.0000	&	0.0000	&	0.0000	&	0.0001	&	-0.0001	\\
	&	$\mathbf{G}	\tXs	{}{}	\left(\omega	{X}\right)	\tYs{}{}	\left(\omega	{Y}\right)	\tZs{}{}	\left(\omega	{Z}\right)$		&	0.0000	&	0.0000	&	0.0000	&	0.0000	&	0.0000	&	0.0000	\\
	&	$\tbXs{}{}\left(\omega	X\right)	\mathbf{B}	\tYs{}{}	\left(\omega	{Y}\right)	\tZs{}{}	\left(\omega	{Z}\right)$				&	0.0000	&	0.0000	&	0.0000	&	-0.0015	&	-0.0017	&	0.0003	\\
	&	$\tbXs{}{}\left(\omega	X\right)	\mathbf{A}^{Y}	\tZs{}{}	\left(\omega	{Z}\right)$							&	1.4187	&	1.3431	&	0.0756	&	1.2441	&	1.1821	&	0.0620	\\
	&	sum												&	1.4187	&	1.3431	&	0.0756	&	1.2426	&	1.1805	&	0.0621	\\
	&													&		&		&		&		&		&		\\
HBr	&	$\mathbf{F}^{X}	\tYs{}{}	\left(\omega	{Y}\right)	\tZs{}{}	\left(\omega	{Z}\right)$						&	0.0000	&	0.0000	&	0.0000	&	0.0000	&	0.0003	&	-0.0003	\\
	&	$\mathbf{G}	\tXs	{}{}	\left(\omega	{X}\right)	\tYs{}{}	\left(\omega	{Y}\right)	\tZs{}{}	\left(\omega	{Z}\right)$		&	0.0000	&	0.0000	&	0.0000	&	0.0000	&	0.0000	&	0.0000	\\
	&	$\tbXs{}{}\left(\omega	X\right)	\mathbf{B}	\tYs{}{}	\left(\omega	{Y}\right)	\tZs{}{}	\left(\omega	{Z}\right)$				&	0.0000	&	0.0000	&	0.0000	&	-0.0015	&	-0.0016	&	0.0000	\\
	&	$\tbXs{}{}\left(\omega	X\right)	\mathbf{A}^{Y}	\tZs{}{}	\left(\omega	{Z}\right)$							&	2.5177	&	2.3886	&	0.1291	&	2.1319	&	2.0286	&	0.1032	\\
	&	sum												&	2.5177	&	2.3886	&	0.1291	&	2.1303	&	2.0274	&	0.1030	\\
	&													&		&		&		&		&		&		\\
HI	&	$\mathbf{F}^{X}	\tYs{}{}	\left(\omega	{Y}\right)	\tZs{}{}	\left(\omega	{Z}\right)$						&	0.0000	&	0.0000	&	0.0000	&	-0.0004	&	0.0006	&	-0.0010	\\
	&	$\mathbf{G}	\tXs	{}{}	\left(\omega	{X}\right)	\tYs{}{}	\left(\omega	{Y}\right)	\tZs{}{}	\left(\omega	{Z}\right)$		&	0.0000	&	0.0000	&	0.0000	&	-0.0004	&	-0.0003	&	-0.0001	\\
	&	$\tbXs{}{}\left(\omega	X\right)	\mathbf{B}	\tYs{}{}	\left(\omega	{Y}\right)	\tZs{}{}	\left(\omega	{Z}\right)$				&	-0.0003	&	-0.0003	&	0.0000	&	0.0161	&	0.0181	&	-0.0020	\\
	&	$\tbXs{}{}\left(\omega	X\right)	\mathbf{A}^{Y}	\tZs{}{}	\left(\omega	{Z}\right)$							&	5.0199	&	4.7611	&	0.2588	&	4.1975	&	3.9698	&	0.2277	\\
	&	sum												&	5.0196	&	4.7607	&	0.2589	&	4.2127	&	3.9881	&	0.2247	\\
\hline															
\end{tabular}
}
    \caption{Breakdown of the contributions (in a.u.) from each of the terms to the dynamic QR-CCSD response functions (after all permutations) appearing in the calculation of the Verdet constants. Results obtained with the $^{2}$DC$^M$ Hamiltonian and correlating eight (valence) electrons virtuals up to 5 a.u.~The column ``singles'' (``doubles'') denote the contributions from the contraction between the singles(doubles) block of the terms in square brackets and the $\tZs{}{}$ tensor. The column ``total'' correspond to the sum of the ``singles'' and ``doubles'' contributions. $\omega = 0.01$ a.u.}
    \label{tab:breakdown-qrf-cc-verdet}
\end{table}

\clearpage

\begin{table}[H]
    \centering
    \setlength{\tabcolsep}{2.2mm}
{
    \begin{tabular}{c|c|rrrr|rrrr}
    \hline
\hline															
& 
& \multicolumn{4}{c}{Im$\langle\langle Y; X, \hat{\mu}_z\rangle\rangle_{\omega,0}$}		
& \multicolumn{4}{c}{Im$\langle\langle Z; Y, \hat{\mu}_x\rangle\rangle_{\omega,0}$}	\\
\hline				
System	& Term & total		& singles		& doubles	& sd+ds 	& total		& singles		& doubles	& sd+sd \\
\hline
HF	&	{-$\bar{t}^X t^y \bar{t}^0 \xi^Z$}			&	0.0000	&	0.0000	&	0.0000	&	0.0000	&	-0.0008	&	-0.0003	&	0.0000	&	-0.0005	\\
	&	{$^{\text{EOM}}\eta^X A^Y t^Z$}			&	0.2764	&	0.2595	&	0.0169	&	0.0000	&	0.2679	&	0.2544	&	0.0134	&	0.0000	\\
	&	{$-t^0 t^Y \bar{t}^Z \xi^X$}			&	0.0000	&	0.0000	&	0.0000	&	0.0000	&	-0.0003	&	-0.0001	&	0.0000	&	-0.0002	\\
	&	sum			&	0.2764	&	0.2595	&	0.0169	&	0.0000	&	0.2668	&	0.2540	&	0.0134	&	-0.0006	\\
	&				&		&		&		&		&		&		&		&		\\
HCl	&	{-$\bar{t}^X t^y \bar{t}^0 \xi^Z$}			&	0.0000	&	0.0000	&	0.0000	&	0.0000	&	-0.0017	&	-0.0010	&	0.0000	&	-0.0007	\\
	&	{$^{\text{EOM}}\eta^X A^Y t^Z$}			&	1.4310	&	1.3542	&	0.0767	&	0.0000	&	1.2585	&	1.1954	&	0.0632	&	0.0000	\\
	&	{$-t^0 t^Y \bar{t}^Z \xi^X$}			&	0.0000	&	0.0000	&	0.0000	&	0.0000	&	-0.0008	&	-0.0004	&	0.0000	&	-0.0004	\\
	&	sum			&	1.4310	&	1.3542	&	0.0767	&	0.0000	&	1.2559	&	1.1939	&	0.0631	&	-0.0011	\\
	&				&		&		&		&		&		&		&		&		\\
HBr	&	{-$\bar{t}^X t^y \bar{t}^0 \xi^Z$}			&	0.0000	&	0.0000	&	0.0000	&	0.0000	&	-0.0034	&	-0.0018	&	-0.0001	&	-0.0015	\\
	&	{$^{\text{EOM}}\eta^X A^Y t^Z$}			&	2.5405	&	2.4093	&	0.1312	&	0.0000	&	2.1599	&	2.0544	&	0.1055	&	0.0000	\\
	&	{$-t^0 t^Y \bar{t}^Z \xi^X$}			&	0.0000	&	0.0000	&	0.0000	&	0.0000	&	-0.0014	&	-0.0008	&	0.0000	&	-0.0007	\\
	&	sum			&	2.5405	&	2.4093	&	0.1312	&	0.0000	&	2.1551	&	2.0518	&	0.1054	&	-0.0022	\\
	&				&		&		&		&		&		&		&		&		\\
HI	&	{-$\bar{t}^X t^y \bar{t}^0 \xi^Z$}			&	0.0000	&	0.0000	&	0.0000	&	0.0000	&	-0.0038	&	-0.0020	&	-0.0001	&	-0.0016	\\
	&	{$^{\text{EOM}}\eta^X A^Y t^Z$}			&	5.0749	&	4.8109	&	0.2640	&	0.0000	&	4.2826	&	4.0484	&	0.2342	&	0.0000	\\
	&	{$-t^0 t^Y \bar{t}^Z \xi^X$}			&	0.0000	&	0.0000	&	0.0000	&	0.0000	&	-0.0019	&	-0.0011	&	0.0000	&	-0.0007	\\
	&	sum			&	5.0749	&	4.8109	&	0.2640	&	0.0000	&	4.2770	&	4.0452	&	0.2341	&	-0.0024	\\
\hline															
\end{tabular}
}
    \caption{Breakdown of the contributions (in a.u.) from each of the terms to the dynamic QR-EOMCCSD response functions (after all permutations) appearing in the calculation of the Verdet constants. Results obtained with the $^{2}$DC$^M$ Hamiltonian and correlating eight (valence) electrons virtuals up to 5 a.u.~The column ``singles'' (``doubles'') denote the contributions from the contraction between the singles(doubles) block of the [$^\text{EOM}\eta^X \mathbf{A}^Y$] and $\tZs{}{}$  tensors. The column ``total'' correspond to the sum of the ``singles'' and ``doubles'' contributions, plus the "sd+ds" column, which corresponds to the cross-terms arising from the 4-tensor products ($[-\tbXs{}{} \tYs{}{}][\bar{t} \xi^Z]$ and $-[\bar{t} \tYs {}{}][\tbZs {}{} \xi^X]$).  $\omega = 0.01$ a.u.}
    \label{tab:breakdown-qrf-eomcc-verdet}
\end{table}

\clearpage

\section{Basis set effects on QR-CCSD and QR-EOMCCSD}

\begin{table}[h]
\begin{center}
    \begin{tabular}{cc rrrrrr}
\hline\hline
basis & aug & \multicolumn{2}{c}{QR-CCSD} & \multicolumn{2}{c}{QR-EOMCCSD} & \multicolumn{2}{c}{$\Delta$} \\
\cline{3-8}
& 
& $\langle\langle Z;Z,Z\rangle\rangle_0$
& $\langle\langle Z;X,X\rangle\rangle_0$
& $\langle\langle Z;Z,Z\rangle\rangle_0$
& $\langle\langle Z;X,X\rangle\rangle_0$
& $\langle\langle Z;Z,Z\rangle\rangle_0$
& $\langle\langle Z;X,X\rangle\rangle_0$ \\
\hline
\multicolumn{8}{c}{HF} \\
\hline
dz & 0 & 13.2229 & 2.3287 & 13.2123 & 2.2415 & 0.0107 & 0.0872\\
   & s & 14.3096 & 2.0325 & 14.1105 & 1.8525 & 0.1991 & 0.1800 \\
   & d &  9.9511 & 0.9452 &  9.5725 & 0.6954 & 0.3785 & 0.2498 \\
   & t &  9.7937 & 0.8674 &  9.3315 & 0.5730 & 0.4622 & 0.2944 \\
   & q &  9.8542 & 0.8798 &  9.3915 & 0.5855 & 0.4627 & 0.2944 \\
\hline
tz & 0 & 12.3787 & 2.7125 & 12.2269 & 2.5902 & 0.1518 & 0.1224 \\
   & s & 11.4831 & 2.0828 & 11.0844 & 1.8557 & 0.3987 & 0.2271 \\
   & d &  9.9831 & 1.5916 &  9.5440 & 1.3522 & 0.4392 & 0.2394 \\
   & t & 10.0285 & 1.6939 &  9.5908 & 1.4553 & 0.4377 & 0.2386 \\
   & q & 10.0244 &     nc &  9.5871 & 1.4604 & 0.4373 &     nc \\
   \hline
qz & 0 & 12.7108 & 2.4288 & 12.4795	& 2.2640 & 0.2313 & 0.1648 \\
   & s & 10.9330 & 2.2018 & 10.5724 & 1.9943 & 0.3606 & 0.2075 \\
   & d &  9.9959 & 1.7325 &  9.6271 & 1.5178 & 0.3688 & 0.2147 \\
   & t & 10.0045 &    nc  &  9.6363 & 1.5598 & 0.3682 &     nc \\
\hline
\multicolumn{8}{c}{HCl} \\
\hline
dz	&	0	&	39.1681	&	2.9262	&	39.0488	&	2.8302	&	0.1193	&	0.0960	\\
	&	s	&	24.0818	&	0.3094	&	22.2638	&	-0.3954	&	1.8180	&	0.7048	\\
	&	d	&	16.0540	&	-0.6990	&	13.8399	&	-1.5290	&	2.2141	&	0.8300	\\
	&	t	&	16.8172	&	-1.1477	&	14.5947	&	-1.9821	&	2.2224	&	0.8344	\\
	&	q	&	16.6875	&	-1.0246	&	14.4602	&	-1.8590	&	2.2273	&	0.8343	\\
\hline
tz	&	0	&	27.2078	&	0.6131	&	26.2926	&	0.2521	&	0.9152	&	0.3610	\\
	&	s	&	14.7684	&	-0.3586	&	12.1485	&	-1.2544	&	2.6199	&	0.8958	\\
	&	d	&	12.1942	&	0.6807	&	9.5459	&	-0.1890	&	2.6482	&	0.8697	\\
	&	t	&	12.2682	&	0.6111	&	9.6206	&	-0.2581	&	2.6476	&	0.8692 \\
	&	q	&	12.2631	&	0.6209	&	9.6009	&	-0.2496	&	2.6622  &	0.8705	\\
\hline
qz  & 0     &   21.2484	& -1.0360   & 19.9914	& -1.4945 & 1.2570	& 0.4584 \\
    & s     &   15.5258	&  0.1262   & 13.4817	& -0.4823 & 2.0441	& 0.6085 \\
    & d     &   12.4599	&  0.9185   & 10.3680	& 0.3229  & 2.0918	& 0.5956 \\
    & t     &       nc  &      nc   & 10.3450	& 0.2745   &    nc  &     nc \\
\hline
\multicolumn{8}{c}{HBr} \\
\hline
dz&	0	&	51.7214	&	1.4814	&	51.5678	&	1.3876	&	0.1536	&	0.0937	\\
	&	s	&	19.1048	&	-2.1136	&	15.0699	&	-3.6473	&	4.0349	&	1.5337	\\
	&	d	&	13.1999	&	-3.2014	&	8.6528	&	-4.8919	&	4.5471	&	1.6904	\\
	&	t	&	14.5226	&	-3.9328	&	9.9731	&	-5.6282	&	4.5495	&	1.6954	\\
	&	q	&	14.5706	&	-3.7002	&	10.0150	&	-5.3961	&	4.5556	&	1.6959	\\
\hline
tz	&	0	&	26.6845	&	-2.5738	&	24.7145	&	-3.4076	&	1.9700	&	0.8339	\\
	&	s	&	12.7200	&	-3.0173	&	7.6611	&	-4.8837	&	5.0589	&	1.8664	\\
	&	d	&	10.9506	&	-1.7600	&	5.8296	&	-3.6023	&	5.1209	&	1.8422	\\
	&	t	&	10.7667	&	-1.7714 &	5.6478	&	-3.6110	&	5.1189	&	1.8395 \\
	&	q	&	10.7543	&	-1.7681	&	5.6029	&	-3.6099	&	5.1514	&	1.8417	\\
\hline
qz  &   0   &  18.3594	& -4.6906 & 15.5807	& -5.7768 & 2.7788	& 1.0862 \\
    &   s   &  15.3353	& -1.7539 & 11.5388	& -3.0491 & 3.7965	& 1.2952 \\
    &   d   &  11.2688	& -1.0713 &  7.3031	& -2.3509 & 3.9657	& 1.2796 \\
    &   t   &       nc  &      nc & 7.4036	& -2.2628 &     nc  &    nc  \\
\hline
    \end{tabular}
    \caption{Basis set dependence of the $\langle\langle Z; Z, Z\rangle\rangle_0$ and $\langle\langle Z; X, X\rangle\rangle_0$ response functions (in a.u.) for the HF, HCl and HBr Systems. Hamiltonian ${}^2$DC${}^{M}$, correlation space  -2 to 5 a.u.}
    \label{tab:InfluenceOfDiffuseSI}
\end{center}
\end{table}

\clearpage

\section{Polarizabilities of HX (X = F-Ts) and noble gas atoms}


\begin{table}[H]
    \centering
    \setlength{\tabcolsep}{2.2mm}
{
    \begin{tabular}{ccccccccc}
    \hline
Property	&	Method	&	HF	&	HCl	&	HBr	&	HI	&	HAt	&	\multicolumn{2}{c}{HTs}\\
\hline
$\mu_z$	&	CC	&	0.6916	&	0.4229	&	0.3011	&	0.1499	&	-0.0880	&	-0.7989	& (-0.2828)\\
	&		&		&		&		&		&		&		& \\
        &  Exp.~\cite{NISTconstants} &   0.7186   &    0.4362      &   0.3254         &  0.1762        &            &   & \\
        &  CCSD(T)\cite{thierfelder_scalar_2009} &          &          &            &          &            &   -0.7644 & \\
	&		&		&		&		&		&		&		& \\
$\alpha_{zz}$	&	CC	&	6.5763	&	18.6242	&	25.5904	&	38.0011	&	46.5183	&	70.1581	& (58.0356)	\\
	&	CC asym	&	6.5763	&	18.6242	&	25.5904	&	38.0011	&	46.5183	&	70.1581	& \\
	&	EOM	&	6.6476	&	18.9614	&	26.0809	&	38.8239	&	47.4742	&	71.5697	& (59.2679)\\
	&	$\Delta$	&	-0.0713	&	-0.3372	&	-0.4905	&	-0.8228	&	-0.9559	&	-1.4116	& (-1.2323)\\
	&		&		&		&		&		&		&		& \\
$\alpha_{xx}$	&	CC	&	5.4560	&	16.9257	&	23.5976	&	35.4944	&	43.5843	&	71.5091	& (56.5022)\\
	&	CC asym	&	5.4560	&	16.9257	&	23.5976	&	35.4944	&	43.5843	&	71.5091	& \\
	&	EOM	&	5.4924	&	17.1394	&	23.9058	&	36.0189	&	44.1797	&	72.4536	& (57.2727)\\
	&	$\Delta$	&	-0.0364	&	-0.2137	&	-0.3082	&	-0.5244	&	-0.5954	&	-0.9445	& (-0.7705)\\
	&		&		&		&		&		&		&		& \\
$\bar{\alpha}$	&	CC	&	5.8295	&	17.4919	&	24.2619	&	36.3300	&	44.5623	&	71.0588	& (57.0134)\\
	&	CC asym	&	5.8295	&	17.4919	&	24.2619	&	36.3300	&	44.5623	&	71.0588	& \\
	&	EOM	&	5.8775	&	17.7467	&	24.6308	&	36.9539	&	45.2779	&	72.1590	& (57.9378) \\
	&	$\Delta$	&	-0.0480	&	-0.2548	&	-0.3690	&	-0.6239	&	-0.7156	&	-1.1002	& (-0.9244)\\
    &  Exp.~\cite{Hohm2013} & 5.6$\pm$0.1 &  17.4$\pm$0.01 & 23.79$\pm$0.5 & 37.1$\pm$0.5 & & \\
	&		&		&		&		&		&		&		\\
$\Delta \alpha$	&	CC	&	1.1203	&	1.6985	&	1.9928	&	2.5067	&	2.9340	&	-1.3510	& (1.5334) \\
	&	CC asym	&	1.1203	&	1.6984	&	1.9929	&	2.5066	&	2.9340	&	-1.3510 & 	\\
	&	EOM	&	1.1551	&	1.8220	&	2.1751	&	2.8051	&	3.2946	&	-0.8839	 & (1.9952)\\
	&	$\Delta$  &	-0.0349	&	-0.1235	&	-0.1823	&	-0.2984	&	-0.3605	&	-0.4671	& (-0.4618) \\
    \hline
\end{tabular}
}
    \caption{CCSD equilibrium dipole moment ($\mu_z$) and static dipole polarizabilities for hydrogen halides obtaied with the symmetric (CC) and asymmetric (aCC) formulations of LR-CCSD, and LR-EOMCCSD (EOM). Apart from components ($\alpha_{zz}, \alpha_{xx}$) we report the mean values and the anisotropy ($\bar{\alpha} = (\alpha_{zz} + 2\alpha_{xx})/3$). $\Delta = \alpha^\text{CC} - \alpha^\text{EOM}$ denotes the difference between properties calculated with CC and EOM.    All calculations correlate eight (valence) electrons virtuals up to 5 a.u., and all values are given in a.u.; The experimental dipole moment values~\cite{NISTconstants} correspond to $\mu_0$ instead of the equilibrium values obtained from the calculations. For HTs, values in parenthesis denote spin-free callculations.}
    \label{tab:polar-hx-table}
\end{table}

\begin{table}[H]
    \centering
    \setlength{\tabcolsep}{2.2mm}
{
    \begin{tabular}{cccccccc}
    \hline
	&		&	\multicolumn{3}{c}{$n$s$n$p}	&	\multicolumn{3}{c}{$(n-1)$d$n$s$n$p}\\
    \cline{2-8}
Atom&	$\omega$	&	CC	&	EOM	&	$\Delta$	&	CC	&	EOM	&	$\Delta$	\\
\hline
Xe	&	0.0000	&	28.1736	&	28.5905	&	-0.4169	&	27.9533	&	28.3785	&	-0.4253	\\
	&	0.0100	&	28.1876	&	28.6047	&	-0.4171	&	27.9670	&	28.3924	&	-0.4254	\\
	&	0.0200	&	28.2297	&	28.6473	&	-0.4176	&	28.0083	&	28.4342	&	-0.4260	\\
	&	0.0300	&	28.3003	&	28.7187	&	-0.4185	&	28.0774	&	28.5043	&	-0.4268	\\
	&	0.0428	&	28.4333	&	28.8534	&	-0.4201	&	28.2078	&	28.6363	&	-0.4285	\\
	&	0.0500	&	28.5293	&	28.9505	&	-0.4212	&	28.3019	&	28.7315	&	-0.4296	\\
	&	0.0656	&	28.7939	&	29.2184	&	-0.4244	&	28.5612	&	28.9941	&	-0.4329	\\
	&	0.0774	&	29.0457	&	29.4732	&	-0.4274	&	28.8079	&	29.2439	&	-0.4359	\\
	&	0.0900	&	29.3720	&	29.8033	&	-0.4313	&	29.1275	&	29.5674	&	-0.4399	\\
	&		&		&		&		&		&		&		\\
Rn	&	0.0000	&	36.4840	&	37.0015	&	-0.5175	&	36.1210	&	36.6936	&	-0.5725	\\
	&	0.0100	&	36.5100	&	37.0278	&	-0.5178	&	36.1461	&	36.7190	&	-0.5729	\\
	&	0.0200	&	36.5883	&	37.1070	&	-0.5187	&	36.2217	&	36.7956	&	-0.5739	\\
	&	0.0300	&	36.7199	&	37.2401	&	-0.5202	&	36.3487	&	36.9243	&	-0.5755	\\
	&	0.0428	&	36.9695	&	37.4925	&	-0.5230	&	36.5896	&	37.1682	&	-0.5787	\\
	&	0.0500	&	37.1506	&	37.6756	&	-0.5250	&	36.7644	&	37.3453	&	-0.5809	\\
	&	0.0656	&	37.6553	&	38.1860	&	-0.5307	&	37.2512	&	37.8384	&	-0.5872	\\
	&	0.0774	&	38.1426	&	38.6787	&	-0.5360	&	37.7211	&	38.3144	&	-0.5933	\\
	&	0.0900	&	38.7851	&	39.3282	&	-0.5431	&	38.3403	&	38.9415	&	-0.6012	\\
	&		&		&		&		&		&		&		\\
Og	&	0.0000	&	62.3732	&	64.9005	&	-2.5273	&	60.4439	&	61.4848	&	-1.0409	\\
	&	0.0100	&	62.4796	&	65.0665	&	-2.5869	&	60.5419	&	61.5844	&	-1.0424	\\
	&	0.0200	&	62.8024	&	65.4570	&	-2.6546	&	60.8392	&	61.8861	&	-1.0469	\\
	&	0.0300	&	63.3526	&	66.0841	&	-2.7315	&	61.3456	&	62.4002	&	-1.0546	\\
	&	0.0428	&	64.4241	&	67.2700	&	-2.8459	&	62.3302	&	63.3996	&	-1.0695	\\
	&	0.0500	&	65.2263	&	68.1455	&	-2.9192	&	63.0659	&	64.1465	&	-1.0805	\\
	&	0.0656	&	67.5821	&	70.6899	&	-3.1079	&	65.2194	&	66.3321	&	-1.1127	\\
	&	0.0774	&	70.0500	&	73.3341	&	-3.2841	&	67.4630	&	68.6090	&	-1.1460	\\
	&	0.0900	&	73.6588	&	77.1808	&	-3.5221	&	70.7189	&	71.9126	&	-1.1937	\\
    \hline
\end{tabular}
}
    \caption{Static and dynamic dipole polarizabilities ($\alpha(\omega)$, in a.u.) for noble gases obtaied with the symmetric (CC) formulation of LR-CCSD and LR-EOMCCSD (EOM). $\Delta = \alpha^\text{CC} - \alpha^\text{EOM}$ denotes the difference between properties calculated with CC and EOM.   All calculations correlate 8 ($n$s$n$p) or 18 (($n$-1)d$n$s$n$p) electrons virtuals up to 5 a.u.}
    \label{tab:polar-ng-table}
\end{table}


\section{Verdet constants for Xe, Rn and Og}

\begin{table}[H]
    \centering
    \setlength{\tabcolsep}{2.2mm}
{
    \begin{tabular}{lrrrrr}
\hline\hline				
 & & \multicolumn{4}{c}{V($\omega$)}\\
\hline
System	&	$\omega$	&	CC	&	EOM	&	$\Delta$ &	\%	\\
\hline											
	Xe	&	0.0100	&	0.1687	&	0.1707	&	-0.0020	&	-1.21		\\
		&	0.0200	&	0.6775	&	0.6857	&	-0.0082	&	-1.21		\\
		&	0.0300	&	1.5347	&	1.5531	&	-0.0185	&	-1.20		\\
		&	0.0428	&	3.1668	&	3.2048	&	-0.0380	&	-1.20		\\
		&	0.0500	&	4.3568	&	4.4090	&	-0.0522	&	-1.20		\\
		&	0.0656	&	7.6928	&	7.7845	&	-0.0917	&	-1.19		\\
		&	0.0774	&	10.9501	&	11.0801	&	-0.1299	&	-1.19		\\
		&	0.0900	&	15.2827	&	15.4629	&	-0.1802	&	-1.18		\\
		&		&		&		&		&			\\
	Rn	&	0.0100	&	0.3210	&	0.3246	&	-0.0036	&	-1.11		\\
		&	0.0200	&	1.2924	&	1.3067	&	-0.0143	&	-1.11		\\
		&	0.0300	&	2.9395	&	2.9720	&	-0.0325	&	-1.10		\\
		&	0.0428	&	6.1134	&	6.1807	&	-0.0673	&	-1.10		\\
		&	0.0500	&	8.4591	&	8.5519	&	-0.0928	&	-1.10		\\
		&	0.0656	&	15.1796	&	15.3447	&	-0.1651	&	-1.09		\\
		&	0.0774	&	21.9529	&	22.1897	&	-0.2368	&	-1.08		\\
		&	0.0900	&	31.2997	&	31.6336	&	-0.3339	&	-1.07		\\
		&		&		&		&		&			\\
	Og	&	0.0100	&	1.4293	&	1.4479	&	-0.0187	&	-1.31		\\
		&	0.0200	&	5.8176	&	5.8936	&	-0.0760	&	-1.31		\\
		&	0.0300	&	13.4833	&	13.6592	&	-0.1759	&	-1.30		\\
		&	0.0428	&	29.0931	&	29.4720	&	-0.3789	&	-1.30		\\
		&	0.0500	&	41.3873	&	41.9256	&	-0.5384	&	-1.30		\\
		&	0.0656	&	80.6199	&	81.6647	&	-1.0448	&	-1.30		\\
		&	0.0774	&	127.1597	&	128.8013	&	-1.6416	&	-1.29		\\
		&	0.0900	&	205.9572	&	208.6017	&	-2.6445	&	-1.28		\\
	\hline												
	\end{tabular}
}
    \caption{Verdet Constant [V($\omega$), in atomic units] for gaseous Xe, Rn and Og at different frequencies $\omega$ [in atomic units], calculated with QR-CCSD (denoted by CC) and QR-EOMCCSD (denoted by EOM) for the $^2$DC$^M$ Hamiltonian upon correlating the $n$s$n$p shells and with virtual threshold at 5 a.u.), as well as the difference in Verdet constant between the two approaches ($\Delta = V^\text{CC} - V^\text{EOM}$) in absolute and relative terms (in \%).}
    \label{tab:verdet-table-scan-nsnp}
\end{table}   

\begin{table}[H]
    \centering
    \setlength{\tabcolsep}{2.2mm}
{
    \begin{tabular}{lrrrrr}
\hline\hline				
 & & \multicolumn{4}{c}{V($\omega$)}\\
\hline
System	&	$\omega$	&	CC	&	EOM	&	$\Delta$ &	\%	\\
\hline											
Xe	&	0.0100	&	0.1654	&	0.1674	&	-0.0021	&	-1.25		\\
	&	0.0200	&	0.6641	&	0.6724	&	-0.0083	&	-1.25		\\
	&	0.0300	&	1.5043	&	1.5231	&	-0.0188	&	-1.25		\\
	&	0.0428	&	3.1039	&	3.1426	&	-0.0387	&	-1.25		\\
	&	0.0500	&	4.2701	&	4.3233	&	-0.0532	&	-1.25		\\
	&	0.0656	&	7.5388	&	7.6323	&	-0.0935	&	-1.24		\\
	&	0.0774	&	10.7297	&	10.8621	&	-0.1324	&	-1.23		\\
	&	0.0900	&	14.9727	&	15.1563	&	-0.1835	&	-1.23		\\
	&		&		&		&		&			\\
Rn	&	0.0100	&	0.3099	&	0.3139	&	-0.0040	&	-1.29		\\
	&	0.0200	&	1.2477	&	1.2637	&	-0.0160	&	-1.29		\\
	&	0.0300	&	2.8375	&	2.8740	&	-0.0364	&	-1.28		\\
	&	0.0428	&	5.9002	&	5.9757	&	-0.0755	&	-1.28		\\
	&	0.0500	&	8.1629	&	8.2671	&	-0.1042	&	-1.28		\\
	&	0.0656	&	14.6423	&	14.8278	&	-0.1855	&	-1.27		\\
	&	0.0774	&	21.1673	&	21.4335	&	-0.2662	&	-1.26		\\
	&	0.0900	&	30.1625	&	30.5382	&	-0.3758	&	-1.25		\\
	&		&		&		&		&			\\
Og	&	0.0100	&	nc	&	nc	&	nc	&	nc		\\
	&	0.0200	&	nc	&	nc	&	nc	&	nc		\\
	&	0.0300	&	nc	&	nc	&	nc	&	nc		\\
	&	0.0428	&	26.6610	&	27.1031	&	-0.4421	&	-1.66		\\
	&	0.0500	&	nc	&	nc	&	nc	&	nc		\\
	&	0.0656	&	73.3725	&	74.5913	&	-1.2187	&	-1.66		\\
	&	0.0774	&	115.0398	&	116.9528	&	-1.9131	&	-1.66		\\
	&	0.0900	&	nc	&	nc	&	nc	&	nc		\\
	\hline												
	\end{tabular}
}
    \caption{Verdet Constant [V($\omega$), in atomic units] for gaseous Xe, Rn and Og at different frequencies $\omega$ [in atomic units], calculated with QR-CCSD (denoted by CC) and QR-EOMCCSD (denoted by EOM) for the $^2$DC$^M$ Hamiltonian upon correlating the $(n-1)$d$n$s$n$p shells and with virtual threshold at 5 a.u.), as well as the difference in Verdet constant between the two approaches ($\Delta = V^\text{CC} - V^\text{EOM}$) in absolute and relative terms (in \%). nc: not calculated}
    \label{tab:verdet-table-scan-n-1dnsnp}
\end{table}

\section{Excitation energies for Xe, Rn and Og}

\begin{table}[H]
    \centering
    \setlength{\tabcolsep}{2.2mm}
{
    \begin{tabular}{ccccccccc}
    \hline
	&		&	\multicolumn{2}{c}{} &	\multicolumn{3}{c}{Exp.\cite{NISTAtomic}}	&	\multicolumn{2}{c}{}	\\
    \cline{5-7}
&	$\Omega$	&	E (eV)(a)	&	E (eV)(b)	&	J	&	E (eV)	&		config.	&	$\Delta$E(a) (eV)&	$\Delta$E(b) (eV)		\\
    \hline
Xe		&	0g	&	0.000	&	0.000	&	0	&	0.000	&		5p$^6$ 	&		&		\\
	&	2u	&	8.076	&	8.101	&	2	&	8.315	&		5p$^5$($^{2}P_{3/2}$)6s	&	-0.239	&	-0.214	\\
	&	1u	&	8.211	&	8.237	&	1	&	8.437	&		5p$^5$($^{2}P_{3/2}$)6s	&	-0.225	&	-0.199	\\
	&	0u	&	9.192	&	9.240	&	0	&	9.447	&		5p$^5$($^{2}P_{1/2}$)6s	&	-0.255	&	-0.208	\\
	&	1u	&	9.323	&	9.372	&	1	&	9.570	&		5p$^5$($^{2}P_{1/2}$)6s	&	-0.247	&	-0.198	\\
	&	1g	&	9.313	&	9.339	&	1	&	9.580	&		5p$^5$($^{2}P_{3/2}$)6p	&	-0.267	&	-0.242	\\
	&	2g	&	9.435	&	9.461	&	2	&	9.686	&		5p$^5$($^{2}P_{3/2}$)6p	&	-0.251	&	-0.225	\\
	&	3g	&	9.464	&	9.492	&	3	&	9.721	&		5p$^5$($^{2}P_{3/2}$)6p	&	-0.257	&	-0.229	\\
	&	1g	&	9.538	&	9.565	&	1	&	9.789	&		5p$^5$($^{2}P_{3/2}$)6p	&	-0.252	&	-0.224	\\
	&	2g	&	9.569	&	9.596	&	2	&	9.821	&		5p$^5$($^{2}P_{3/2}$)6p &	-0.253	&	-0.225	\\
	&		&		&		&		&		&			&		&		\\
Rn	&	0g	&	0.000	&	0.000	&	0	&	0.000	&		6p$^6$  	&		&		\\
	&	2u	&	6.542	&	6.575	&	2	&	6.772	&		6p$^5$($^{2}P_{3/2}$)7s 	&	-0.230	&	-0.197	\\
	&	1u	&	6.724	&	6.763	&	1	&	6.942	&		6p$^5$($^{2}P_{3/2}$)7s 	&	-0.217	&	-0.179	\\
	&	1g	&	7.948	&	8.024	&	1	&	8.213	&		6p$^5$($^{2}P_{3/2}$)7p 	&	-0.265	&	-0.189	\\
	&	2g	&	8.017	&	8.096	&	2	&	8.271	&		6p$^5$($^{2}P_{3/2}$)7p 	&	-0.253	&	-0.175	\\
	&	0u	&	8.179	&	8.259	&	0	&	8.419	&		6p$^5$($^{2}P_{3/2}$)6d 	&	-0.240	&	-0.161	\\
	&	3g	&	8.170	&	8.245	&	3	&	8.436	&		6p$^5$($^{2}P_{3/2}$)7p 	&	-0.266	&	-0.191	\\
	&	1g	&	8.211	&	8.287	&	1	&	8.472	&		6p$^5$($^{2}P_{3/2}$)7p 	&	-0.262	&	-0.185	\\
	&	2g	&	8.268	&	8.345	&	2	&	8.529	&		6p$^5$($^{2}P_{3/2}$)7p 	&	-0.260	&	-0.183	\\
	&	1u	&	8.305	&	8.386	&	1	&	8.541	&		6p$^5$($^{2}P_{3/2}$)6d 	&	-0.236	&	-0.155	\\
	&		&		&		&		&		&			&		&		\\
Og	&	0g	&	0.000	&	0.000			&		&			&		&		\\
	&	2u	&	3.996	&	4.068			&		&			&		&		\\
	&	1u	&	4.383	&	4.466			&		&			&		&		\\
	&	1g	&	5.841	&	5.971			&		&			&		&		\\
	&	2g	&	5.879	&	6.012			&		&			&		&		\\
	&	0u	&	6.392	&	6.502			&		&			&		&		\\
	&	3g	&	6.397	&	6.515			&		&			&		&		\\
	&	1g	&	6.410	&	6.529			&		&			&		&		\\
    \hline
\end{tabular}
}
    \caption{Excitation energies for the some of the lowest-lying electronic states for Xe, Rn and Og, upon correlating the (a) $n$s$n$p and (b) $(n-1)$d$n$s$n$p electrons and virtuals up to 5 a.u.~A comparison to the experimental data from NIST~\cite{NISTAtomic} is provided for Xe and Rn, with $\Delta$E values indicating the difference between our calculations and experiment.}
    \label{tab:excitation-ng}
\end{table}   

\clearpage

\section{Sample timings for quadratic response evaluation}
\begin{table}[H]
    \centering
        \caption{Comparison of timings for the quadratic response evaluation step for the $\beta_{zzz}$ component of the dipole hyperpolarizability (wall time, in hh:mm:ss format) between for QR-CC and QR-EOMCC, and the ratio between them. Calculations with the ${}^2$DC$^M$ Hamiltonian. O: number of correlated electrons, V: number of virtual spinors. QRF: quadratic response function evaluation step; total: all steps (QR, solution of linear systems etc.)} 
    \begin{tabular}{crrlcrr}
    \hline\hline
system & O & V & Method & total walltime  &  ratio  & QRF walltime$^\dagger$\\
\hline
\multicolumn{7}{c}{valence electrons correlated} \\
\hline
HF  &   8 & 196 &  CC    & 0:34:07  &        & 0:13:51 \\    
    &     &     &  EOMCC & 0:18:51	&   0.6  & 0:00:12 \\
    &     &     &        &          &        & \\  
HCl &   8 & 226 &  CC    & 0:37:37  &        & 0:21:36\\    
    &     &     &  EOMCC & 0:16:18  &   0.4  & 0:00:06 \\
    &     &     &        &          &        & \\ 
HBr &   8 & 226 &  CC    & 1:05:55  &        & 0:26:31\\
    &     &     &  EOMCC & 0:39:35  &   0.6  & 0:00:08 \\
\hline
\multicolumn{7}{c}{all electrons correlated} \\
\hline
HF  &  10 & 282 &  CC    & 0:57:44  &        & 0:17:31 \\    
    &     &     &  EOMCC & 0:41:06	&   0.7  & 0:00:11 \\
    &     &     &        &          &        & \\  
HCl &  18 & 300 &  CC    & 1:56:59  &        & 0:28:44\\    
    &     &     &  EOMCC & 1:32:24  &   0.8  & 0:00:33\\
    &     &     &        &          &        & \\
HBr &  36 & 402 &  CC    & 8:36:03  &        & 2:24:50\\
    &     &     &  EOMCC & 6:14:54  &   0.7  & 0:02:36\\
    \hline
    \end{tabular}
    \label{tab:timings-all-electron-zzz}
\end{table}

\clearpage

\bibliography{SortedBiblio.bib}